\documentclass{emulateapj}

\usepackage{apjfonts}
\usepackage{amsmath}
\usepackage{graphicx}
\newcommand{\etal}{et al.}
\newcommand{\hbeta}{H{$\beta$}}
\newcommand{\halpha}{H{$\alpha$}}
\newcommand{\CIV}{C{\sevenrm IV}}

\newcommand{\MgII}{Mg{\sevenrm II}}

\newcommand{\lgL}{\log \left({L_{\rm bol} \over {\rm erg\,s^{-1}}}\right)}

 \font\sevenrm=cmr7 scaled 1000

\slugcomment{Submitted to ApJ on September 19, 2007}

\begin{document}
\title{Biases in Virial Black Hole Masses: An SDSS Perspective}


\shorttitle{QUASAR BLACK HOLE MASSES}

\shortauthors{SHEN ET AL.}

\author{Yue Shen\altaffilmark{1}, Jenny E. Greene\altaffilmark{1,2}
, Michael A. Strauss\altaffilmark{1}, Gordon T.
Richards\altaffilmark{3}, Donald
  P. Schneider\altaffilmark{4}}

\altaffiltext{1}{Princeton University Observatory, Princeton, NJ
08544.}

\altaffiltext{2}{Hubble Fellow and Princeton-Carnegie Fellow.}

\altaffiltext{3}{Department of Physics, Drexel University, 3141
Chestnut Street, Philadelphia, PA 19104.}

\altaffiltext{4}{Department of Astronomy and Astrophysics, 525
Davey Laboratory, Pennsylvania State University, University Park,
PA 16802.}

\begin{abstract}
We compile black hole (BH) masses for $\sim 60,000$ quasars in the
redshift range $0.1 \lesssim z \lesssim 4.5$ included in the Fifth
Data Release of the Sloan Digital Sky Survey (SDSS), using virial
BH mass estimators based on the \hbeta, \MgII, and \CIV\ emission
lines. We find that: (1) within our sample, the widths of the
three lines follow log-normal distributions, with means and
dispersions that do not depend strongly on luminosity or redshift;
(2) the \MgII- and \hbeta-estimated BH masses are consistent with
one another; and (3) the \CIV\ BH mass estimator may be more
severely affected by a disk wind component than the \MgII\ and
\hbeta\ estimators, giving a positive bias in mass correlated with
the \CIV-\MgII\ blueshift. Most SDSS quasars have virial BH masses
in the range $10^8-10^{10}\ M_\odot$.  There is a clear upper mass
limit of $\sim 10^{10}\ M_\odot$ for active BHs at $z \gtrsim 2$,
decreasing at lower redshifts. Making the reasonable assumptions
that the underlying BH mass distribution decreases with mass and
that the Eddington ratio distribution at fixed true BH mass has
non-zero width, we show that the measured virial BH mass
distribution and Eddington ratio distribution within finite
luminosity bins are subject to Malmquist bias if the scatter in
luminosity at fixed true BH mass is uncorrelated with the scatter
in line width. Given the current versions of virial calibrations
and their uncertainties, we present a model which reproduces the
observed virial mass distribution, quasar luminosity function, and
line width distribution of our sample; it has an underlying BH
mass distribution which is a power-law with slope $\gamma_M\sim
-2.6$, and a true Eddington ratio distribution at fixed BH mass
which is a log-normal with mean dependent on BH mass ($\sim
10^{-1.2}$ for typical $10^8\ M_\odot$ BHs) and with dispersion
$0.4$ dex. In this model, the observed virial mass distribution
for the SDSS sample is biased high by $\sim 0.6$ dex within finite
luminosity bins, and the Eddington ratio distribution is biased
low by the same amount. A radio quasar subsample (with
$1.5\lesssim z\lesssim 2.3$) has mean virial BH mass larger by
$\sim 0.12$ dex than the radio-quiet sample matched in luminosity
and redshift. A broad absorption line (BAL) quasar subsample (with
$1.7\lesssim z\lesssim 2.2$) has a virial mass distribution
identical to that of the non-BAL quasar sample matched in
luminosity and redshift, with no mean offset.
\end{abstract}

\section{Introduction}
There has been increasing interest in recent years in the role that
supermassive black holes (SMBHs) play in galaxy formation, primarily
because of the discovery that most, if not all, present-day
massive galaxies harbor a SMBH in their nuclei (e.g., Kormendy \&
Richstone 1995; Richstone \etal\ 1998), and that the mass of the
nuclear SMBH is related to the bulge mass/luminosity (e.g.,
Magorrian \etal\ 1998) and even more tightly to the bulge stellar
velocity dispersion (e.g., Ferrarese \& Merritt 2000; Gebhardt
\etal\ 2000a; Tremaine \etal\ 2002). These tight correlations imply
that the formation of galaxies and the growth of the central SMBHs
are ultimately connected (e.g., Silk \& Rees 1998; Kauffmann \&
Haehnelt 2000; Wyithe \& Loeb 2003; Di Matteo \etal\ 2005; Hopkins
\etal\ 2006).

On the other hand, it has long been suggested that active galactic
nuclei (AGN) or quasars are SMBHs in the process of accretion and
growth (e.g., Salpeter 1964; Zel'dovich \& Novikov 1964;
Lynden-Bell 1969). Indeed, the integrated luminosity density of
optically-selected AGNs, which represents the accretion history of
black holes, is consistent with the mass density in the local
dormant SMBH population (the So{\l}tan [1982] argument; e.g.,
Salucci \etal\ 1999; Yu \& Tremaine 2002). This is particularly
encouraging, since with ever larger AGN/quasar samples in modern
surveys, we can hope to understand the cosmic evolution of SMBHs
within the framework of hierarchical structure formation, and shed
light on the coevolution of SMBHs and their host galaxies. A
central issue in this regard is to measure the masses of both
inactive SMBHs and active AGNs/quasars.

Perhaps the most reliable way to measure the black hole mass is
via stellar/gas dynamics in the vicinity of the SMBH where its
gravity dominates the dynamics (e.g., Richstone \etal\
1998), but the relevant scales can be resolved only for the
nearest galaxies. A more indirect mass indicator uses the observed
tight correlations between the SMBH mass $M_{\rm BH}$ and the
stellar bulge velocity dispersion $\sigma$ (the $M-\sigma$
relation), or the bulge luminosity (the $M-L$ relation). Accurate
measurements of the bulge velocity dispersion or bulge luminosity
are difficult in luminous quasars, where the AGN light overwhelms
that of the host galaxy; these relations are mostly applied to
measure the local dormant SMBH mass function. More seriously, the
$M-\sigma$ and $M-L$ relations have not been measured directly at
the most massive end, and the high-mass end of the BH mass
functions estimated using the $M-\sigma$ and $M-L$ relations
differ by as much as one order of magnitude (Tundo \etal\ 2007;
Lauer \etal\ 2007). Thus there are still unsettled issues on the
usage of these techniques.

A third method, reliant on AGN physics rather than galaxy
properties, is reverberation mapping, which uses the temporal offset
between continuum and emission line variability to determine the
distance $R$ from the central engine to the broad emission line
region (BLR; e.g., Blandford \& McKee 1982; Peterson
1993; Kaspi \etal\ 2000; Peterson \etal\ 2004). Using the observed
line width $V$ and assuming that the BLR is virialized allows
determination of the black hole mass: $M_{\rm BH}\backsimeq
G^{-1}RV^2$. In a handful of cases, measurements of $V$ and $R$
for several lines in the same object have shown $V\propto
R^{-1/2}$, consistent with the virial hypothesis (e.g., Peterson
\& Wandel 2000). Furthermore, for the few cases in which we have
more than one mass indicator, the reverberation mapping mass is
also consistent with the dynamical mass or that derived from the
$M_{\rm BH}-\sigma$ relation (e.g., Gebhardt \etal\ 2000b;
Ferrarese \etal\ 2001; Nelson \etal\ 2004; Onken \etal\ 2004,
2007; Greene \& Ho 2006; Davies \etal\ 2006). The reverberation
mapping method is very time consuming, and we have reliable masses
with this method for only about three dozen AGNs.

However, reverberation mapping data has revealed a correlation
between BLR size $R$ and luminosity (Kaspi \etal\ 2000, 2005),
which allows us to estimate BH masses based on single-epoch
spectra. Using this $R-L$ correlation and reverberation mapping
masses, various empirical scaling relations (which we term virial
BH mass estimators from now on) have been derived using \hbeta\
(Kaspi \etal\ 2000; Vestergaard 2002; McLure \& Jarvis 2002;
Vestergaard \& Peterson 2006), \halpha\ (Greene \& Ho 2005),
\MgII\ (McLure \& Jarvis 2002; McLure \& Dunlop 2004), and \CIV\
(e.g., Vestergaard 2002; Vestergaard \& Peterson 2006). This is
the only practical method for measuring black hole masses for
large numbers of objects at high redshifts and luminosities (where
variability timescales are too long to measure easily; see the
discussion in Kaspi \etal\ 2007), but the $R-L$ relation has only
been established in the luminosity range $\sim 10^{42}-10^{46}
{\rm\ erg\ s^{-1}}$ and for redshifts $z<0.3$.

The virial BH mass estimators are usually expressed as:
\begin{equation}\label{eqn:virial_estimator}
\log \left({M_{\rm BH,vir} \over M_\odot}\right)
=a+b\log\left({\lambda L_{\lambda} \over 10^{44}\,{\rm
erg\,s^{-1}}}\right)+2\log\left({\rm FWHM\over km\,s^{-1}}\right)\
,
\end{equation}
where $\lambda L_{\lambda}$, the surrogate for the BLR size $R$,
is the continuum luminosity near the emission line (5100\AA\ for
\hbeta, 3000\AA\ for \MgII\ and 1350\AA\ for \CIV), and FWHM is
the full-width-at-half-maximum of the line. Some authors prefer to
use other quantities in their virial estimates. For example, Wu
\etal\ (2004) suggest using recombination line luminosities
instead of continuum luminosities because the latter might be
contaminated by jet emission or host galaxy starlight (also see
Greene \& Ho 2005). Peterson \etal\ (2004) and Collin \etal\
(2006) argue that the second moment of the line profile
$\sigma_{\rm line}$ is a better quantity to characterize the
emission-line widths than the FWHM. As we will show below,
different measurement techniques yield systematic differences in
derived quantities, and so it is important to use the original
definitions of line widths and luminosities for whichever
calibration is used.

The UV \CIV\ virial mass estimator can be used for $z\gtrsim 2$
quasars, as the \CIV\ line is the only relevant line available in
optical spectra in this redshift range. However, it has been
suggested by a few authors that the \CIV\ line might be a less
secure mass estimator than \MgII\ or \hbeta. The \CIV\ line tends
to be asymmetric and blueshifted with respect to lower-ionization
lines such as \hbeta\ or \MgII\ (e.g., Gaskell 1982; Tytler \& Fan
1992), and the most blueshifted objects tend to have the largest
FWHMs (Richards \etal\ 2002b). These features suggest that \CIV\
might be more severely affected than the other lines by a
non-virialized gas component, and the measured line width could
depend on the viewing angle, biasing the BH mass estimates.
Indeed, using a sample of $\sim 80$ $z\le 0.5$ Palomar-Green
(Green, Schmidt, \& Liebert 1986) quasars, Baskin \& Laor (2005)
showed that the \CIV\ FWHM is poorly correlated with the \hbeta\
FWHM (e.g., see their figure 3), suggesting that the two lines
have different origins. On the other hand, the \MgII\ FWHM is
well-correlated with that of \hbeta, and these two low-ionization
line estimators usually give consistent virial masses (e.g.,
McLure \& Dunlop 2004; Salviander \etal\ 2007). One of the
purposes of this paper is to explore how the \CIV\ estimator might
be biased relative to low-ionization line estimators such as
\MgII\ using a large sample of quasars from the Sloan Digital Sky
Survey (SDSS; York \etal\ 2000).

Despite the caveats in these virial estimators, virial BH masses
have been measured for various AGN/quasar samples covering a wide
range of redshifts and luminosities. For example, McLure \& Dunlop
(2004) measured virial BH masses for $12698$ quasars with $0.1\le
z\le 2.1$ from the SDSS DR1 quasar catalog (Schneider \etal\ 2003)
using \hbeta\ and \MgII. Vestergaard (2004) used \hbeta\ and \CIV\
to measure virial BH masses for a hybrid sample including 87 $z\le
0.5$ Bright Quasar Survey (BQS, Schmidt \& Green 1983) quasars,
114 $1.5\lesssim z\lesssim 3.5$ quasars, and $\sim 150$ $z>3.5$
SDSS quasars. Greene \& Ho (2007) present the BH mass function for
$\sim 8500$ broad-line AGNs with $z < 0.35$ from the SDSS. Fine
\etal\ (2006) used composite spectra to measure the redshift
evolution of the mean BH mass for the 2QZ quasar sample (Croom
\etal\ 2004) from $z\sim 0.5$ to $z\sim 2.5$, and Kollmeier \etal\
(2006) presented virial BH masses for a sample of 407 AGN with
$z\sim 0.3-4$ selected from the AGN and Galaxy Evolution Survey
(AGES; Kochanek \etal\ 2004) and focused on the distribution of
Eddington ratios. Yet by far the largest quasar sample is the
recently published SDSS DR5 quasar catalog (Schneider \etal\
2007), containing more than $77,000$ quasars, about half of which
are homogeneously selected (e.g., Richards \etal\ 2006a; Shen
\etal\ 2007).   In this paper we present virial BH mass estimates
for this optical quasar sample and explore possible biases in
their estimation. Though the dynamical range in luminosity and
color of SDSS quasars is limited, the large size of the sample
provides unprecedent statistics.

The paper is organized as follows: in \S~\ref{sec:sample} we
describe our quasar sample.  We present our procedures of
estimating virial BH masses in \S~\ref{sec:BH_esti}, where we
compare results from different lines in detail.  The distributions
of black hole masses and Eddington ratios are described in
\S~\ref{sec:result}, and we discuss our results as well as some
general issues with virial estimators in \S\ref{sec:discussion}.
We summarize our results in \S~\ref{sec:conclusion}. Throughout
this paper we adopt a flat $\Lambda$CDM cosmology:
$\Omega_M=0.26$, $\Omega_\Lambda=0.74$ and $h=0.71$ (Spergel
\etal\ 2007).

\section{The Sample}\label{sec:sample}
The SDSS uses a dedicated 2.5-m wide-field telescope (Gunn \etal\
2006) with a drift-scan camera with 30 $2048 \times 2048$
CCDs (Gunn \etal\ 1998) to image the sky in five broad bands
($u\,g\,r\,i\,z$; Fukugita \etal\ 1996).  The imaging data are
taken on dark photometric nights of good seeing (Hogg \etal\
2001), are calibrated photometrically (Smith \etal\ 2002; Ivezi\'c
\etal\ 2004; Tucker \etal\ 2006) and astrometrically (Pier \etal\
2003), and object parameters are measured (Lupton \etal\ 2001;
Stoughton \etal\ 2002). Quasar candidates (Richards \etal\ 2002a)
for follow-up spectroscopy are selected from the imaging data
using their colors, and are arranged in spectroscopic plates
(Blanton \etal\ 2003) to be observed with a pair of double
spectrographs. The quasars observed through the Third Data Release
(Abazajian \etal\ 2005) have been cataloged by Schneider \etal\
(2005), while Schneider \etal\ (2007) extend this catalog to the
DR5 (Adelman-McCarthy \etal\ 2007a).

Our parent sample is the published SDSS DR5 quasar catalog
(Schneider \etal\ 2007), which contains 77,429 bona fide quasars
that have luminosities larger than $M_i=-22.0$ (using a slightly
different cosmology in that paper) and have at least one broad
emission line (${\rm FWHM}>1000 {\rm\ km\ s^{-1}}$) or have
interesting/complex absorption features. About half of the quasars
in this catalog were selected from a uniform algorithm (as
described in Richards \etal\ 2002a), which is flux limited
to\footnote{There are a few $i>19.1$ quasars at $z\lesssim 3$
which were selected by the high-z ($griz$) branch of the targeting
algorithm (Richards \etal\ 2002b). The fraction of these objects
is tiny ($\lesssim 2\%$). Although these objects have ${\rm
UNIFORM\ FLAG}=1$ in Table 1, they will be rejected when
constructing a flux-limited sample (see \S\ref{sec:MC}).} $i=19.1$
at $z\lesssim 3$ and $i=20.2$ at $z\gtrsim 3$ (magnitudes are
corrected for Galactic extinction using the Schlegel, Finkbeiner
\& Davis 1998 map). These objects are flagged with the ${\rm
UNIFORM\ FLAG}=1$ in the DR5 quasar catalog, and they can be used
to construct a statistically homogeneous sample (e.g., Richards
\etal\ 2006a; Shen \etal\ 2007).

There are also flags indicating whether or not a quasar is
detected in the {\em FIRST} (Becker \etal\ 1995) or {\em ROSAT}
(Voges \etal\ 1999) surveys, which we will use to define our radio
quasar subsample (see \S\ref{sec:BAL}).  Most of our analysis will
avoid broad absorption line quasars (BALs), whose line widths are
problematic to measure; we will come back to the BALs in
\S\ref{sec:BAL}.  We have identified $\sim 4200$ BALs (with
$1.7\le z\le 4.2$ for \CIV\ and $0.5\le z\le 1.9$ for \MgII; Shen
\etal\ 2008) in the
DR5 quasar catalog using traditional BAL criteria (Weymann \etal\
1991). This list of BALs is by no means complete, and an official
DR5 BAL catalog is forthcoming (Hall et al., in preparation; see
Trump \etal\ 2006 for the DR3 BAL catalog).

All spectra are reduced with the new version of the SDSS
spectroscopic reduction pipeline, as described in the DR6 paper
(Adelman-McCarthy \etal\ 2007b); the flux scale of these spectra
is higher than that of previous releases by roughly 38\%. Thus in
this paper all luminosities measured from spectra will be
typically larger by $\sim 0.14$ dex than previous values for the
same objects.

A comprehensive catalog of our measurements of the spectral
properties of each quasar is given in the electronic version of
this paper; the columns are described in Table 1. This catalog can
be regarded as an extension to the published DR5 quasar catalog.

\begin{deluxetable*}{lcl}
\tablecolumns{3}\tablewidth{0.8\textwidth} \tablecaption{Catalog
Format} \tablehead{Column & Format & Description} \startdata
1  & A18   & SDSS DR5 designation hhmmss.ss$+$ddmmss.s (J2000.0)\\
2  & F11.6 & Right ascension in decimal degrees (J2000.0) \\
3  & F11.6 & Declination in decimal degrees (J2000.0) \\
4  & F7.4  & Redshift \\
5  & F8.3  & $M_i (z=2)$ ($h=0.71$, $\Omega_M=0.26$,
$\Omega_\Lambda=0.74$, $K$-corrected to $z=2$, following Richards
\etal\ 2006a)\\
6  & F7.3  & Bolometric luminosity ($\log (L_{\rm bol}/\rm erg\,
s^{-1})$)\\
7  & I5    & Spectroscopic plate number \\
8  & I5    & Spectroscopic fiber number\\
9  & I6    & MJD of spectroscopic observation\\
10  & I12   & Target selection flag when the spectrum was taken
(i.e., using TARGET photometry)\\
11  & I3    & {\em FIRST} selection flag (0 or 1)\\
12 & I3    & {\em ROSAT} selection flag (0 or 1)\\
13 & I3    & Uniform selection flag (0 or 1)\\
14 & I3    & BAL flag (0 or 1)\\
15 & I7    & \hbeta\ FWHM (${\rm km\ s^{-1}}$)\\
16 & F8.3  & Monochromatic luminosity $\lambda L_\lambda$ at
$5100$\AA\
($10^{44}{\rm\ erg\ s^{-1}}$)\\
17 & F7.3  & Virial BH mass estimated using \hbeta\ ($\log
(M_{\rm BH,vir}/M_\odot)$)\\
18 & I7    & \MgII\ FWHM (${\rm km\ s^{-1}}$)\\
19 & F8.3  & Monochromatic luminosity $\lambda L_\lambda$ at
$3000$ \AA\
($10^{44}{\rm\ erg\ s^{-1}}$)\\
20 & F7.3  & Virial BH mass estimated using \MgII\ ($\log
(M_{\rm BH,vir}/M_\odot)$)\\
21 & I7    & \CIV\ FWHM (${\rm km\ s^{-1}}$)\\
22 & F8.3  & Monochromatic luminosity $\lambda L_\lambda$ at
$1350$\AA\
($10^{44}{\rm\ erg\ s^{-1}}$)\\
23 & F7.3  & Virial BH mass estimated using \CIV\ ($\log
(M_{\rm BH,vir}/M_\odot)$)\\
24 & F7.3 & Virial BH mass (using \hbeta\ for $z<0.7$; \MgII\ for
$0.7<z<1.9$ and \CIV\ for $z>1.9$)\\
25 & I7 & \CIV-\MgII\ blueshift (${\rm km\ s^{-1}}$)\\
26 & F8.3 & Mean spectrum S/N (signal-to-noise ratio)\\
\enddata
\tablecomments{(1) Objects in this table are in the same order as
in the DR5 quasar catalog (Schneider \etal\ 2007); (2)
$K-$corrections are the same as in Richards \etal\ (2006a); (3)
Bolometric luminosities are computed using bolometric corrections
in Richards \etal\ (2006b) using one of the $5100$\AA, $3000$\AA,
or $1350$\AA\ monochromatic luminosities depending on redshift;
(4) Entries reading $-9999$ for FWHM, luminosity, BH mass or
\CIV-MgII\ blueshift measurement indicate that this quantity was
not measureable from the SDSS spectrum, either because it fell
outside of the SDSS spectral coverage or because of low S/N. }
\end{deluxetable*}

\section{Black hole mass estimates}\label{sec:BH_esti}
To estimate BH masses using virial estimators one needs two
parameters: the width of an emission line and the corresponding
continuum luminosity. However, different authors have used
different definitions of line width in their calibrations of the
virial mass estimators. In what follows, we take care to define
line width in a way consistent with each calibration we use. We
also note that the assumed virial coefficient $f$, which accounts
for our ignorance of the BLR geometry, is different in different
calibrations. For instance, $f=1$ in McLure \& Dunlop (2004) and
$f\sim 5.5$ (e.g., Onken \etal\ 2004) in Vestergaard \& Peterson
(2006). Thus different versions of calibration for the same line
do not necessarily yield the same results, especially when
extrapolated to high luminosity quasars. We will discuss this further
in \S\ref{sec:disc1}.

We focus on nonBALs throughout this section and
\S\ref{sec:BH_z_eov}-\ref{sec:mass_fun} because of the ambiguities
of line width measurement for BALs; we return to the issue of BALs
in \S\ref{sec:BAL}.

\subsection{Line widths and continuum luminosities}

\begin{figure}
  \centering
    \includegraphics[width=0.45\textwidth]{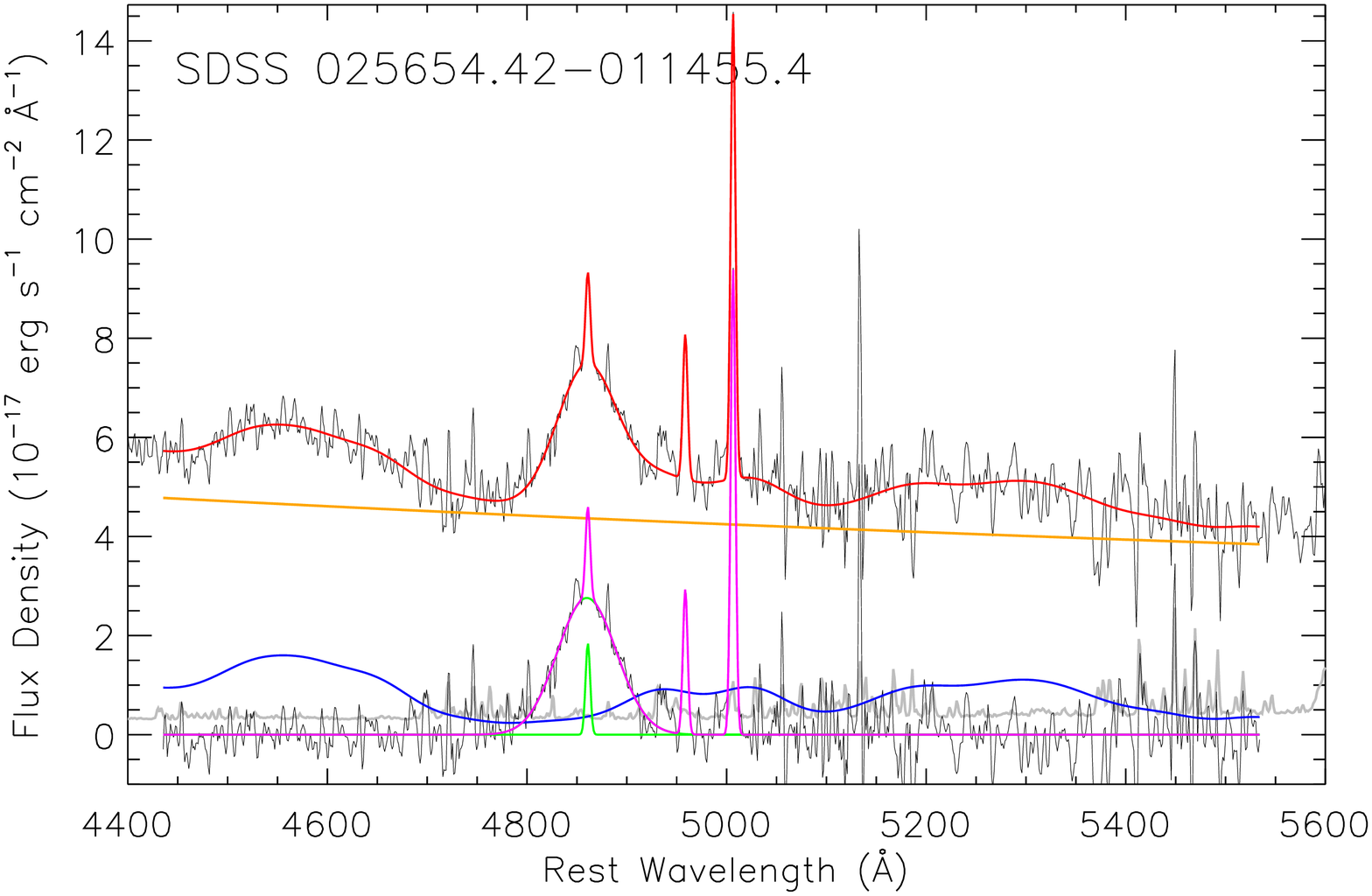}
    \includegraphics[width=0.45\textwidth]{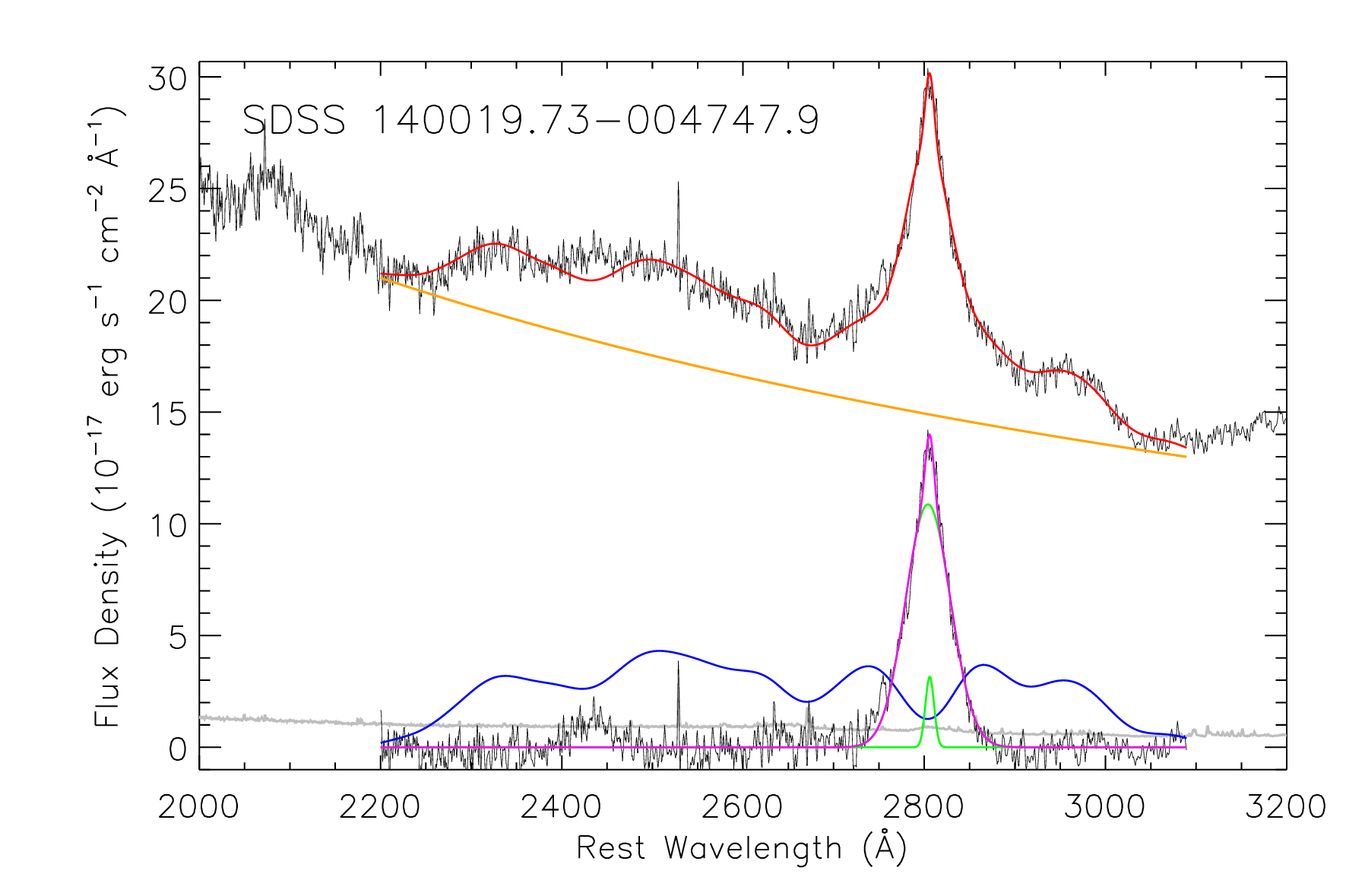}
    \includegraphics[width=0.45\textwidth]{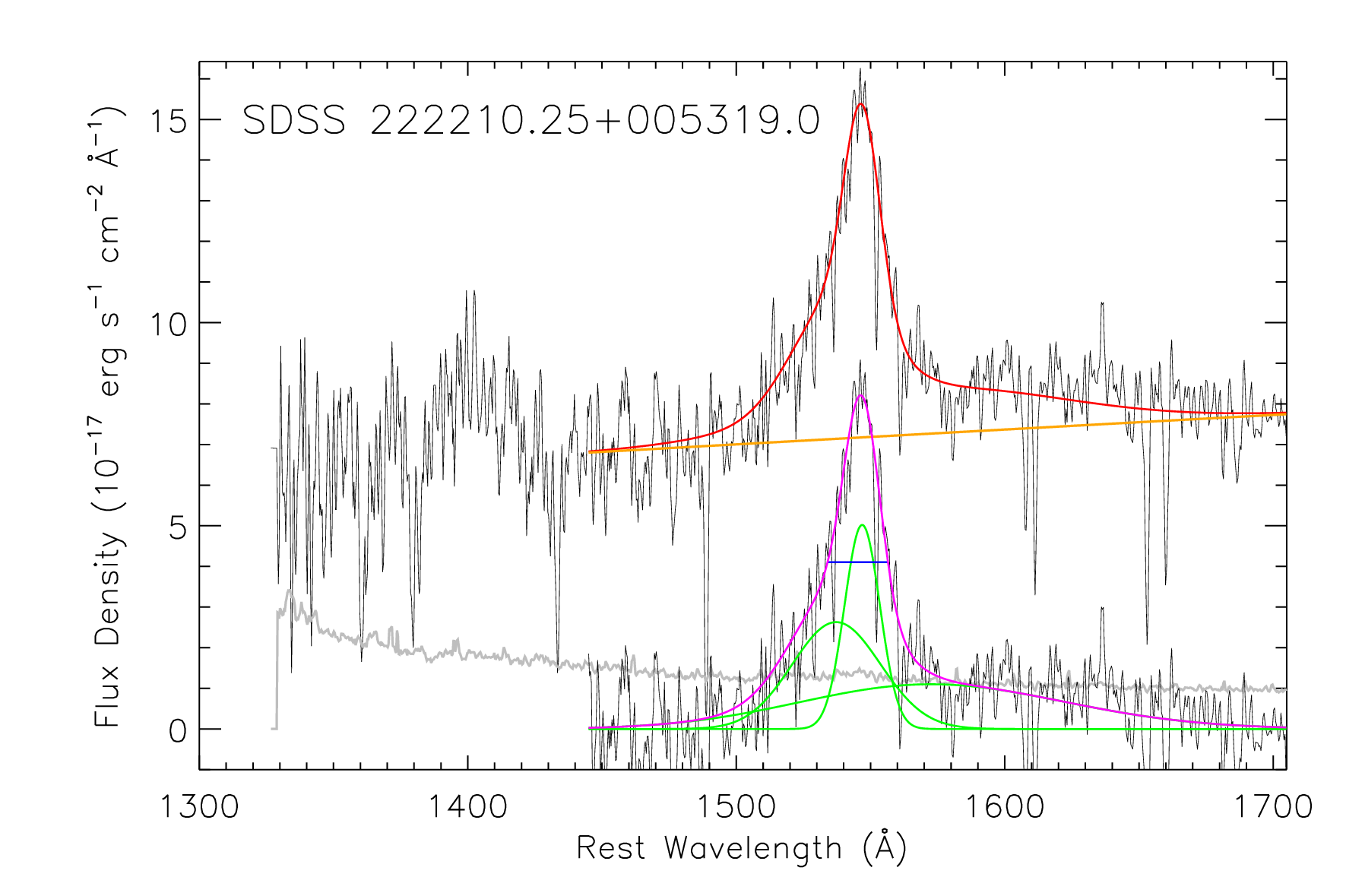}
    \caption{Examples of spectral fits. From top to bottom: \hbeta, \MgII, \CIV. In each panel,
     the upper and lower black solid lines shows the original and continuum$+$Fe subtracted spectra;
     the gray lines show the flux density errors; the red lines show the full fits; the orange lines show
     the fitted power-law continuum; the lower blue lines in the upper and middle panels show
      the iron template fits; the magenta lines show the fits for the emission lines and green lines show
      the Gaussian components. The blue horizontal segment in the bottom panel marks the FWHM for \CIV.}
    \label{fig:linefits}
\end{figure}

\subsubsection{\hbeta\ and \MgII}
For the \hbeta\ and \MgII\ estimators, we have adopted the
calibrations in McLure \& Dunlop (2004), hence $a=0.672$, $b=0.61$
for \hbeta\ and $a=0.505$, $b=0.62$ for \MgII\ in equation
(\ref{eqn:virial_estimator}). The \hbeta\ estimator has been
calibrated directly with reverberation mapping, while the $R_{\rm
BLR}-\lambda L_\lambda$ relation for \MgII\ at 3000\AA\ is
empirically determined in McLure \& Jarvis (2002) based on \hbeta\
reverberation mapping masses. We note that the calibration for
\MgII\ in McLure \& Dunlop (2004) is an updated version of that in
McLure \& Jarvis (2002), where in the former only objects with
$10^{44}{\rm\ erg\ s^{-1}}<\lambda L_\lambda<10^{47}{\rm\ erg\
s^{-1}}$ were included in their fitting. This luminosity range is
suitable for the SDSS quasar sample studied here.

We follow McLure \& Dunlop (2004, Appendix A) to measure the line FWHM
for \hbeta\ and \MgII\ with slight changes in details:
\begin{enumerate}
\item[1)] a power-law continuum
and an iron emission line template are simultaneously fitted to
the spectrum with the \hbeta\ or \MgII\ emission line regions
excluded. The fitting ranges are [4435,4700]\AA\ and
[5100,5535]\AA\ for \hbeta, [2200,2675]\AA\ and [2925,3090]\AA\
for \MgII. The iron template for \hbeta\ is taken from Boroson \&
Green (1992) and the iron template for \MgII\ is a modified version
of the Vestergaard \& Wilkes (2001) template (Salviander \etal\
2007) which extends under the \MgII\ line itself. In the fitting,
the normalization and velocity broadening of the iron template are
left as free parameters.

\item[2)] The best-fit continuum and iron emission are then
subtracted from the spectrum, and two Gaussians are fitted to each
emission line, one for the broad line component and the other for
the narrow line component. The line fitting ranges are
[4700,5100]\AA\ for \hbeta\ and [2700,2900]\AA\ for \MgII. The
FWHM of the narrow component is constrained to be less than 1200
${\rm km\ s^{-1}}$, and that of the broad component is constrained
to be larger than 1200 km s$^{-1}$ (e.g., Hao \etal\ 2005). In the
case of \hbeta, two additional Gaussians whose FWHMs are tied to
that of the narrow \hbeta\ component are fitted simultaneously for
{[O{\sevenrm III}]\ }4959\AA\ and {[O{\sevenrm III}]\ }5007\AA.
The FWHM of the broad Gaussian-component of \hbeta\ and \MgII\ is
then taken as the line width to be inserted in equation
(\ref{eqn:virial_estimator}).
\end{enumerate}

It is not entirely clear if there exists a strong narrow line
component for \MgII, thus the way to measure the \MgII\ FWHM in
McLure \& Dunlop (2004) and here is not justified. If we measure
the \MgII\ FWHM from the two-Gaussian fitted flux, it will
typically be smaller by $\sim 0.15$ dex. However, in order to use
their calibration we have to follow the same procedure to measure
the \MgII\ FWHM. As we will see later in \S\ref{sec:hbeta_MgII},
the fitted broad line FWHMs for \MgII\ and \hbeta\ are quite
similar, and as a result, both line estimators yield consistent
virial masses.

For continuum luminosities $\lambda L_\lambda$ at 5100\AA\ for
\hbeta\ and at 3000\AA\ for \MgII, we have used the fitted
continuum flux density at the corresponding wavelength and
corrected for Galactic extinction using the Schlegel, Finkbeiner
\& Davis (1998) map.

The calibration of the \hbeta\ mass estimator is not unique, and
other authors have determined slightly different versions (e.g.,
Vestergaard 2002; Vestergaard \& Peterson 2006). We find other
forms of the \hbeta\ estimator yield comparable but not identical
results. In particular, the Vestergaard \& Peterson (2006) \hbeta\
calibration gives a systematic $\sim 0.15$ dex offset compared
with the McLure \& Dunlop (2004) calibration. We will come back to
this point in \S\ref{sec:discussion}.

\begin{figure*}
  \centering
    \includegraphics[width=0.45\textwidth]{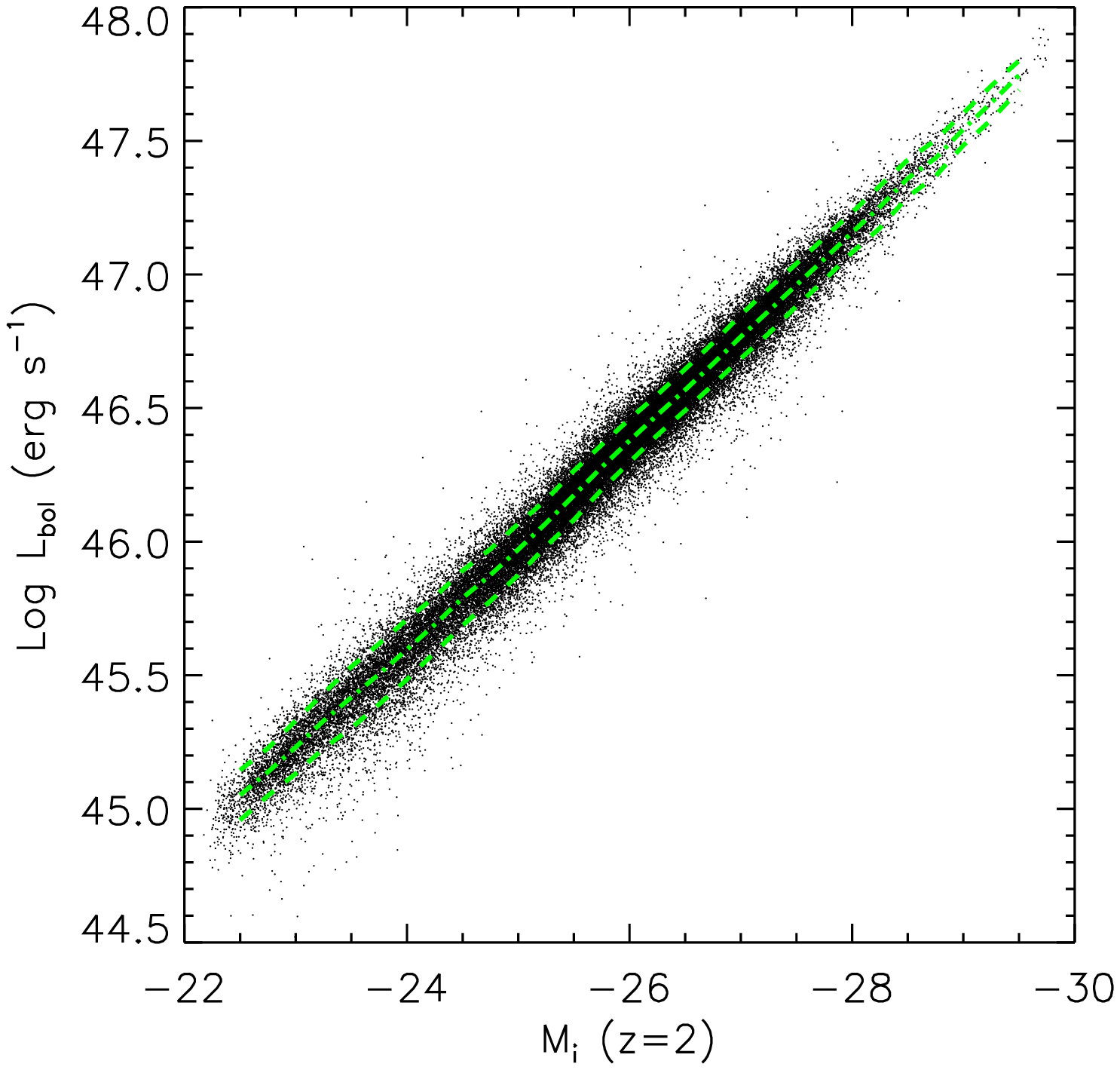}
    \includegraphics[width=0.45\textwidth]{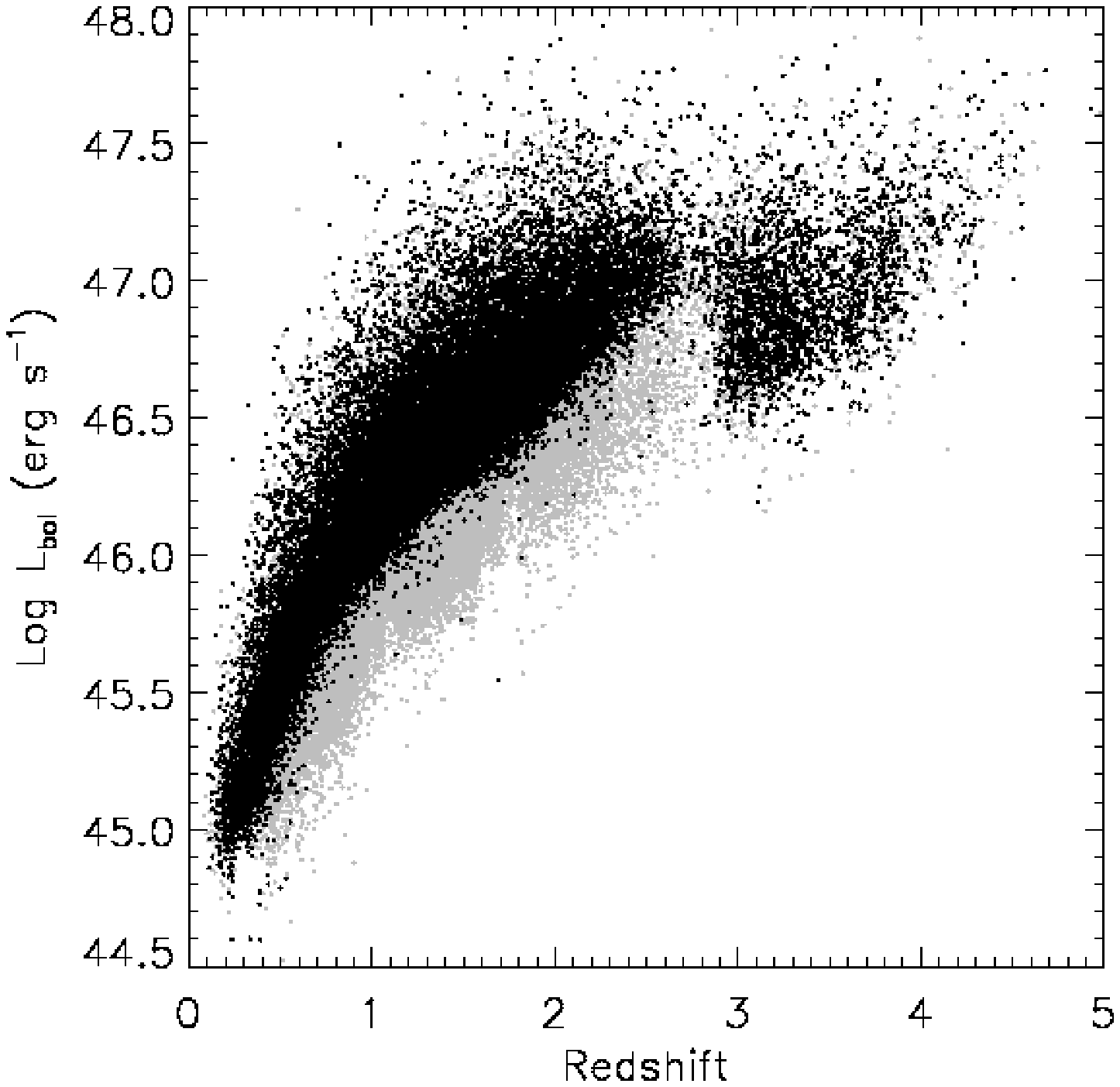}
    \caption{{\em Left}: correspondence between $M_{i} (z=2)$ and
      $L_{\rm bol}$. Lines show median values and standard
      deviations.  The absolute magnitude has been K-corrected to the
      $i$-band measured at $z=2$ (i.e., a rest-frame wavelength of
      $\sim 2500$ \AA), following Richards \etal\ (2006a).  {\em
    Right}: redshift distribution
    of $L_{\rm bol}$ for our sample, where black points are uniformly selected quasars while gray points
    are quasars selected with special targeting algorithms.}
    \label{fig:L_bol}
\end{figure*}

\begin{figure*}
  \centering
    \includegraphics[width=0.45\textwidth]{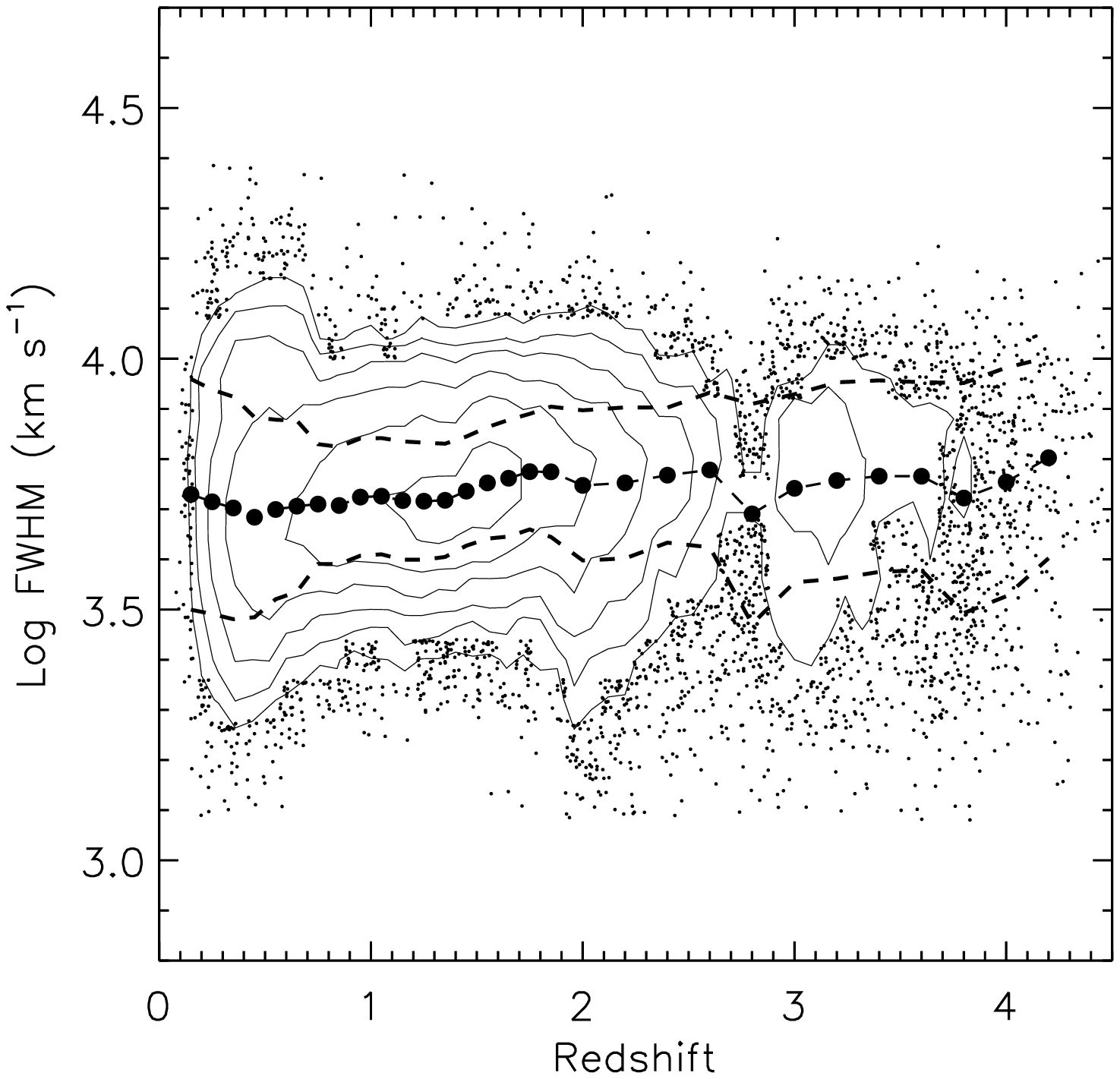}
    \includegraphics[width=0.45\textwidth]{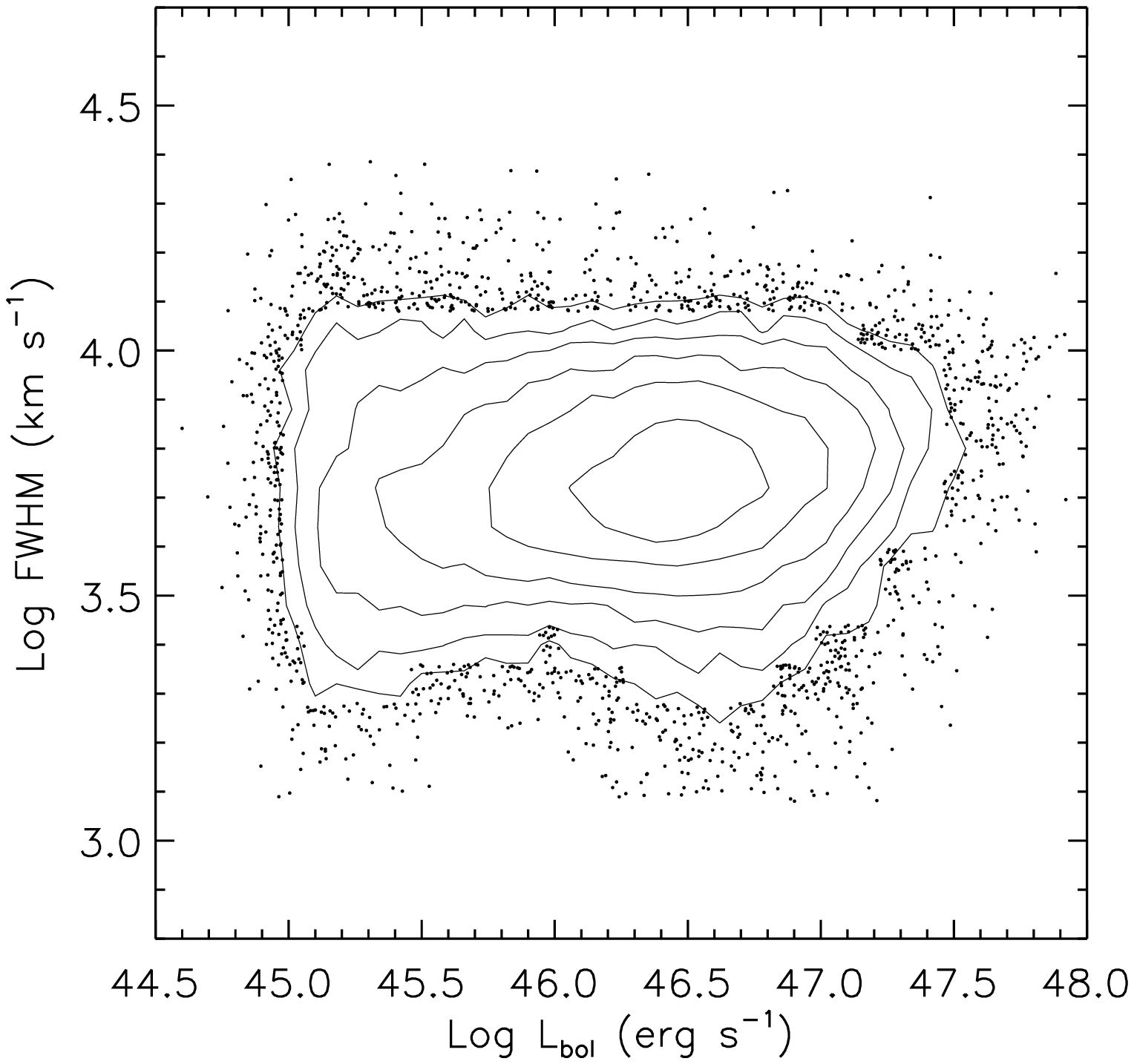}
    \caption{Distributions of FWHMs as functions of redshift and luminosity. The FWHMs for the three lines
     are plotted together. Contours are local point density contours, drawn to guide the eye. In the left panel, we also
     show the mean and $1-\sigma$ of the FWHM distribution as a
     function of redshift. FWHMs show little or no dependence on
     either redshift or luminosity, and their distributions follow a log-normal at all redshifts
     and luminosities (see Fig.~\ref{fig:FWHM_dist_more}).}
    \label{fig:FWHM_dist}
\end{figure*}

\begin{figure*}
  \centering
    \includegraphics[width=1\textwidth]{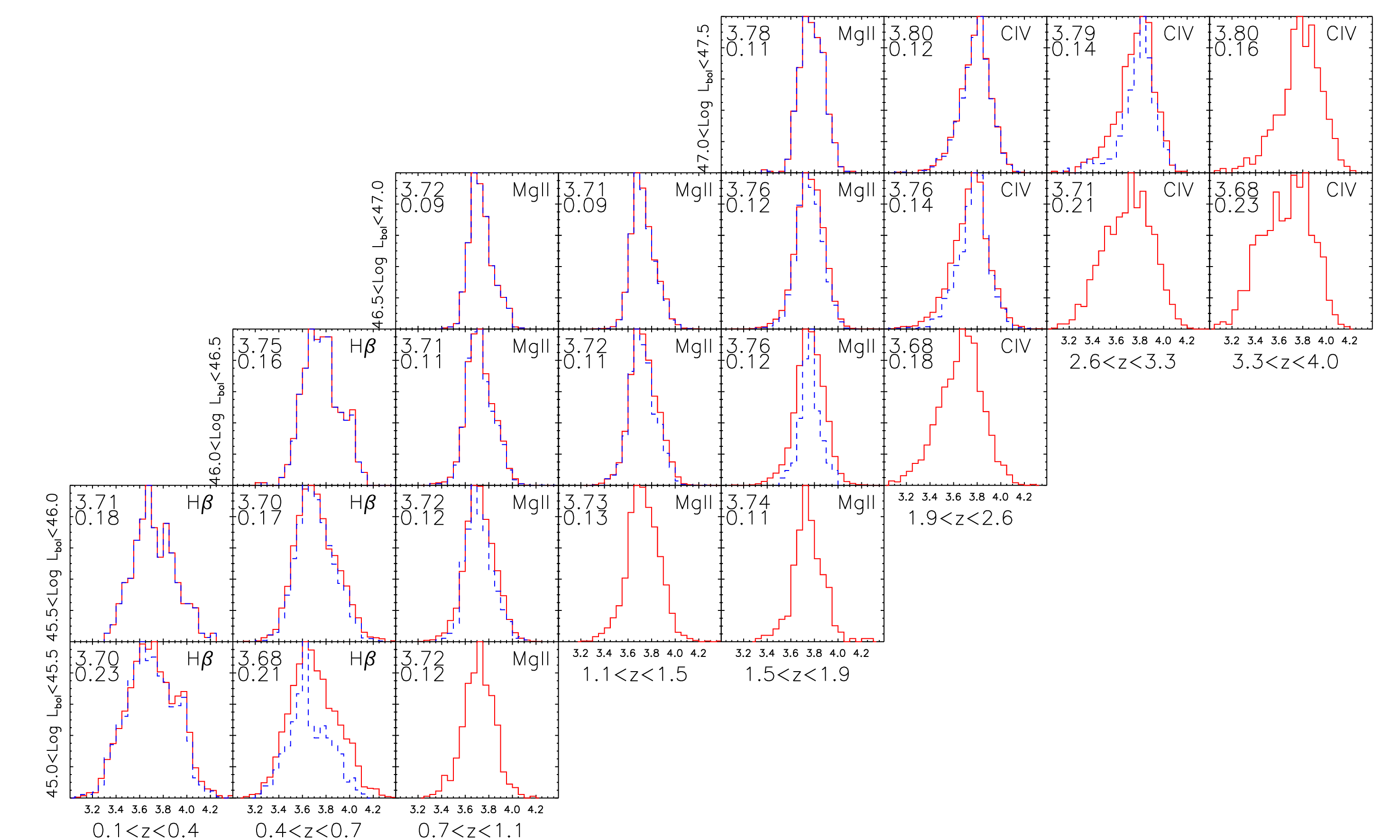}
    \caption{Distributions of log FWHM (in km s$^{-1}$) in different redshift (horizontally arranged) and luminosity (vertically arranged) bins for \hbeta, \MgII\ and \CIV.
     Solid histograms show the results without a ${\rm S/N}$ cut and dashed histograms show the results
     with ${\rm S/N}>10$. We only plot a histogram if there are
     more than 200 objects (no S/N cut) or 50 objects (with S/N
     cut) in each bin. The number of quasars drops rapidly when we impose the S/N cut, but the solid and
     dashed histograms show almost identical
     distributions. The means and dispersions of a fitted log-normal
     to the distributions are shown in the upper-left corners of each panel.
     }
    \label{fig:FWHM_dist_more}
\end{figure*}

\subsubsection{\CIV}
For the \CIV\ estimator, we use the calibration in Vestergaard \&
Peterson (2006), which has $a=0.66$ and $b=0.53$ in equation
(\ref{eqn:virial_estimator}).

Iron emission contamination is not a problem for \CIV, so we
have simply fitted a power law continuum to regions in the vicinity
of \CIV, namely [1445,1465]\AA\
and [1700,1705]\AA\ . This continuum is then subtracted from the
spectrum for the line fitting which follows.

The FWHM for \CIV\ as defined in Vestergaard \& Peterson (2006) is
indeed the ``{\em full-width-at-half maximum}'' of the full line
profile, whether or not the line profile is Gaussian. Since many
\CIV\ lines have profiles ``peakier'' than a Gaussian, the FWHM
obtained from a single Gaussian fit overestimates its formal
definition. For this reason, following the standard approach in
the literature (e.g., Baskin \& Laor 2005; Fine \etal\ 2006), we
use a set of three Gaussians to fit the \CIV\ line region
[1500,1600]\AA.  The FWHM is then measured from the fitted model
flux, providing better measurements of FWHM for noisy spectra. The
\CIV\ line in high redshift quasars may be biased by associated
absorption lines. To reduce their effect, a second line fit is
performed which excludes pixels that are $1.5\sigma$ below the
first fitted model, where $\sigma$ is the estimated error in flux
density. The model is replaced by the second fit if the reduced
$\chi^2$ is smaller than in the first fit. Also, if one or two of
the three Gaussians have integrated flux less than $5\%$ of the
total model flux, they are removed from the model flux; this step
ensures that there are no unphysically narrow peaks to bias the
FWHM measurement. As in the cases of \hbeta\ and \MgII, we require
${\rm FWHM}>1200{\rm\ km\ s^{-1}}$.

The \CIV\ FWHMs measured in this way are systematically narrower by
$\sim 0.15$ dex than what results from a single Gaussian fit.
Clearly, we would get systematic offsets in the \CIV\ based BH
masses if we used the single Gaussian FWHM.

The 1350\AA\ continuum luminosity $\lambda L_\lambda$ is measured
using the mean flux density within the wavelength range
[1340,1360]\AA, and corrected for Galactic extinction. Given the
wavelength coverage of the SDSS spectra, this implies that we
cannot estimate \CIV\ black hole masses for quasars with redshift
below $z\sim 1.8$.

\begin{figure*}
  \centering
    \includegraphics[width=0.45\textwidth]{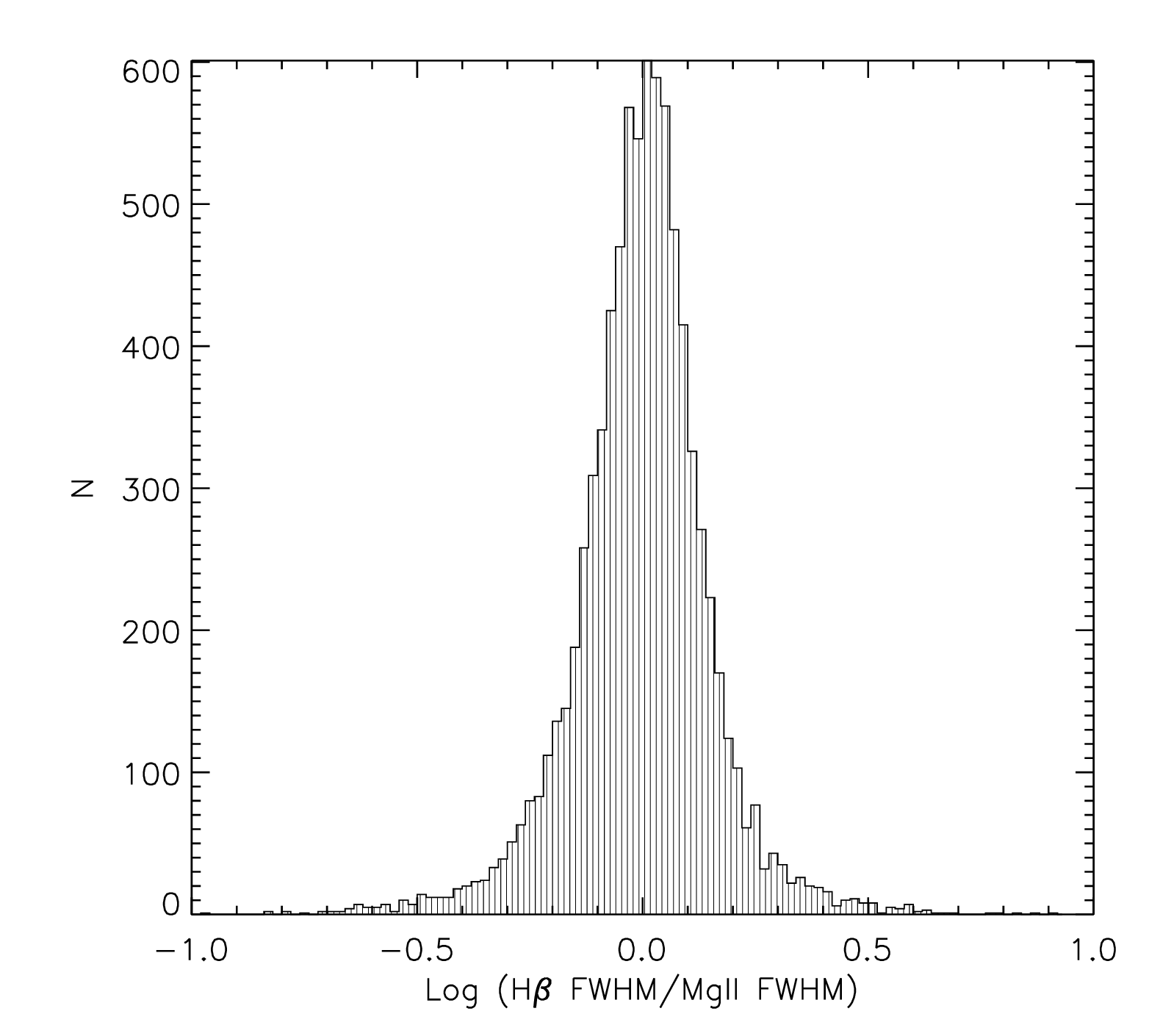}
    \includegraphics[width=0.45\textwidth]{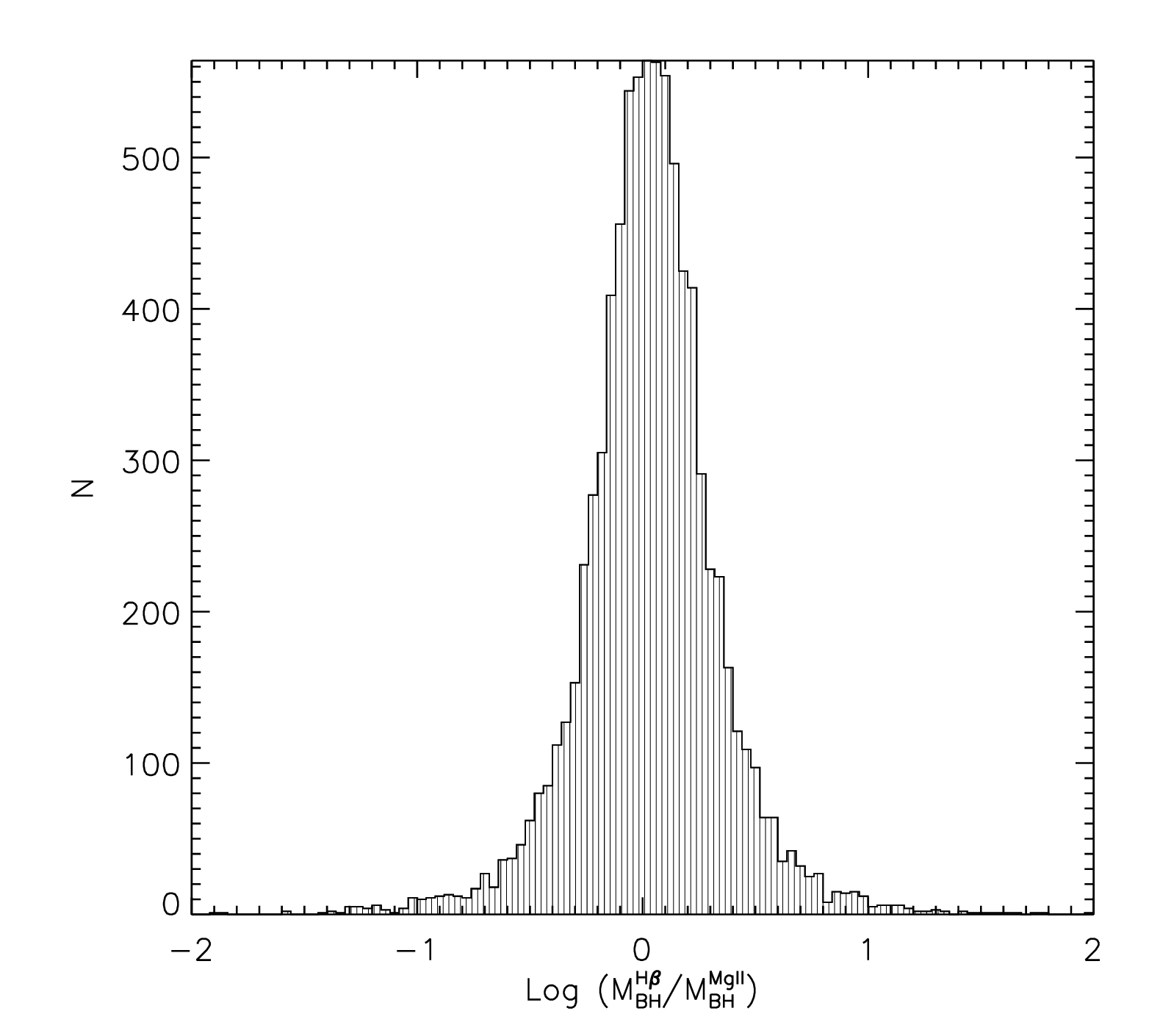}
    \caption{Comparisons between the \hbeta\ and the \MgII\ estimators. {\em left}: histogram of the FWHM ratios.
    {\em right}: histogram of the virial BH mass ratios.}
    \label{fig:Hbeta_MgII_hist}
\end{figure*}

\subsection{Bolometric Luminosities}
For each object in our sample that has a measurable virial mass,
we use its continuum luminosity $\lambda L_\lambda$ (at 1350\AA\
for \CIV, 3000\AA\ for \MgII\ and 5100\AA\ for \hbeta) to estimate
the bolometric luminosity. The bolometric correction factors (BC)
are computed using the composite SED for a sample of SDSS-DR3
quasars constructed by Richards \etal\ (2006b), and are slightly
different from the commonly adopted values in Elvis \etal\ (1994).
In particular, ${\rm BC}_{1350\AA}=3.81$, ${\rm
BC}_{3000\AA}=5.15$, ${\rm BC}_{5100\AA}=9.26$. When two continuum
luminosities are available, they give quite similar bolometric
luminosities. The typical error in bolometric corrections
assuming a universal SED is $\sim 0.1$ dex for optically-selected
SDSS quasars (Richards \etal\ 2006b), but it might be slightly
larger for ${\rm BC}_{1350\AA}$. The left panel of
Fig.~\ref{fig:L_bol} shows the relation between $L_{\rm bol}$ and
the $z = 2$ $i-$band absolute magnitude calculated using the $K$-correction in
Richards \etal\ (2006a), while the right panel shows $L_{\rm bol}$
versus redshift for our sample (the darker points are the
uniformly selected subsample).

Fig.~\ref{fig:FWHM_dist} shows the distribution of log FWHM in
bins of redshift and bolometric luminosity. The FWHMs of all three
emission lines are only weakly dependent on either redshift or
luminosity within our sample, and they follow a log-normal
distribution with typical dispersion $\sim 0.1-0.2$ dex. This
width is real and is not dominated by noise; we find essentially
identical distributions when restricting our sample to objects
with S/N per pixel greater than 10, or greater than 20. To further
demonstrate that the FWHM distributions do not evolve,
Fig.~\ref{fig:FWHM_dist_more} shows the FWHM distributions for the
three lines in different redshift and luminosity bins. Our
measured distributions of FWHMs are consistent with other
investigations (e.g., Baskin \& Laor 2005; Salviander \etal\
2007), with similar widths and a $\sim 0.1$ dex difference in the
mean values\footnote{The difference of $\sim 0.1$ dex in FWHM will
result in $\sim 0.2$ dex mean offset in virial masses. We do not
consider this substantial, since the zero-point of the virial
calibrations has uncertainties of the same level. } (which might
be caused by small differences in the techniques to determine
FWHMs). The observed distribution of FWHMs has profound effects on
our determination of the distribution of virial BH masses, as will
be further explored in section \ref{sec:MC}.

\begin{figure*}
  \centering
    \includegraphics[width=0.45\textwidth]{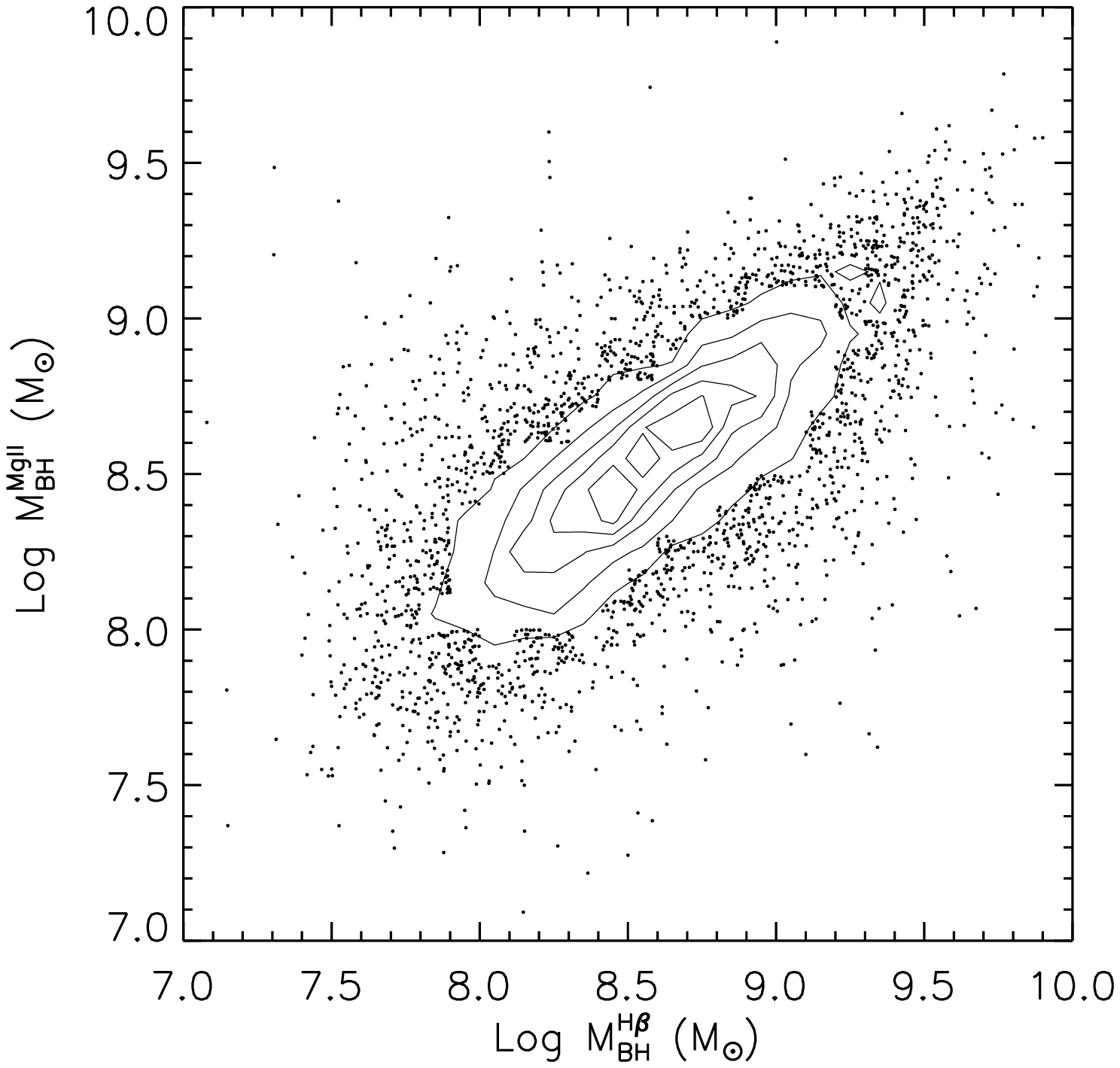}
    \includegraphics[width=0.45\textwidth]{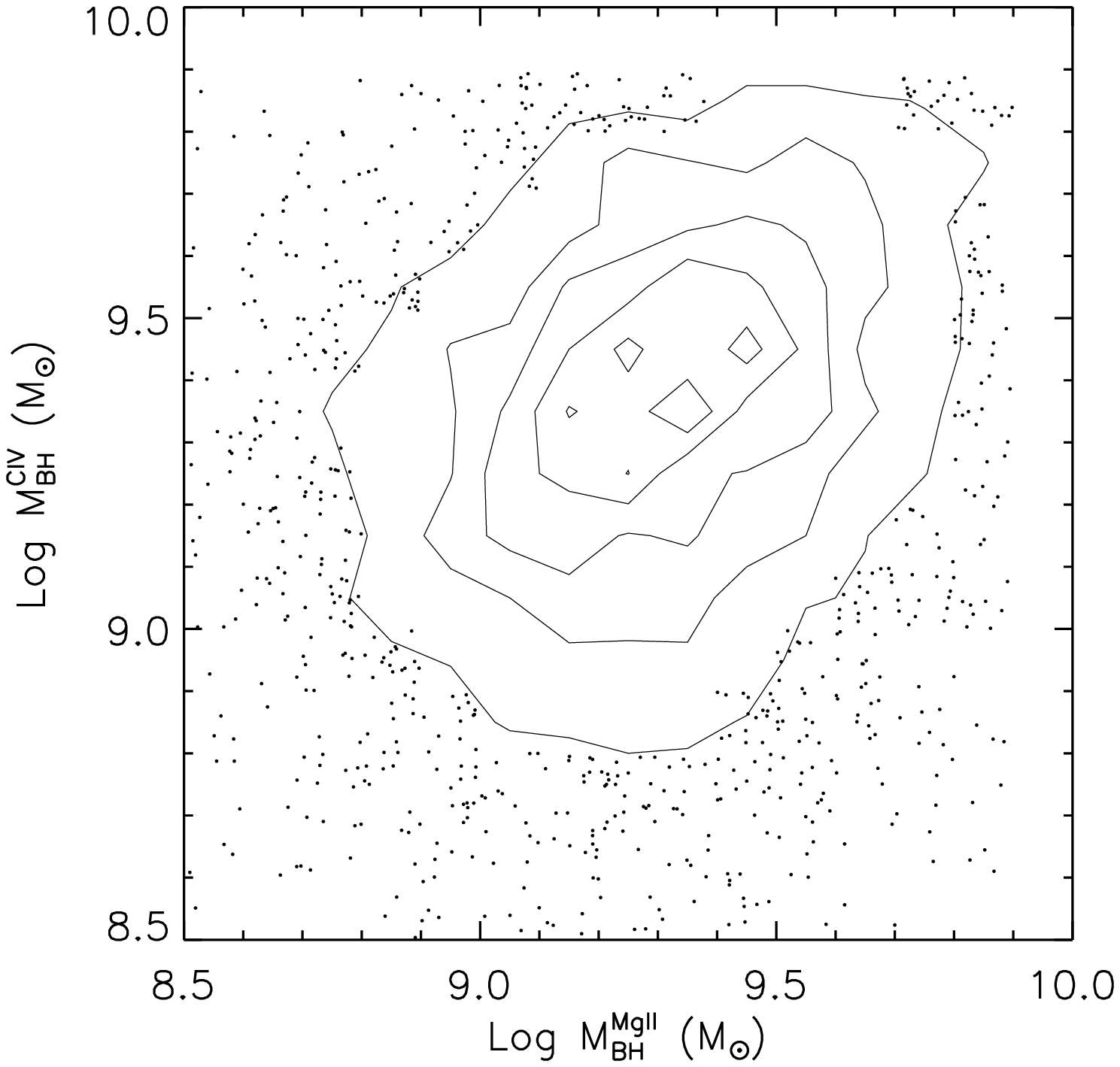}
    \caption{A one-to-one comparison between two virial mass estimators. {\em left}: \hbeta\
    versus \MgII; {\em right}: \MgII\ versus \CIV. The correlation is much better between \hbeta\
    and \MgII\ than the one between \MgII\ and \CIV. Contours are local point density contours,
    drawn to guide the eye.}
    \label{fig:one2one}
\end{figure*}

\begin{figure*}
  \centering
    \includegraphics[width=0.33\textwidth]{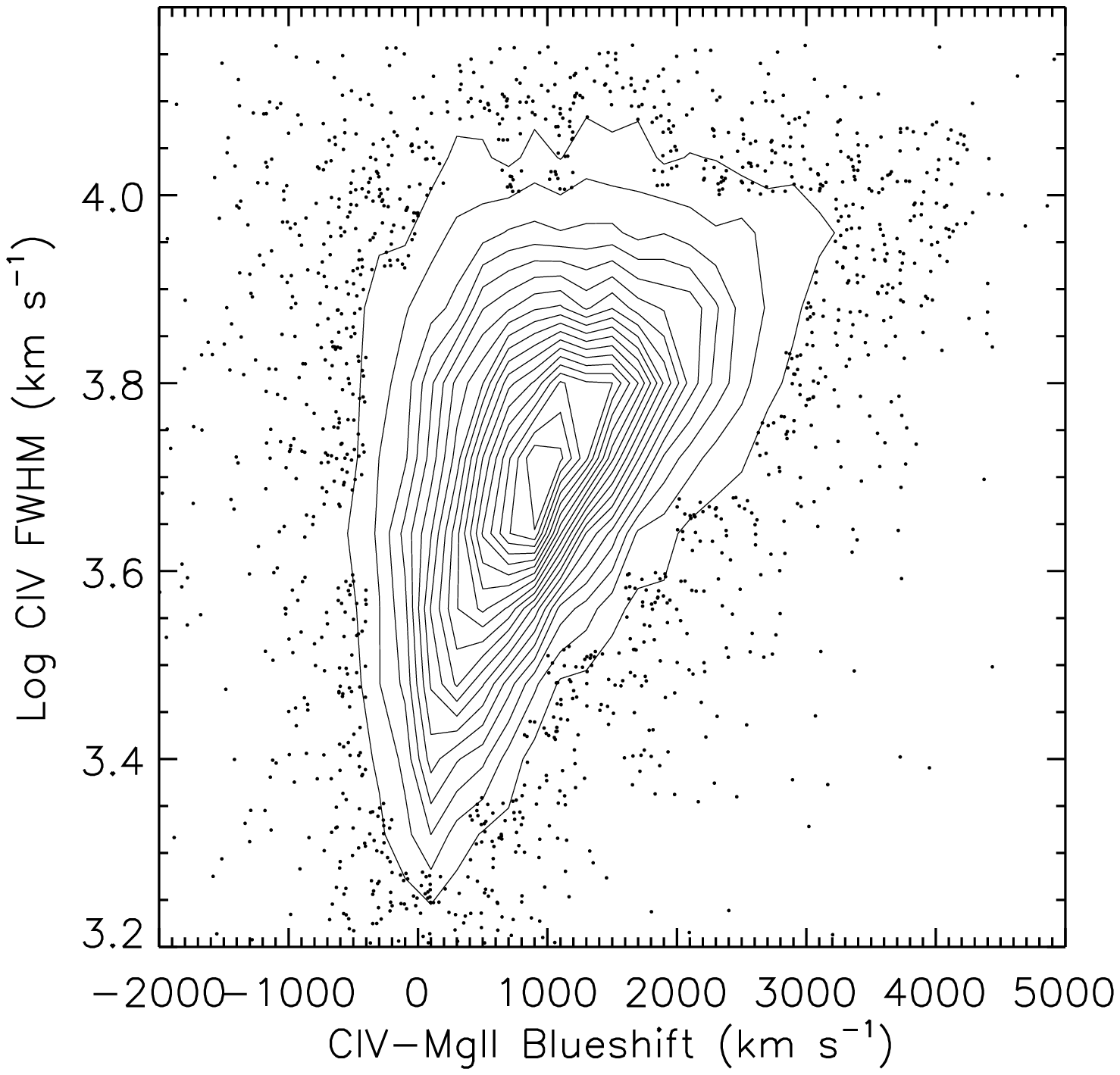}
    \includegraphics[width=0.33\textwidth]{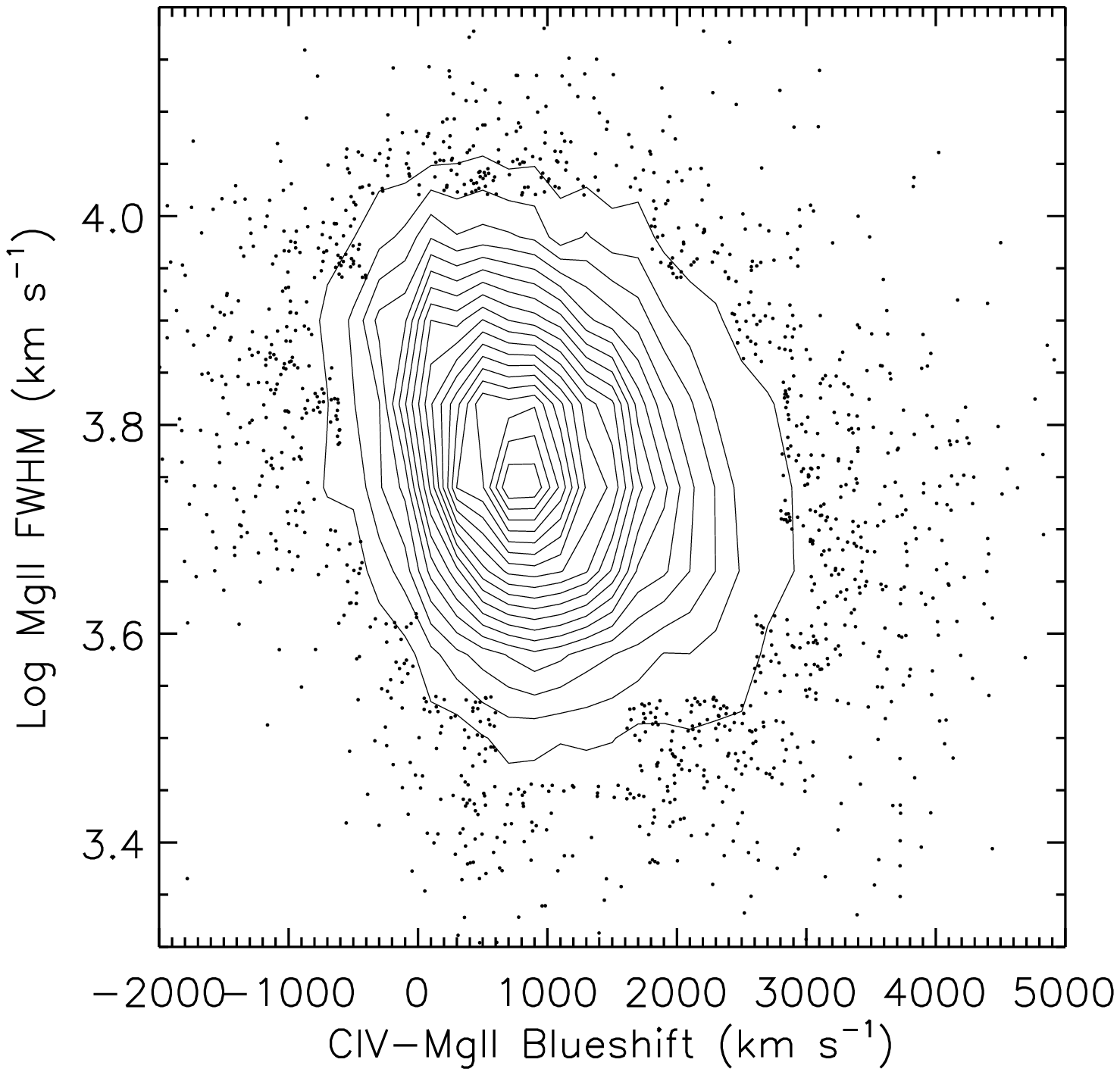}
    \includegraphics[width=0.33\textwidth]{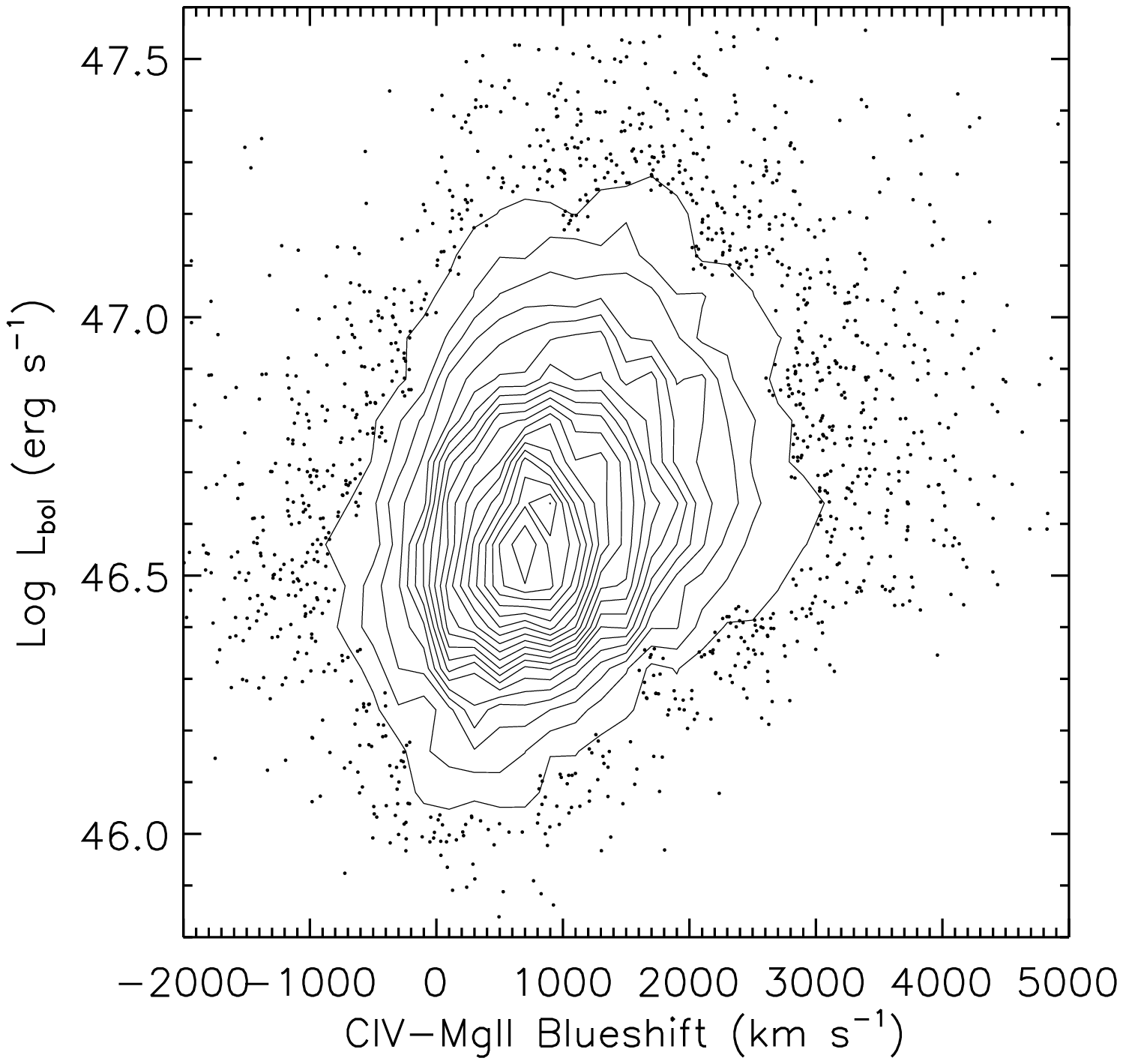}
    \caption{Various properties against the \CIV-\MgII\ blueshift.
    Contours are local point density contours, drawn to guide the
    eye. {\em Left}: \CIV\ FWHM against blueshift.
    The mean value of FWHMs rises with increasing blueshift. For those objects with the largest blueshift,
    their FWHMs rarely drop below $4000 {\rm\ km\ s^{-1}}$. {\em Middle}: \MgII\ FWHM against blueshift. In contrary to
    the \CIV\ case, the \MgII\ FWHM shows a mild decrease with increasing blueshift, but it is also consistent with no correlation
    at all considering the huge scatter. {\em Right}: Bolometric luminosity against blueshift. The mean luminosity increases with blueshift,
    but again the dispersion is large. }
    \label{fig:blueshift}
\end{figure*}

\subsection{Comparison of Emission Lines}
There are subsamples of quasars for which we have two lines
available; given the SDSS spectroscopic coverage, we observe both
\hbeta\ and \MgII\ for $0.4\lesssim z \lesssim
0.9$, and both \MgII\ and \CIV\ for $1.7\lesssim z\lesssim 2.2$.
Therefore we can check if different BH mass estimators give
consistent results.

\subsubsection{\hbeta\ vs \MgII}
\label{sec:hbeta_MgII}

For a subsample of $\sim 8,000$ quasars with redshift $0.4\lesssim
z \lesssim 0.9$ we have both \hbeta\ and \MgII\ linewidth
measurements. The distribution of $\log($\hbeta\ FWHM/\MgII\
FWHM$)$ is shown in the left panel of
Fig.~\ref{fig:Hbeta_MgII_hist}, which follows a Gaussian with mean
0.0062 and dispersion 0.11 dex. The distribution of $\log(M_{\rm
BH}^{\rm H\beta}/M_{\rm BH}^{\rm MgII})$ is shown in the right
panel of Fig. \ref{fig:Hbeta_MgII_hist}, which follows a Gaussian
with mean 0.034 and dispersion 0.22 dex. The small discrepancy in
the mean offsets between our results and those in McLure \& Dunlop
(2004, where the mean FWHM ratio and BH mass ratio offsets are
$-0.004$ and $0.013$ dex, respectively) is perhaps due to
different samples, but is totally negligible considering the $\sim
0.3-0.4$ dex uncertainties of either of the BH mass estimators.
Therefore, the \MgII\ and \hbeta\ estimators give consistent BH
masses.

A plot of the \hbeta\ FWHM against the \MgII\ FWHM is very similar
to Fig.~3 in Salviander \etal\ (2007); the two FWHMs are
correlated, with a slope differing slightly from unity.

We plot the
\hbeta\ against the \MgII\ virial masses in the left-hand panel of
Figure~\ref{fig:one2one}.  The two mass estimators correlate strongly
with one  other over two orders of magnitude in mass.

\begin{figure}
  \centering
    \includegraphics[width=0.45\textwidth]{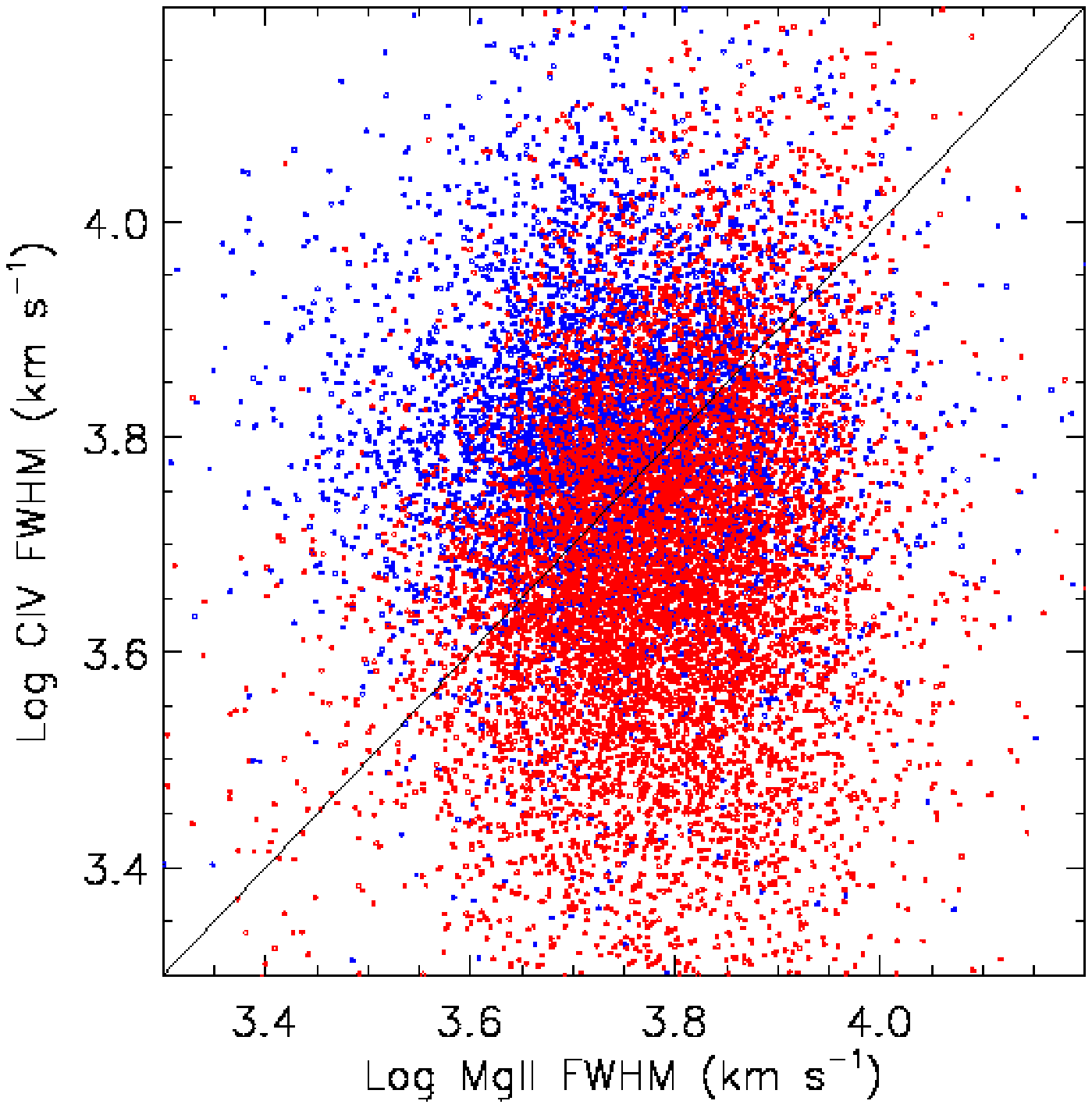}
    \includegraphics[width=0.45\textwidth]{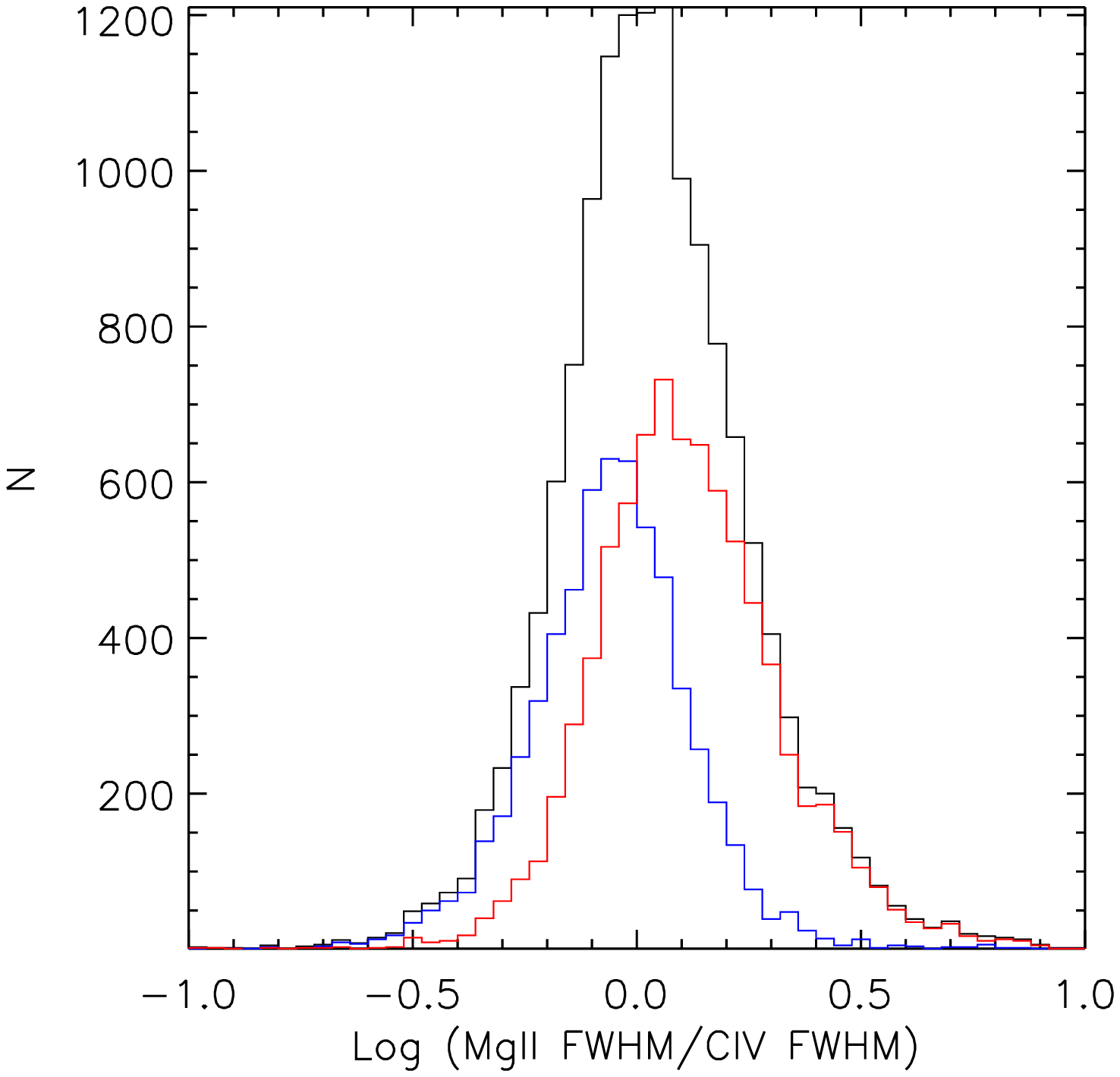}
    \caption{{\em Top}: \CIV\ FWHMs against \MgII\ FWHMs for a sample of quasars with both
    FWHMs available. Blue and red dots show those with \CIV-\MgII\ blueshift greater and less than 1000 ${\rm km\ s^{-1}}$;
    the solid line shows the unity relation.
    {\em Bottom}: histograms of the ratios of the \MgII\ FWHM to the \CIV\ FWHM. Black histogram
    is for the whole sample; blue and red ones are for objects with \CIV-\MgII\ blueshift greater and
    less than $1000\ {\rm km\ s^{-1}}$.
    }
    \label{fig:CIV_MgII_FWHM}
\end{figure}

\begin{figure}
  \centering
    \includegraphics[width=0.45\textwidth]{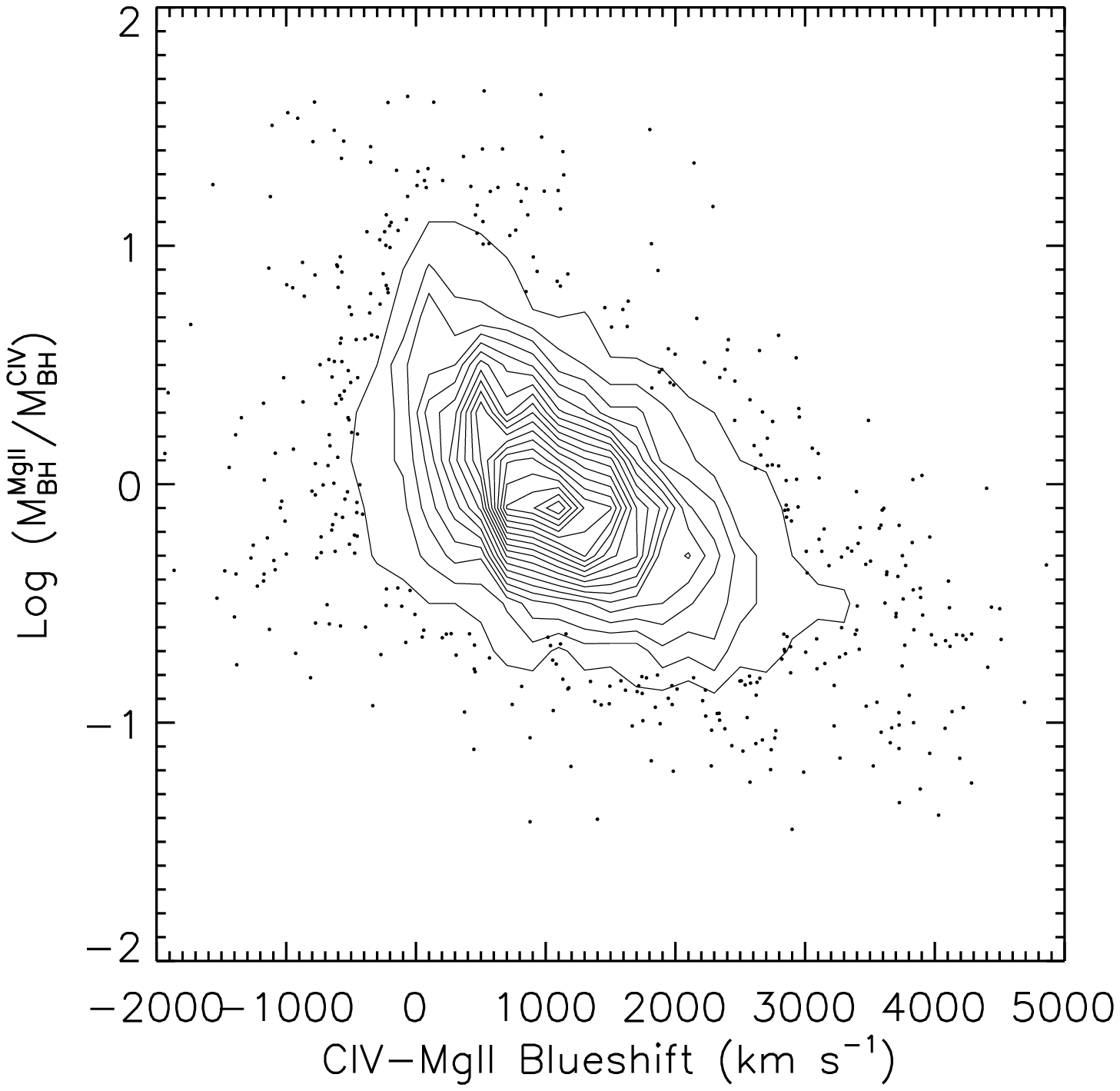}
    \includegraphics[width=0.45\textwidth]{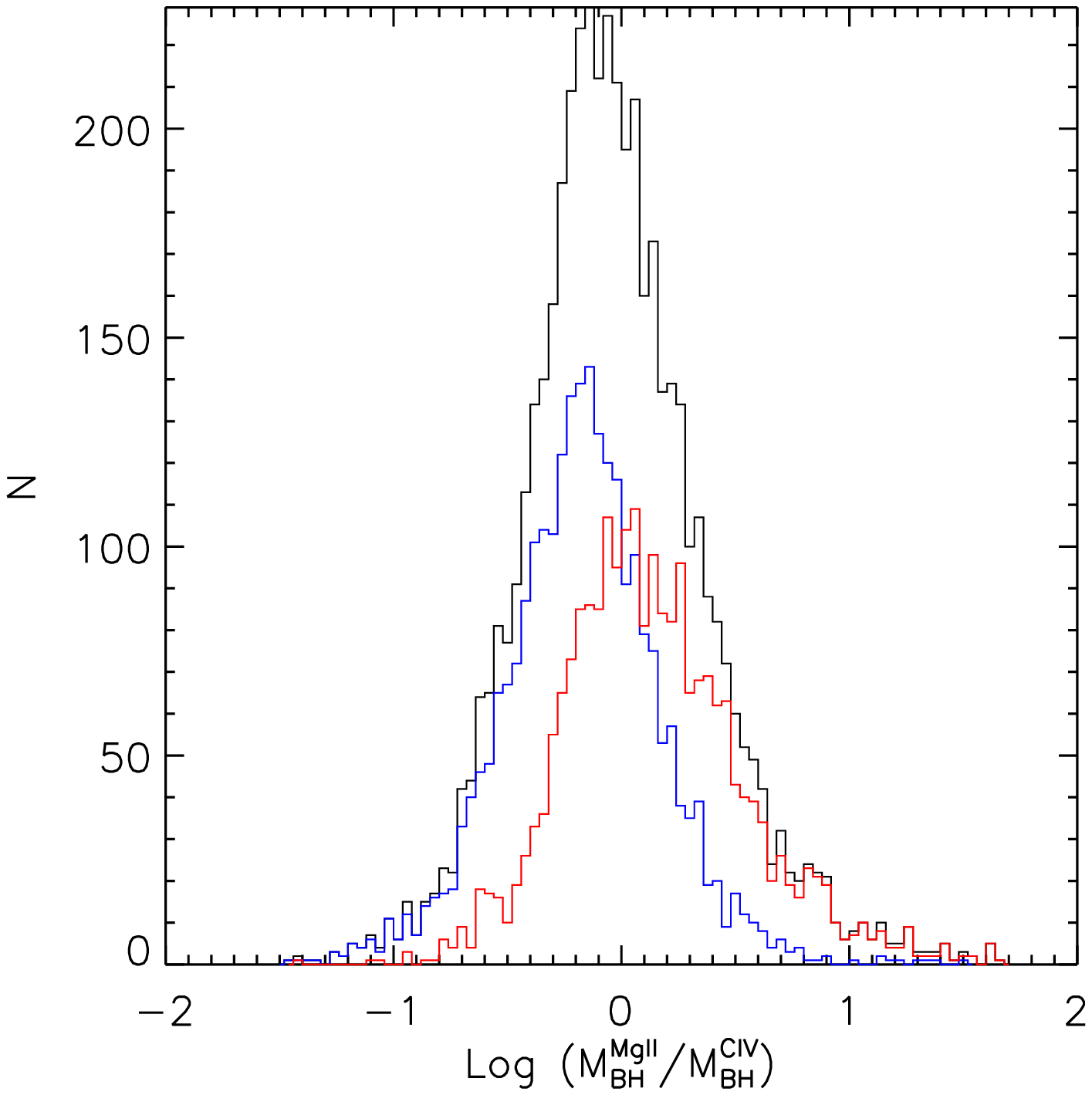}
    \caption{{\em Top}: Difference in the \MgII\ and \CIV\ based virial BH masses as function of the \CIV-\MgII\ blueshift. For those with
    small (large) blueshifts, the \CIV\ estimator systematically underestimates (overestimates) the BH mass as
    compared to the \MgII\ based virial masses. Contours are local point density contours, drawn to guide the
    eye. {\em Bottom}: Distributions of the ratios of the two virial mass estimates. The black histogram
    is for the entire sample; blue and red ones are for objects with \CIV-\MgII\ blueshift greater and
    less than $1000\ {\rm km\ s^{-1}}$.}
    \label{fig:CIV_MgII_BH}
\end{figure}

\subsubsection{\MgII\ vs \CIV}
\label{sec:MgII-CIV}
Linewidths of both \MgII\ and \CIV\ are available for
a subsample of $\sim 15,000$ quasars with redshift $1.7\lesssim
z\lesssim 2.2$, of which $\sim 5,000$ quasars have both \MgII\
and \CIV\ virial masses available. However, before we compare the
results, we review some of the unique characteristics of the \CIV\
line.

First, the definitions of FWHM are quite different for \MgII\ and
\CIV. While the broad component of \MgII\ is well-described as a
Gaussian in our fits, the vast majority of \CIV\ lines are by no
means Gaussian. The difference in line shapes suggests that the
broad \MgII\ and \CIV\ might have different physical origins. Second, it is
well known that high ionization lines such as \CIV\ often show
systematic blueshifts with respect to low ionization lines such as
\MgII\ (e.g., Gaskell 1982; Tytler \& Fan 1992; Richards \etal\
2002b). Richards \etal\ (2002b) used $\sim 800$ SDSS quasars to
show that \CIV\ is systematically blueshifted with respect to
\MgII\ by $\sim 900{\rm\ km\ s^{-1}}$ on average. They interpreted
this apparent blueshift as due to the absorption of the \CIV\ red
wing, as might be expected if the broad-line region gas is
actually a radiatively driven, equatorial outflow.  Such disk-wind
models are compelling for a variety of reasons, particularly their
success in explaining the properties of broad absorption line
quasars (e.g., Murray \etal\ 1995; Proga \etal\ 2000; Elvis 2000).
Richards \etal\ also see a trend of increasing average \CIV\ FWHM
with increasing \CIV-\MgII\ blueshift, which again suggests that
the broad \CIV\ line is dominated by a disk wind component; in
this case the \CIV-\MgII\ blueshift reflects the orientation of
the line-of-sight (LOS). There is additional evidence supporting
this disk wind/orientation scheme, including the anti-correlation
between the \CIV-\MgII\ blueshift and radio loudness, and the
correlation between the \CIV-\MgII\ blueshift and BAL fraction
(see also Gallagher \etal\ 2005).  These properties are also seen
in a sample of low redshift broad line AGNs (Sulentic \etal\
2007). Of course, alternative interpretations of these features
are possible, which depend more on the structure of the wind than
on the LOS orientation (see Richards 2006).

If the broad \CIV\ is indeed contaminated or even dominated by a
disk wind component, then the measured \CIV\ FWHM does not
represent the actual virial velocity of the BLR. If the LOS is
along the disk wind, then the FWHM will overestimate the virial
velocity, while if the LOS is along the pole direction, the FWHM
will probably underestimate the virial velocity. Although the
\CIV\ estimator is calibrated in such a way that it gives, on
average, BH masses consistent with those estimated from \hbeta\
(e.g., Vestergaard 2002; Vestergaard \& Peterson 2006), there will
be biases in the \CIV-based virial masses if the above picture is
correct.

We explore these issues by plotting various quantities against the
\CIV-\MgII\ blueshift in Fig.~\ref{fig:blueshift}. The left panel
shows the correlation between the \CIV\ FWHM and the \CIV-\MgII\
blueshift. We see the rise of the average \CIV\ FWHM as the
blueshift increases, as reported by Richards \etal\ (2002b). In
particular, very few objects with large blueshifts have FWHM less
than $4000\ {\rm km s^{-1}}$, while the distribution of FWHM is
broader for objects with small blueshifts. The middle panel shows
a mild anti-correlation between the \MgII\ FWHM and the
\CIV-\MgII\ blueshift, and the right panel shows a mild
correlation between the bolometric luminosity $L_{\rm bol}$ and
the blueshift, which has been found before (e.g., Richards \etal\
2002b). The cause of the anti-correlation between \MgII\ FWHM and
\CIV-\MgII\ blueshift is not clear at this point. It might reflect
inclination effects as well. The different correlations of the
\CIV\ and \MgII\ line widths with blueshift suggest that the two
lines have different origins, as we argued above.

Since the \CIV\ estimator is calibrated using \hbeta\
reverberation mapping masses (e.g., Vestergaard 2002; Vestergaard
\& Peterson 2006), there should be no mean offset in either the
FWHMs or the estimated BH masses, as we find (see the black solid
histograms in the bottom panels in Figs. \ref{fig:CIV_MgII_FWHM}
and \ref{fig:CIV_MgII_BH}). While the log of the ratio of \MgII\
to \CIV\ FWHMs follows a Gaussian with mean 0.027 and dispersion
0.18 dex, the {\em correlation} between the two line widths is
very weak (top panel of Fig. \ref{fig:CIV_MgII_FWHM}); indeed the
width of 0.18 dex is larger than the dispersion in the \MgII\ FWHM
distribution itself. Correspondingly, the log of the ratio of
\MgII\ to \CIV\ BH masses follows a Gaussian with mean $-0.06$ dex
and dispersion 0.34 dex. Baskin \& Laor (2005) similarly found
weak correlations between the \CIV\ and \hbeta\ FWHMs at low
redshift, while Netzer \etal\ (2007) report a similar effect at $z
\approx 2$ (but see Kelly \& Bechtold 2007 for an alternate view).
Moreover, this scatter is systematic with blueshift.  The blue
points and histogram in Figure~\ref{fig:CIV_MgII_FWHM} refer to
objects with blueshift larger than 1000 km s$^{-1}$, while the red
points have smaller blueshift.

The \CIV\ and \MgII\ virial mass estimators are plotted against
one another in the right-hand panel of Figure~\ref{fig:one2one}.
The correlation is much weaker than we found for the \MgII-\hbeta\
comparison.  The ratio of these two masses is correlated with the
\CIV-\MgII\ blueshift, as is shown in Fig.~\ref{fig:CIV_MgII_BH}
(top panel).  Given the similarity between the properties of the
\MgII\ and \hbeta\ lines, and the evidence that the \CIV\ line may
be affected by a disk wind, we argue that \hbeta\ and \MgII\ are
better BH mass estimators than is \CIV. In particular, we
emphasize the danger of using \CIV\ as a BH mass estimator for
individual objects and small samples. For example, Kurk \etal\
(2007) and Jiang \etal\ (2007a) have obtained observations of both
\MgII\ and \CIV\ in four $z\sim 6$ quasars. In the three quasars
in which \CIV-\MgII\ blueshifts are observed, the \CIV\ virial
mass is $\sim 2$ times larger than the \MgII\ mass, while they are
in good agreement for the one object with a \CIV-\MgII\ redshift,
consistent with the trends we see in Figure~\ref{fig:CIV_MgII_BH}.
Finally we point out that the ratio of \MgII\ and \CIV\ virial
masses does not depend on luminosity in the range where both line
estimators are available, excluding luminosity-based biases.

Of course we cannot measure the blueshift, and thereby correct the
\CIV\ mass estimate for the bias of Figure~\ref{fig:CIV_MgII_BH},
of objects for which our spectral coverage only includes \CIV.
However, Richards \etal\ (2002b) show that the \CIV-\MgII\
blueshift correlates broadly with the \CIV\ line asymmetry,
suggesting that we might use this asymmetry as a surrogate to
estimate the mass estimate bias. However, the correlation between
the \CIV-\MgII\ blueshift and the \CIV\ line asymmetry shows a
large amount of scatter, making it difficult to correct for this
bias using the \CIV\ line asymmetry.

It is beyond the scope of this paper to carry out a thorough
investigation of the BLR and disk geometry and dynamics
needed to understand the physics behind the trends we've found. Therefore, in
the following sections, we simply use the \CIV\ estimator to
measure BH masses at $z> 1.9$. Our confidence in the \CIV\
estimator is built on the large sample studied here and the belief
that there is no systematic offsets {\em in the mean} using the
\CIV\ estimator.

\section{Distribution of Black Hole Masses}\label{sec:result}

We present virial BH masses for the SDSS DR5 quasars with
$z\lesssim 4.5$. Excluding those objects for which we cannot
obtain reliable FWHM or continuum luminosity measurements due to
poor spectra or redshift limitations, we have 56,872 objects
(nonBALs) with measurable virial BH masses. The masses are based
on the \hbeta\ estimator for $z < 0.7$, \MgII\ for $ 0.7 < z <
1.9$, and \CIV\ for $1.9<z<4.5$, to avoid using measurements which
are too close to the red and blue ends of the spectra.  We should
note that the quasars in this catalog were not all uniformly
targeted; in particular, special targeting algorithms were used to
select quasar candidates to fainter magnitudes than the main
quasar survey (Richards \etal\ 2002a). The completeness of these
fainter quasars is poor.  We start by neglecting this effect; we
take all the quasars in the DR5 quasar catalog with measurable
masses and present the redshift evolution of virial BH masses
(\S\ref{sec:BH_z_eov}) and Eddington ratio distributions
(\S\ref{sec:Eddington_dist}). We restrict ourselves to the complete
subsample in \S\ref{sec:MC} and \S\ref{sec:MC-results}, where we
model the observed black hole mass distribution.  We describe the
active BH mass function in \S\ref{sec:mass_fun}, and discuss
radio-loud and BAL quasars in \S\ref{sec:BAL}.

\subsection{Redshift evolution of virial BH masses}
\label{sec:BH_z_eov}

\begin{figure*}
  \centering
    \includegraphics[width=1\textwidth]{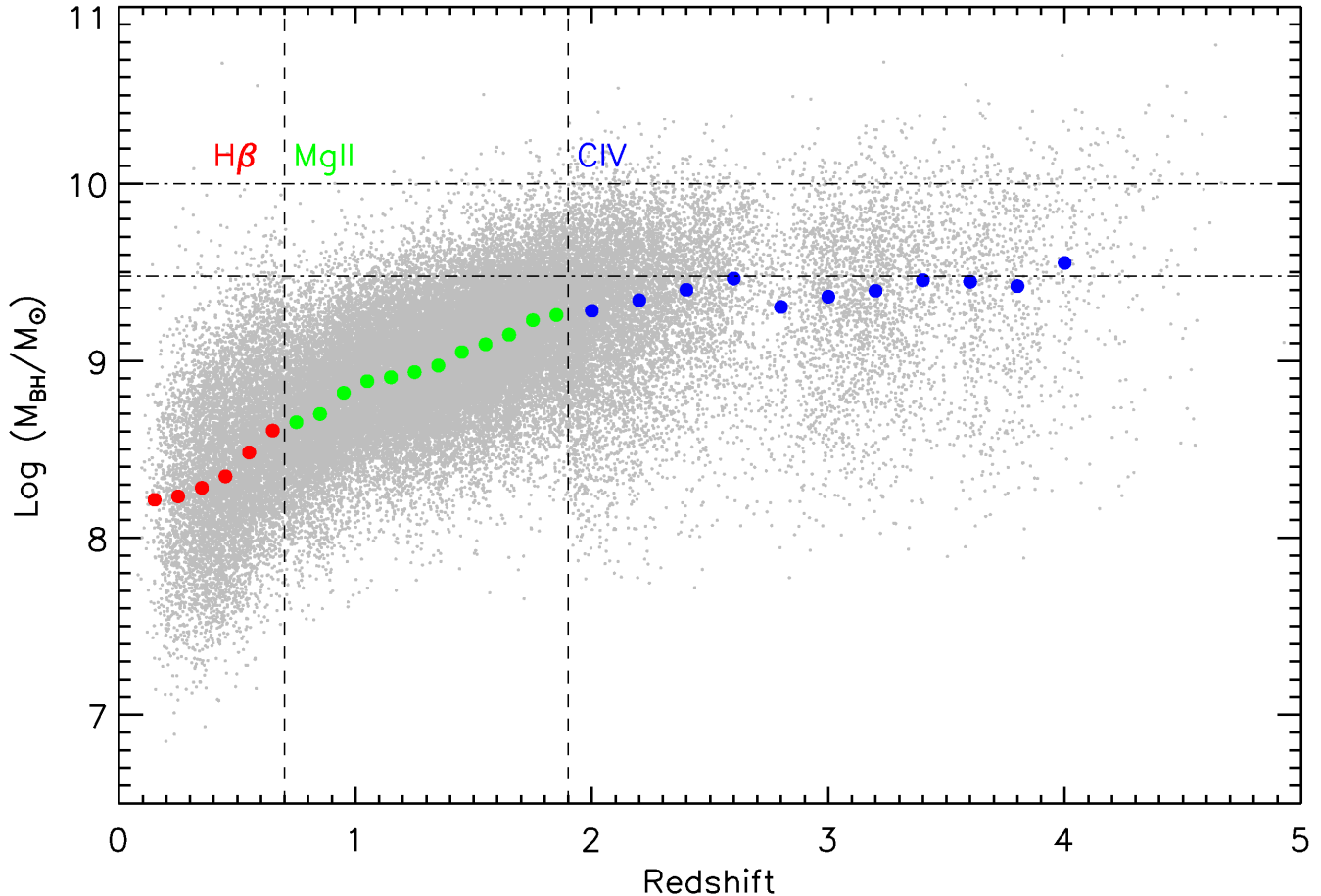}
    \caption{The cosmic evolution of virial BH masses for our entire DR5 quasar catalog with
    measurable virial masses. The two dashed vertical lines mark the
    transitions from one estimator to another. The filled circles show
    the mean value in bins of redshift of width $\Delta z = 0.1$ for
    $z \le 1.9$ and $\Delta z = 0.2$ at higher redshit.  The gaps
    around $z\sim 2.8$ and $z \sim 3.5$ are caused by the inefficiency
    of selecting quasars around these redshifts (e.g., Richards \etal\
    2002a).
    The upper envelope flattens out at $z\gtrsim 2$ but decreases
    towards lower redshift, which suggests that the most massive SMBHs are no longer shining as
    optical quasars at lower redshift. There is clearly an upper limit at all redshifts of black hole masses between $3\times 10^9$
    and $10^{10}\ M_\odot$, shown as the two dash-dotted horizontal lines.}
    \label{fig:BH_evo}
\end{figure*}

Fig.~\ref{fig:BH_evo} shows the dependence of virial BH masses on
redshift, where gray dots represent individual objects and the
filled circles show the mean value at each redshift. Because of
the incompleteness of the sample at the faint end, the ``mean'' BH
masses here are weighted towards high luminosity objects at all
redshifts.  Nevertheless, the mean BH mass increases with redshift
below $z\sim 2$, and then flattens out above $z\sim 2$; the most
massive BHs seem to turn off around this epoch.  There is clear
evidence of a limiting BH mass $\sim 3\times 10^9-10^{10}\
M_\odot$, consistent with the conclusions of McLure \& Dunlop
(2004) for $z\le 2.1$ SDSS DR1 quasars.  The trend of this figure
is quite similar to fig. 4 of Vestergaard (2004), except that we
have filled the gaps in their figure with our much larger sample.

\subsection{Distribution of Eddington ratios}
\label{sec:Eddington_dist}

\begin{figure*}
  \centering
    \includegraphics[width=1\textwidth]{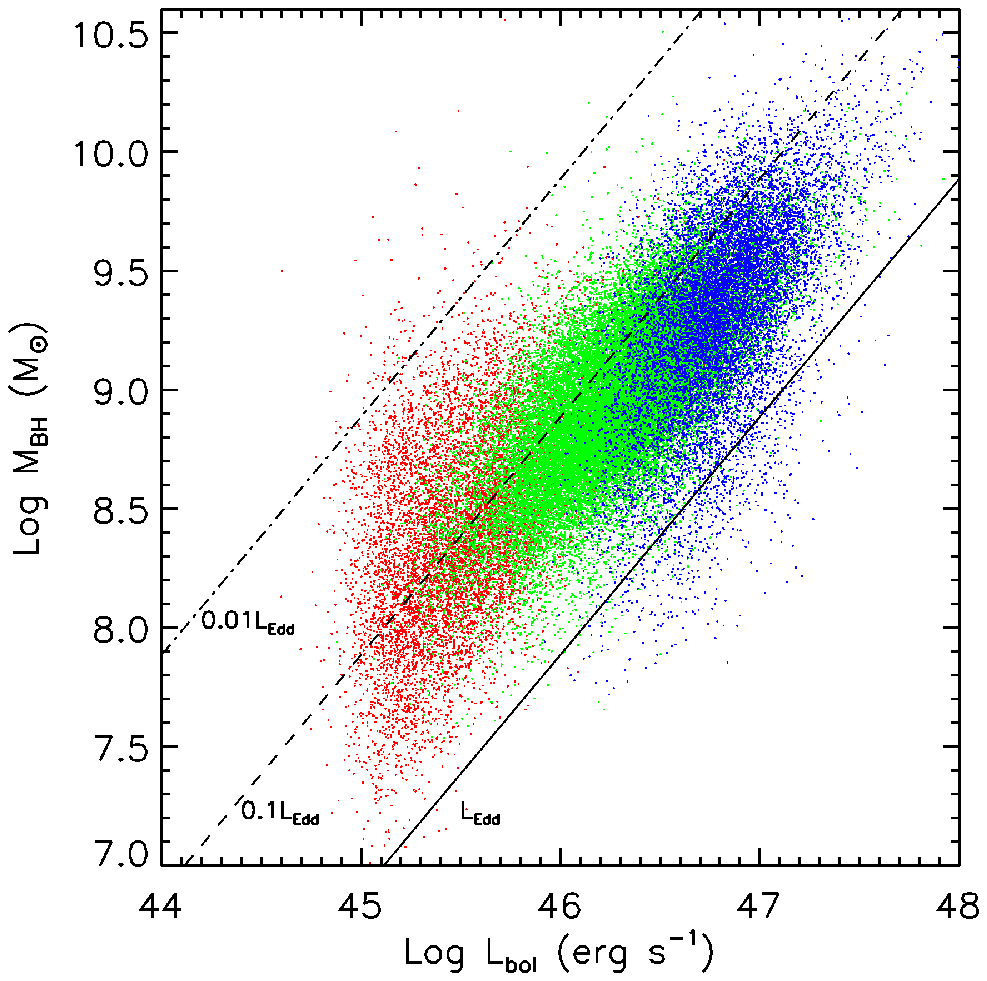}
    \caption{Distribution of quasars from the whole sample in the mass-luminosity diagram, where different colors
    show BH masses using different virial estimators (red for \hbeta, green for \MgII\ and blue for \CIV).
    The three diagonal lines (from top to bottom) show 0.01, 0.1 and $L_{\rm Edd}$ respectively.
    Most quasars are bounded by $L_{\rm bol}/L_{\rm Edd}=0.01$ and $1$.}
    \label{fig:Edd_dist}
\end{figure*}

The relation between virial BH mass and bolometric luminosity is
shown in Fig.~\ref{fig:Edd_dist}.  The figure shows objects using
the three virial estimators in different colors (red for \hbeta,
green for \MgII\ and blue for \CIV), while the diagonal lines
indicate various Eddington ratios $L/L_{\rm Edd}$, where $L_{\rm
Edd}=1.3\times 10^{38}(M_{\rm BH}/M_\odot) {\rm\ erg\ s^{-1}}$ is
the Eddington luminosity. The vast majority of quasars in our
sample are bounded by $L_{\rm Edd}$ and $0.01L_{\rm Edd}$, with a
geometric mean value of $\sim 0.1L_{\rm Edd}$. We are not
surprised to see that objects using these virial relations occupy
a stripe in the mass-luminosity plane, with a slope $\sim b\sim
0.5$ and scatter $\sim 0.2-0.3$ dex around the mean. This simply
reflects the fact that the FWHM distribution is almost independent
on luminosity, and has a very narrow width of $\sim 0.1-0.15$ dex
at fixed luminosity (see Fig. \ref{fig:FWHM_dist_more}). This also
explains the remarkably similar appearance of the $M_{\rm
BH}-L_{\rm bol}$ diagram in several other investigations even
though the underlying samples are quite different (e.g., Woo \&
Urry 2002; Kollmeier et al. 2006), simply because similar virial
estimators are used to estimate BH masses, and the FWHM
distributions of different samples are similar.

\begin{figure*}
  \centering
    \includegraphics[width=1\textwidth]{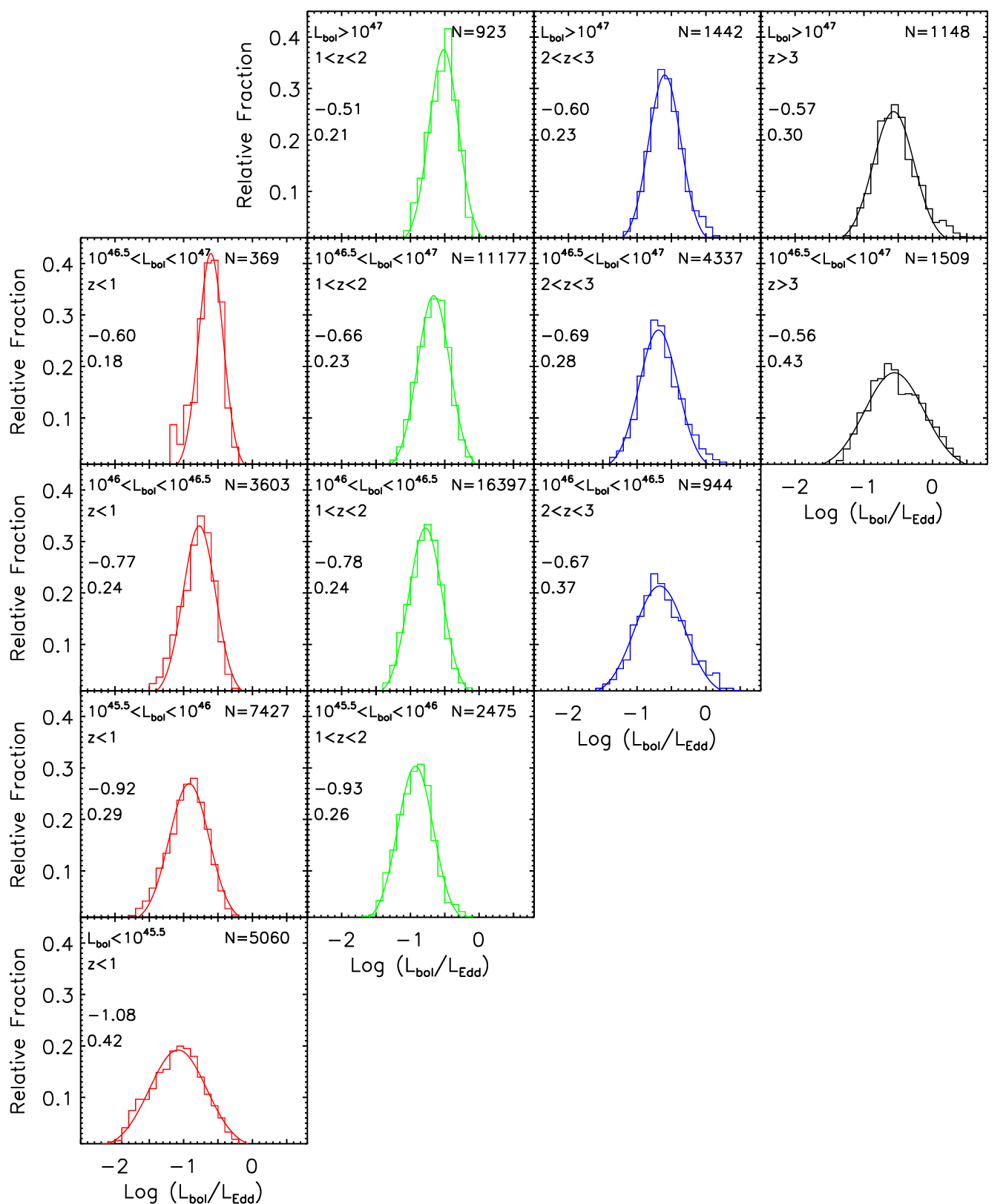}
    \caption{Apparent distributions of Eddington ratios $\log (L_{\rm bol}/L_{\rm Edd})$ based on
     virial BH masses in various redshift and luminosity bins. Also plotted are the fitted Gaussians, with
     the means and dispersions shown in the middle-left in each panel. The number of objects in
     each bin is shown in the upper-right corner.}
    \label{fig:Edd_hist}
\end{figure*}

Fig.~\ref{fig:Edd_hist} shows the distributions of $L_{\rm
bol}/L_{\rm Edd}$ in five luminosity and four redshift bins.
The distribution of Eddington ratios is log-normal in almost all
of the bins. No appreciable redshift evolution is seen for objects
with the same luminosity. The widths of the distributions are as
small as $\sim 0.3$ dex, similar to those in Kollmeier \etal\
(2006). Since virial mass is approximately proportional to the
square root of luminosity, the widths of the observed Eddington
ratio and the virial BH mass distributions in each
luminosity/redshift bin should be comparable, as is seen. The
width of the virial BH mass distributions within each
luminosity/redshift bin is thus $\sim 0.3$ dex or less.

At face value, the very narrow widths of the virial BH mass
distribution in each luminosity bin seem somewhat surprising,
since many studies have quoted significantly larger scatter
between single-epoch virial masses and both reverberation-mapped
masses on the one hand and stellar velocity dispersions on the
other. For instance, McLure \& Jarvis (2002) find an rms scatter
of 0.4 dex of virial \MgII\ masses around reverberation-mapping
masses, while Vestergaard \& Peterson (2006) find scatters of 0.3
and 0.5 dex of the \CIV\ and \hbeta\ virial masses about the RM
masses respectively. An independent estimate may be derived by
comparing virial masses with stellar velocity dispersions; Greene
\& Ho (2006) find an rms scatter of 0.4 dex of virial \halpha\
masses around the expectation from the $M-\sigma$ relation. In all
of these cases, it is important to remember that the objects in
question are typically much less luminous (and at much lower
redshift) than the quasars we study here, but these are the
measurements at our disposal at the present time.

However, the observed width of the virial mass distribution within
a given luminosity/redshift bin is {\em not} the uncertainty in
the virial mass estimators themselves. Although we do not have a
deterministic model for the origin of the uncertainty in the
virial relations, it must result both from the imperfect relation
between luminosity and BLR radius on the one hand, and from the
imperfect relation between FWHM and BLR virial velocity on the
other. Thus the uncertainty in the virial mass estimators comes
from the uncorrelated rms scatters in both luminosity and FWHM.
When we impose luminosity cuts to our sample, i.e., either by
working within luminosity bins or because of the flux limits of
our sample, we will artificially narrow the observed distribution
of virial masses.

At the same time, if the underlying BH mass distribution is such
that there are more BHs towards lower masses/lower luminosities,
then more objects will scatter from low-mass bins to high than the
other way around, giving rise to a Malmquist-type bias (e.g.,
Eddington 1913; Malmquist 1922) in the observed virial black hole
mass and Eddington ratio. In the next section, we use Monte Carlo
simulations to investigate the impact of luminosity cuts and
Malmquist bias on our observed distributions of luminosity, line
width, virial BH masses, and Eddington ratios. The results are
shown in \S~\ref{sec:MC-results}, where we also present an
analytic derivation of the Malmquist bias.

\subsection{Monte Carlo simulations}\label{sec:MC}

We ask whether we can reproduce the observed distributions given a
set of reasonable assumptions for the true underlying
distributions of black hole masses and Eddington ratios. In
particular, our model should be able to reproduce the observed
distributions of FWHMs in each redshift-luminosity bin, and a
reasonable amount of uncertainty in virial relations. Also, the
model should produce a luminosity distribution consistent with the
observed quasar luminosity function. Since the BH mass and
Eddington ratio distributions may evolve with redshift, we
consider two redshift ranges: $0.7<z<1.0$, where the \MgII\
estimator is used throughout, and $1.9<z<2.1$, where the \CIV\
estimator is used. We only consider the subset of
uniformly-selected quasars in this section, and we restrict
ourselves to objects with $\lgL>45.9$ and $\lgL>46.6$,
respectively, above which our sample is close to complete in our
two redshift ranges (e.g., Richards et al. 2006a).

In outline, we begin by specifying a model for the true black hole
mass distribution. For a given mass, we will assume a single
underlying central Eddington ratio (and thus bolometric
luminosity) and a single underlying central FWHM. To get observed
quantities (continuum luminosities and FWHM), we add scatter {\em
independently} to the central Eddington ratio and FWHM. Our model
is thus characterized by the assumed black hole mass function, the
assumed relation between underlying Eddington ratios and black
hole mass, and the scatters which take us to the observed
luminosities and FWHM. We will present a model with reasonable
(although not necessarily unique) choices for these parameters,
which fits the data quite well.

The observed quasar luminosity function has a power-law form at
the bright end (Richards \etal\ 2006a; Hopkins \etal\ 2007).  We
therefore start by assuming a power-law black hole mass
distribution.  For a given true black hole mass $M_{\rm BH,true}$, we
assume a central underlying Eddington ratio:
\begin{equation}\label{eqn:sim1}
\left \langle \log \left({L_{\rm bol} \over L_{\rm Edd}}\right)\right \rangle = C_1+C_2\log \left({M_{\rm
BH, true}\over M_\odot}\right)\ ,
\end{equation}
where $C_1$ and $C_2$ are constants to be set by the model. Given
this Eddington ratio and our fiducial bolometric correction, we
can determine a central continuum luminosity. Given this
luminosity and the true black hole mass, we {\em assume} that the
virial estimator, equation~(\ref{eqn:virial_estimator}), with the
observed values of $a$ and $b$ appropriate for the sample we're
simulating, holds exactly for the central value of FWHM. That is,
this central FWHM is given by:
\begin{equation}
\begin{split}
&\left\langle\log \left({{\rm FWHM} \over {\rm
km\,s^{-1}}}\right)\right\rangle =\\
&{1
  \over 2} \left[\log
\left({M_{\rm BH,true} \over M_\odot}\right) - a -
b\left\langle\log\left({\lambda L_{\lambda} \over 10^{44}\,{\rm
erg\,s^{-1}}}\right)\right\rangle\right]\ .
\end{split}
\label{eq:meanFWHM}
\end{equation}

This does not give us observed quantities yet. To go from the
central to {\em observed} bolometric and continuum luminosities,
we add a scatter to each simulated object selected from a
log-normal distribution of width $\sigma_{\rm E}$, which is then
also the width of the Eddington ratio distribution at this fixed
true BH mass\footnote{Here we have assumed that bolometric
luminosity is always perfectly proportional to continuum
luminosity, because this is how we determined the {\em observed}
bolometric luminosities in previous sections.}. And to go from the
central to the observed emission-line FWHM, we add a scatter to
each simulated object selected from a log-normal distribution of
width $\sigma_{\rm FHWM}$. Given the observed quantities, we can
simulate an ``observed'' virial black hole mass. By construction,
then, at a given true mass, the observed virial mass is unbiased
in the mean, and the uncertainty of virial estimator at this fixed
true mass is:
\begin{equation}\label{eqn:sig_relation}
\sigma_{\rm vir}=\sqrt{(b\sigma_{\rm E})^2+(2\sigma_{\rm
FWHM})^2}\ .
\end{equation}

The two scatters we have added, $\sigma_{\rm E}$ and $\sigma_{\rm
FWHM}$, are statistically independent. We discuss the physical
interpretation of these scatters, the effects of any possible
correlated terms, and how these quantities relate to the physical
quantities of true BLR size and BLR virial velocity at fixed true
BH mass, in \S~\ref{sec:disc3}.

Once we have simulated luminosities for each object we can impose
various luminosity cuts and make comparisons with observations. We
have the freedom to vary the normalization ($C_1$) and mass
dependence ($C_2$) of the central Eddington ratio in eqn.
(\ref{eqn:sim1}), and the magnitude of the uncorrelated scatters
in both FWHM ($\sigma_{\rm FWHM}$) and Eddington ratio
($\sigma_{\rm E}$), as well as the slope of the underlying
power-law BH mass distribution.

The observed quasar luminosity function has a power-law form at
the bright end with slope $\gamma_L\sim -2$ (Richards \etal\
2006a; Hopkins \etal\ 2007). We therefore start by assuming a
power-law BH mass function with slope $\gamma_M$.  The slope of
the simulated bolometric luminosity distribution will be $\sim
\gamma_M/(1+C_2)$ in the limit of small $\sigma_{\rm E}$, which is
required to match with $\gamma_L\sim -2$. Although there is an
observed break in the luminosity function (and possibly in the
mass function as well), we do not include such a break in our
modeling, but return to that possibility below.

The parameters $C_1$, $C_2$, and $\sigma_{\rm FWHM}$ are jointly
constrained by the observed FWHM distribution in our sample: if
the central Eddington ratio is too large or too small, the central
value of FWHM will be correspondingly under(over)estimated; if the
slope $C_2$ between $\log M_{\rm BH, true}$ and $\langle\log
L_{\rm bol}/L_{\rm Edd}\rangle$ is too small, the central FWHM
will show a systematic trend with BH mass and luminosity stronger
than the one we actually observe; if $\sigma_{\rm FWHM}$ is too
large, then the observed FWHM distribution will be too broad. The
comparison between the observed FWHM distribution and the
simulated one must be made in each luminosity bin separately. Once
we are able to reproduce the FWHM distribution and the luminosity
distribution, we automatically reproduce the virial BH mass and
apparent Eddington ratio distributions.

Finally, $\sigma_{\rm E}$ and $\sigma_{\rm FWHM}$ are jointly
constrained by the requirement $\sigma_{\rm
vir}=\sqrt{(b\sigma_{\rm E})^2+(2\sigma_{\rm FWHM})^2}\gtrsim 0.3$
dex. If both $\sigma_{\rm E}$ and $\sigma_{\rm FWHM}$ are too
small, we will predict an uncertainty of $\sigma_{\rm vir}<0.3$
dex, an accuracy below that of reverberation mapping. We will
discuss this point further in \S\ref{sec:disc3}.

\begin{deluxetable}{lllllll}
\tablecolumns{7}\tablewidth{0.45\textwidth} \tablecaption{Model
parameters} \tablehead{Redshift & $\gamma_M$ & $C_1$ & $C_2$  &
$\sigma_{\rm E}$  & $\sigma_{\rm FWHM}$  & $\sigma_{\rm
vir}$ \\
& & & & (dex) & (dex) & (dex)} \startdata
$0.7<z<1.0$    & $-2.6$ & $-3.60$ & $0.3$  & $0.4$  & $0.11$ & $0.33$ \\
$1.9<z<2.1$  & $-2.6$ & $-2.88$ & $0.2$  & $0.4$  & $0.11$ &
$0.31$
\enddata
\tablecomments{See the text for symbol meanings. }
\label{tab:para}
\end{deluxetable}

\subsection{Results}
\label{sec:MC-results} We carry out the comparison between our
simulation for a given set of model parameters and the observed
distributions at three luminosity bins in each redshift bin. We
systematically and manually search the parameter space to find
values where the simulated distributions agree with observations,
judged by eye. The best-fit parameters are listed in Table
\ref{tab:para}. We have not searched the parameter space
exhaustively enough to claim unequivocally that these solutions
are unique, but the arguments we gave at the end of the previous
section suggest it is unlikely that there is another region of
parameter space which will satisfy all the constraints.  Of
course, more complicated models (e.g., deviations from a pure
power law in the mass distribution) could also be found with
different parameters.  Note also that we do not quote formal
errors on the parameters, given the way in which we found the best
solution.

Given the sets of parameters in Table~\ref{tab:para},
eq.~\ref{eqn:sig_relation} yields an uncertainty of the virial
masses at fixed true BH mass of $\sigma_{\rm vir}=0.33$ dex and
$0.31$ dex for the \MgII\ and \CIV\ cases
respectively\footnote{These values are slightly smaller than the
observed values (e.g., McLure \& Jarvis 2002; Vestergaard \&
Peterson 2006), but our $\sigma_{\rm vir}$ is the scatter at fixed
BH mass, while the quoted uncertainties are determined from a
sample of reverberation mapping objects covering a range of BH
masses.}. The central Eddington ratio for a typical quasar with
$M_{\rm BH,true}=10^8\ M_\odot$ is $\langle\log L_{\rm bol}/L_{\rm
Edd}\rangle =-1.2$ and $-1.3$ for the two cases; the
(insignificant) difference arises because of the difference in the
virial relations and in the observed FWHM distributions.  These
distributions of Eddington ratios must be interpreted with care,
since they are heavily dependent on the exact forms and scatter of
the virial relations, which are not well-understood at this time.

\begin{figure}
  \centering
    \includegraphics[width=0.45\textwidth]{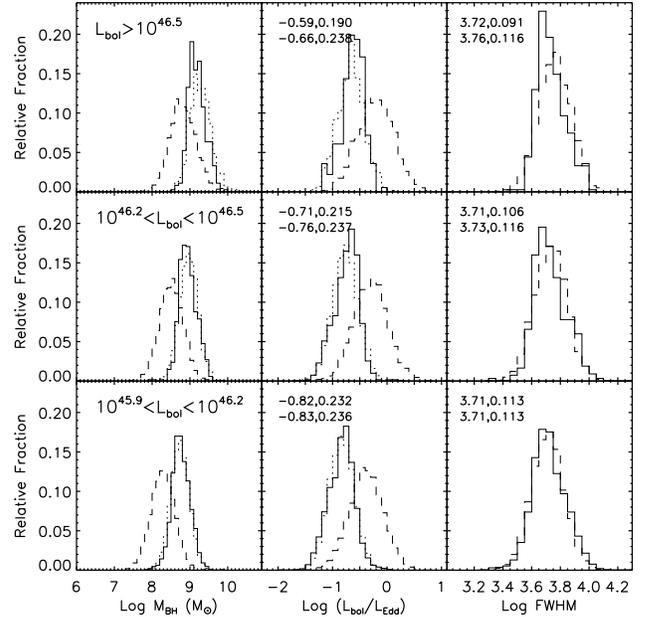}
    \caption{Comparison between observed and simulated distributions for \MgII\ at $0.7<z<1.0$.
     The assumed underlying true BH mass distribution is a power-law with slope $\gamma_M=-2.6$.
     The assumed Eddington ratio distribution at fixed true BH mass is $\log L_{\rm bol}/L_{\rm Edd}=-3.6+0.3\log M_{\rm BH,true}+\sigma_{\rm E}$, where
     $\sigma_{\rm E}=0.4$ dex. The survey is complete above $\lgL=45.9$, and the
     observed slope
     $\gamma_L\sim -2$ in the power-law distribution of bolometric luminosities is reproduced. Each comparison is done for
     three luminosity bins arranged in rows. In each panel, the solid histograms are observed distributions; the dotted histograms are
     simulated distributions for virial masses and apparent Eddington ratios based on virial masses. In the BH mass and Eddington ratio
     panels, we also show the distributions of the true underlying BH masses and Eddington
     ratios in dashed histograms. The simulated FWHM distributions in
     the third column are shown as dashed histograms. In the second and third column,
     the numbers show the mean and $\sigma$ of the
     fitted log-normal for both the observed (upper) and simulated (lower) distributions.}
    \label{fig:sim1}
\end{figure}

\begin{figure}
  \centering
    \includegraphics[width=0.45\textwidth]{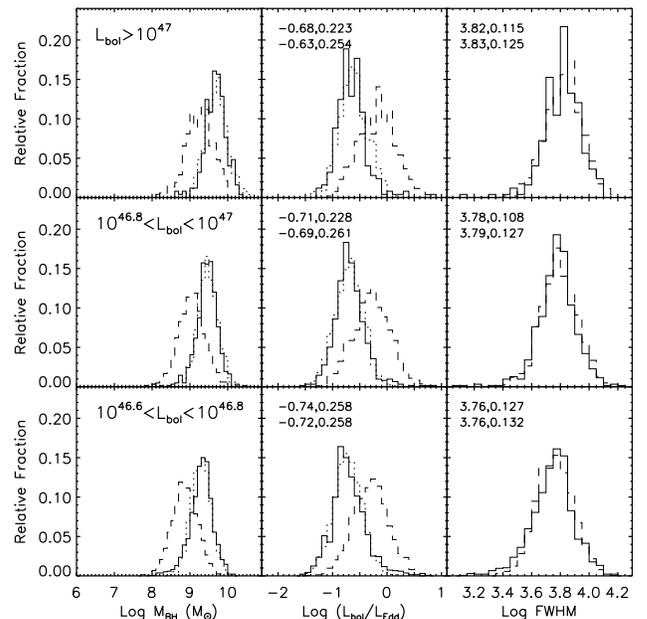}
    \caption{Comparison between observed and simulated distributions for \CIV\ at $1.9<z<2.1$.
     The assumed underlying true BH mass distribution is a power-law with slope $\gamma_M=-2.6$.
     The assumed Eddington ratio distribution at fixed true BH mass is $\log L_{\rm bol}/L_{\rm Edd}=-2.88+0.2\log M_{\rm BH,true}+\sigma_{\rm E}$
     with $\sigma_{\rm E}=0.4$ dex. The survey is complete above $\lgL=46.6$, and the
     observed slope
     $\gamma_L\sim -2$ in the power-law distribution of bolometric luminosities is reproduced. Notations are similar to Fig.~\ref{fig:sim1}.}
    \label{fig:sim2}
\end{figure}

The comparison of the simulated and observed results are displayed
in three luminosity bins in Figs.~\ref{fig:sim1} (for \MgII) and
\ref{fig:sim2} (for \CIV). The figures show the distributions of
BH masses (left), Eddington ratios (middle) and FWHMs (right).
Our simulations are very successful at reproducing the observed
distributions of virial BH masses, Eddington ratios and FWHMs. The
virial BH masses (solid and dotted histograms) on average
overestimate the true BH masses (dashed histograms) in each
luminosity bin because of the Malmquist bias caused by the
finite scatter of the black hole mass estimator (the part of
scatter in virial masses that originates from luminosity) and the
fact that there are more low-mass than high-mass black holes.

The widths of the observed BH mass and Eddington ratio distributions
in each luminosity bin are smaller than those of the true quantities,
and the Eddington ratios are biased low.  However, although there are
more quasars
at the faint luminosity end, these faint objects are rarely
scattered into high-luminosity bins, because of the finite width
of the true Eddington ratio distribution. Running our model in
alternate redshift bins ($1.5<z<1.7$ for \MgII\ and $3.0<z<3.2$
for \CIV) yields similar parameters. Thus our model is consistent
with a non-evolving true Eddington ratio distribution, at least at
the current levels of observations.

There are objects with $L_{\rm
bol}/L_{\rm Edd}>1$ in Figs. \ref{fig:sim1} and \ref{fig:sim2},
caused by the high-luminosity tail in the scatter $\sigma_{\rm
E}=0.4$ dex at individual BH masses.  However, the fraction of
these super-Eddington objects is small and the formal Eddington
limit is only approximate.  Enforcing a cutoff in the model
Eddington ratio distribution has little effect on our conclusions.

Malmquist bias will be present independent of any luminosity cut on the
sample,  given the scatter between
the true and ``observed'' masses and a bottom-heavy true BH mass
distribution. This Malmquist-type bias is important in analyses
that are based on ``observed'' BH masses, such as the active BH
mass function (e.g., McLure \& Dunlop 2004) and the
redshift-evolution of the $M-\sigma$ relation (Lauer \etal\ 2007).

We have demonstrated the Malmquist bias with simulations, but the
form of the bias can be shown analytically.  We start by deriving
the bias at fixed observed virial mass, and then generalize to the
bias at fixed observed luminosity. Following Lynden-Bell \etal\
(1988), we model the probability distribution of $m_{e}\equiv\ln
M_{\rm BH,vir}$ given $m\equiv\ln M_{\rm BH,true}$ as a Gaussian;
that is:
\begin{equation}\label{eqn:bias_1}
p_0(m_e|m)=(2\pi \Delta_m^2)^{-1/2}\exp
\left[-\frac{1}{2}(m-m_e)^2/\Delta_m^2\right]\ ,
\end{equation}
where $\Delta_m=(\ln 10)\sigma_{\rm vir}$ is the Gaussian
$\sigma$. Let $n\,d \ln M_{\rm BH,true}$ be the number of BHs with
masses between $\ln M_{\rm BH,true}$ and $\ln M_{\rm BH,true}+d\ln
M_{\rm BH,true}$. For a power-law distribution of the underlying
true BH masses, $n \propto M_{\rm BH,true}^{\gamma_M}$, we have:
\begin{equation}\label{eqn:bias_2}
n\frac{dM_{\rm BH,true}}{M_{\rm BH,true}}\equiv n\ dm\propto
M_{\rm BH,true}^{\gamma_M-1}\ dM_{\rm BH, true}\propto
e^{\gamma_Mm}\ dm\ ,
\end{equation}
where $\gamma_M$ is our assumed slope in the underlying power-law
BH mass distribution. Then the distribution of $m$ for a given
$m_e$ is
\begin{equation}\label{eqn:bias_3}
\begin{split}
p_1(m|m_e)&=p_0e^{\gamma_M m}\left[\int p_0e^{\gamma_Mm}dm\right]^{-1}\\
&=(2\pi\Delta_m^2)^{-1/2}\exp
\left\{-\frac{1}{2}[m-(m_e+\gamma_M\Delta_m^2)]^2/\Delta_m^2\right\}\
.
\end{split}
\end{equation}
This is a Gaussian distribution, centered on the mass $\langle m
\rangle= m_e + \gamma_M\Delta_m^2$ and therefore the bias in
$\log$ mass is $-\gamma_M\Delta_m^2/\ln10$. Inserting
$\gamma_M=-2.6$, $\sigma_{\rm vir}\approx 0.3$ and $\Delta_m=(\ln
10)\sigma_{\rm vir}\approx 0.7$ produces a bias in $\log$ mass of
$0.55$ dex.

However, the Malmquist bias we derived above is the bias at fixed
virial mass, not the bias at fixed luminosity
(Figs.~\ref{fig:sim1} and \ref{fig:sim2}). To calculate the
latter, we proceed similarly.

Let $l\equiv \ln L_{\rm bol}$ and $m\equiv \ln M_{\rm BH, true}$,
then the probability distribution of $l$ at fixed $m$ is:
\begin{equation}
\begin{split}
p_0(l|m)=(2\pi \Delta_l^2)^{-1/2}\exp
\left[-\frac{[l-(C_3+C_4m)]^2}{2\Delta_l^2}\right]\ ,
\end{split}
\end{equation}
where the constants $C_3\equiv [C_1+\log(1.3\times 10^{38})]\ln
10$, $C_4\equiv 1+C_2$ and $\Delta_l\equiv (\ln10)\sigma_{\rm E}$
following equation (\ref{eqn:sim1}).

The probability distribution of $m$ at fixed $l$ is then
\begin{equation}
\begin{split}
p_1(m|l)&=p_0e^{\gamma_Mm}\left[\int
p_0e^{\gamma_Mm}dm\right]^{-1}\\
&=\left(2\pi\frac{\Delta_l^2}{C_4^2}\right)^{-1/2}\exp\left\{-\frac{C_4^2}{2\Delta_l^2}\left[m-\frac{C_4(l-C_3)+\gamma_M\Delta_l^2}{C_4^2}\right]\right\}\
,
\end{split}
\end{equation}
which is a Gaussian distribution with dispersion $\Delta_l/C_4$
centered on $\langle
m\rangle=(l-C_3)/C_4+\gamma_M\Delta_l^2/C_4^2$. A similar analysis
for the distribution of the ``observed'' BH mass $m_{e}\equiv\ln
M_{\rm BH,vir}$ at fixed luminosity gives a Gaussian distribution
with mean $\langle m_e\rangle=(l-C_3)/C_4$ and dispersion
$2\,\sigma_{\rm FWHM}\,\ln 10$.  The Malmquist bias, i.e., the
difference in the two means is $-\gamma_M\Delta_l^2/(C_4^2\ln 10)
\approx 0.6$ dex when plugging in numbers.  The distribution of
log true masses at fixed luminosity has a width of
$\Delta_l/(C_4\ln10)=0.32$ dex; while the distribution of log
virial masses at fixed luminosity has a width of $\sigma_{\rm
vir}^\prime=2\,\sigma_{\rm FWHM}= 0.22$ dex. These results are in
excellent agreement with what we see in Figs. \ref{fig:sim1} and
\ref{fig:sim2}.

The Malmquist bias is mainly caused by the bottom-heavy true BH
mass distribution and the scatter $\sigma_{\rm vir}$ in the virial
estimators (or the uncorrelated scatter $\sigma_{\rm E}$ in the
case of fixed luminosity). However, we emphasize that the exact
magnitude and form of the bias depends in detail on the partition
and behavior of the scatter in the virial relations. We have used
$\sigma_{\rm E}=0.4$ dex and $\sigma_{\rm FWHM}= 0.11$ dex in
order to reproduce the observed uncertainty in virial BH
estimators of $\sigma_{\rm vir}\gtrsim 0.3$ dex and the observed
FWHM distributions, but a smaller scatter of virial BH mass at
fixed true mass would reduce the Malmquist bias. Also, the values
of $\sigma_{\rm E}$ and $\sigma_{\rm FWHM}$ could depend on
luminosity.  Our data do not allow us to consider such refinements
in detail. Since the Malmquist bias is proportional to
$\gamma_{M}$, we expect a smaller bias for objects fainter than
the break in the luminosity function, such as those in the AGES
sample (Kollmeier \etal\ 2006).

Although our model of the underlying true BH mass
and Eddington ratio distributions can roughly reproduce the
observations, we have made a few simplifications. In particular,
the largest uncertainty arises from the assumed virial relations
and their scatters, for which there is no consensus yet. Changes
in these virial relations (e.g., a different slope in the $R-L$
relation, etc., see Bentz \etal\ 2006; Netzer \etal\ 2007) will
certainly change the values of our model parameters. We also note
that there might be other systematic errors associated with virial
BH masses or reverberation mapping masses in general (Krolik
2001). Given all the limitations and simplifications of our model,
we do not wish to overinterpret our modeled distributions at this
time, but rather highlight the danger that the virial masses for a
population of quasars could be subject to serious systematic
biases. Until we have both better observations and a theoretical
understanding of the systematic behavior of the virial estimators
with BH properties, we urge caution in the interpretation of BH
mass functions based on virial estimators.

\subsection{Active BH mass function in quasars}\label{sec:mass_fun}

Our model explicitly provides the true BH mass distribution and
connects it to the bolometric luminosity distribution via our
modelled Eddington ratio distribution, and thus provides an
estimate of the underlying BH mass function in quasars based on
our knowledge of the bolometric luminosity function.

An alternative method using optically selected samples uses the
observed luminosity function and the fraction of BHs above a
certain mass threshold within some luminosity range, to place
lower limits on the number density of active BHs above the mass
threshold (McLure \& Dunlop 2004). The measurement is a lower
limit simply because it neglects BHs scattered out of the
luminosity range being considered. Using this method, McLure \&
Dunlop (2004) found that the majority of $>10^{8.5}\ M_\odot$
SMBHs are already in place at $z\sim 2$ when geometric obscuration
and quasar duty cycles were taken into account. However, their
results were based on virial BH masses, and they did not take into
account the Malmquist bias that we have discussed here.

\begin{deluxetable}{lll}
\tablecolumns{3}\tablewidth{0.45\textwidth} \tablecaption{quasar
black hole mass function} \tablehead{$\bar{z}$ & $\Phi_{L_{\rm
bol}>10^{46}\ {\rm erg\ s^{-1}}}\ {\rm (Mpc^{-3})}$ &
$\Phi_{M_{\rm BH}>10^{9.5}\ M_\odot}\ {\rm (Mpc^{-3})}$}
\startdata
$0.85$ & $2.87\times 10^{-6}$   & $1.0\times 10^{-8}$ \\
$2.0$  & $1.20\times 10^{-5}$   & $7.4\times 10^{-8}$ \\
$3.1$  & $8.08\times 10^{-6}$   & $5.0\times 10^{-8}$
\enddata
\label{tab:mass_fun}
\end{deluxetable}

Here we use our model to estimate the cumulative BH mass function
in quasars above a certain mass threshold. Notice that the break
in the bolometric luminosity function (Hopkins \etal\ 2007) is at
$\lgL\sim 46$ at $z\sim 1-3$, corresponding to $\log M_{\rm BH,
true}\sim 8.8-9.0$ in our model (e.g., equation \ref{eqn:sim1}).
Thus our single power-law distribution of the underlying BH masses
is inaccurate below this mass threshold\footnote{Our conclusions
on the distributions of FWHMs and virial BH masses are insensitive
to this detail. However, the Malmquist bias will be smaller below
the break, as mentioned above.}, and will affect the mapping from
the BH mass function to the bolometric luminosity function. We
will explore the effects of a break in a future paper.

We focus on the cumulative mass function above $10^{9.5}\
M_\odot$, i.e., the most massive SMBHs. The estimation is simple:
we take our simulated catalogs at different redshifts and count
the number of BHs with mass greater than $10^{9.5}\ M_\odot$ and
the number of BHs with $\lgL>46$; the latter number is
compared to the cumulative bolometric luminosity function (Hopkins
\etal\ 2007, the ``full'' model in their table 3) with $\log
L_{\rm bol}>46$ and normalized; then we use this normalization to
estimate the number density of active BHs with $\log M_{\rm BH,
true}>9.5$. Our results are listed in Table \ref{tab:mass_fun},
for three redshift bins: $z=0.7-1.0$, $1.9-2.1$, $3.0-3.2$. The
number density of BHs with mass $>10^{9.5}\ M_\odot$ in optical
quasars peaks around $z\sim 2$, and then decreases at lower
redshift.

The numbers above are lower limits on the total (active and
inactive) SMBH population because of quasar duty cycles and the
presence of obscured populations of active BHs at high redshift.
Surveys in other wave bands, especially in the X-ray and infrared,
are starting to reveal those missing active SMBHs (e.g., Treister
\etal\ 2006; see also Zakamska \etal\ 2003), and will provide better
constraints of the BH mass
function at high redshift, given our knowledge of the local
dormant BH mass function (e.g., Marconi \etal\ 2004; Merloni
2004). We note that the numbers derived here are fully consistent
with current constraints on the local BH mass function at the
high-mass end (e.g., Tundo \etal\ 2007; Lauer \etal\ 2007), and
allow an additional boost factor of $\sim 10-20$ due to quasar
duty cycles and geometric obscuration at $z\sim 2$.

\subsection{Results for radio loud quasars and BALs}\label{sec:BAL}

In this section, we determine virial BH masses for radio-loud
quasars and BALs, to explore whether their black hole properties
are distinctive from ``ordinary'' quasars of similar luminosity
and redshift. The Eddington ratio and BH mass distributions will
be affected by Malmquist bias, but the bias should be similar for
each individual subsample, allowing this kind of comparison to be
made.

\begin{figure}
  \centering
    \includegraphics[width=0.45\textwidth]{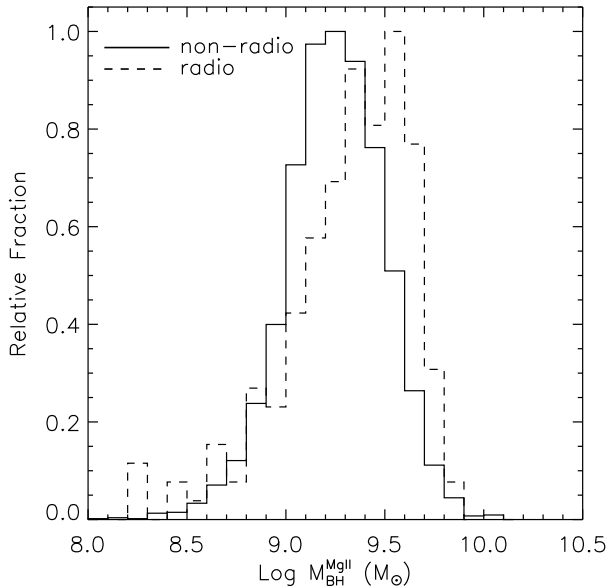}
    \caption{Distributions of \MgII\ based virial masses for the radio-undetected (solid
      histogram) and radio-loud (dashed histogram) quasars
    in subsamples matched in redshift and luminosity distributions;
    see text.}
    \label{fig:RQ_RL_FWHM}
\end{figure}

For the radio analysis, we carve out a narrow range in
redshift-luminosity space, $\lgL=46.6-46.75$ and $z=1.5-2.3$, over
which we have \MgII\ virial masses for each source, and the {\em
FIRST}-detected and undetected quasars have similar redshift and
optical luminosity distributions. At these redshifts, the {\em
FIRST} radio flux limit corresponds to a radio luminosity density
of $\sim 3 \times 10^{25}$ Watts/Hz, which is radio-loud by all
standard definitions.  This region contains 3360 radio-undetected
quasars and 170 radio-loud quasars. The radio-loudness, defined as
${\cal R}=f_{\rm 6\ cm}/f_{2500}$ for these {\em FIRST}-detected
objects ranges from $\sim 5$ to $\sim 4600$ with median value
$\sim 100$, based on a subset of DR3 quasars studied by Jiang
\etal\ (2007b). Fig.~\ref{fig:RQ_RL_FWHM} shows that the \MgII\
based virial mass distributions for the radio-undetected and
radio-loud quasars are slightly different, with radio-loud quasars
having slightly larger median value by $\sim 0.12$ dex, comparable
to the results in McLure \& Jarvis (2004). Radio-loud quasars in
our sample have smaller Eddington ratios than radio-undetected
quasars on average, as has been found in previous analyses (e.g.,
Ho 2002).

For BALs (most of which are \CIV\ BALs), the situation is more
complicated. First, BALs in general are more reddened than nonBALs
(e.g., Reichard \etal\ 2003), affecting the bolometric luminosity
estimates. Second, the broad absorption troughs can severely
impact the reliability of FWHM measurements; the concerns about
disk winds affecting the measurement of the \CIV\ line (discussed in
\S~\ref{sec:MgII-CIV}) are
especially important for BALs.

\begin{figure}
  \centering
    \includegraphics[width=0.45\textwidth]{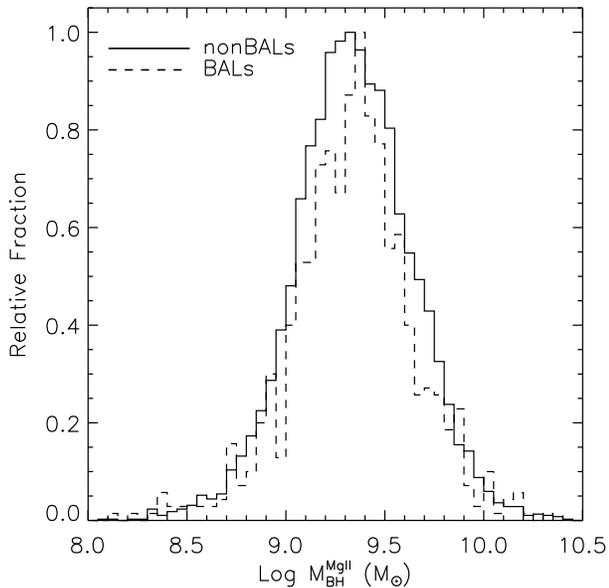}
    \caption{Distributions of \MgII\ based virial masses for nonBALs (solid histogram) and BALs (dashed histogram)
    with $1.7\le z\le 2.2$, which are within the uniformly-selected quasar sample. The redshift and luminosity distributions of
    nonBALs and BALs are identical.}
    \label{fig:BAL_BH}
\end{figure}

Fortunately, for $1.7\lesssim z\lesssim 2.2$ we have both \MgII\
and \CIV, thus it becomes feasible to compare the difference in
virial mass for high-ionization BALs (which appear in the \CIV\
line only) and nonBALs, based on the \MgII\ estimator. For the
uniformly-selected quasar sample the redshift and luminosity
distributions are almost identical for \CIV\ BALs and nonBALs at
$1.7\lesssim z\lesssim 2.2$ (Shen \etal\ 2008). We choose
uniformly-selected quasars that have redshift $1.7\le z\le 2.2$
and have \MgII\ based virial masses; the sample includes 5408
nonBALs and 796 BALs. Fig.~\ref{fig:BAL_BH} shows that their
\MgII\ virial mass distributions are quite similar, with median
mass $\log M_{\rm BH,vir}= 9.34$ for both. This result is
consistent with recent findings by Ganguly \etal\ (2007) based on
5088 quasars from SDSS DR2, although the mean virial BH masses are
different due to difference in the calibrations used and line
measurements, as we discuss below. This similarity in virial BH
masses for BALs and nonBALs supports the idea that BALs are
intrinsically no different from nonBALs, and the BAL phenomenon is
caused by different viewing angles with respect to a disk wind
(e.g., Weymann \etal\ 1991; Elvis 2000). This picture is also
consistent with our recent findings that BALs show similar
large-scale clustering to nonBALs (Shen \etal\ 2008).

\section{Discussion}\label{sec:discussion}

\subsection{General issues with virial estimators}\label{sec:disc1}
The virial BH mass measurements presented in this paper are
consistent with, but not identical to results from a variety of papers (e.g.,
McLure \& Dunlop 2004; Dietrich \& Hamann 2004; Kollmeier \etal\
2006; Ganguly \etal\ 2007). The difference in the mean of the
distribution can be due to a variety of effects: different
samples, different virial estimators and different calibrations
for the same estimator, as well as different procedures to measure
luminosities and line widths used in these virial estimators. We
elaborate below how these various choices lead to differences in
virial masses even for a given underlying sample.

\subsubsection{\em Different virial calibrations for the same line
estimator.}

There are several versions in the literature of a given virial
estimator, and they do not generally yield the same virial mass
when applied to SDSS or other quasar/AGN samples. For example, the
\hbeta\ calibrations in McLure \& Dunlop (2004) and in Vestergaard
\& Peterson (2006) differ by $\sim 0.17$ dex given the mean
luminosity $\lambda L_{\lambda,5100}\sim 10^{44.6}\ {\rm erg\
s^{-1}}$ in our sample. Likewise, for the \MgII\ estimator, the
version in McLure \& Dunlop (2004) yields a larger virial mass by
$\sim 0.3$ dex for $\lambda L_{\lambda,3000}=10^{46.1}\ {\rm erg\
s^{-1}}$ (which corresponds to SDSS quasars at $z\sim 2$), than
the older version in McLure \& Jarvis (2002). The latter case
partially explains the discrepancy between our measurements and a
subset of our sample measured by Ganguly \etal\ (2007). While we
have used the most recent calibrations available (McLure \& Dunlop 2004;
Vestergaard \& Peterson 2006), it is quite likely that those
calibrations will change when the updated $R-L$ relation and
virial coefficient (zero-point offset) are incorporated (Onken \etal\
2004; Bentz \etal\ 2006; Kaspi \etal\ 2007).

\subsubsection{\em Different ways to measure luminosity and FWHM.}

Even for the same calibration, different authors have used a
variety of ways to measure luminosity and line width parameters,
which will lead to systematically different results. For example,
the mean \MgII\ FWHM taken from the broad component of a
two-Gaussian fit is larger by $\sim 0.15$ dex than the
non-parametric FWHM measured from the modelled spectrum which best
reproduces the line profile. This contributes to the
underestimation of the \MgII\ based virial masses (e.g., Dietrich
\& Hamann 2004; Ganguly \etal\ 2007). For \CIV, on the other hand,
it is more reasonable to use the FWHM measured from the modelled
flux that best reproduces the \CIV\ line profile because that is
how the \CIV\ relation is calibrated (e.g., Vestergaard \&
Peterson 2006). Although some argue that a narrow-line component
is present in the \CIV\ line (e.g., Marziani \etal\ 1996) and
recipes have been provided for subtracting it (cf., Bachev \etal\
2004), we cannot perform such an exercise until the \CIV\
estimator is re-calibrated with narrow-line-subtracted line
widths. Indeed, it is {\em crucial} to use the methods of
measuring FWHMs used in the original calibrations of these virial
relations if one wants to obtain unbiased results. We feel
confident about our choices, as we found no mean offset between
virial masses based on different lines
(Figs.~\ref{fig:Hbeta_MgII_hist} and \ref{fig:CIV_MgII_BH}).

\subsubsection{\em Other issues with our measurements.}

There are some additional factors that might lead to differences
between our measurement and others for the same sample of quasars.
First, the spectroscopic flux scale used in this study is higher
than previous SDSS data releases (DR5 and prior) by $\sim 0.14$
dex, giving rise to a systematic $\sim 0.07$ dex increase in
virial masses. Second, we have separated the broad and narrow
components of \hbeta\ and \MgII\ at a FWHM of 1200 km s$^{-1}$,
following Hao \etal\ (2005). It is possible that we are biasing
the mean virial mass by doing so. Nevertheless, these effects are
minor considering the uncertainties in the zero-point offsets of
these virial estimators.

As we discussed in \S\ref{sec:Eddington_dist}, given the
independence of the observed FWHM distribution with luminosity,
objects in the $M_{\rm BH,vir}-L_{\rm bol}$ plane tend to lie on a
narrow stripe with a slope $b\sim 0.5$ (see Fig.
\ref{fig:Edd_dist}). The lower-luminosity high redshift quasars in
the Kollmeier \etal\ (2006) sample would also fall on the stripe.
The mean apparent Eddington ratio $\langle \log (L_{\rm
bol}/L_{\rm Edd})\rangle_{\rm vir}\propto (1-b)\log L_{\rm bol}$.
Thus the expected $\langle \log (L_{\rm bol}/L_{\rm
Edd})\rangle_{\rm vir}$ would be $\sim -1.75$ for $\lgL=44$, $\sim
-2.25$ for $\lgL=43$, and $\sim -3.25$ for $\lgL=41$ for $b\sim
0.5$. These Eddington ratio values are in good agreement with the
typical Eddington ratios from X-ray selected AGN samples which
cover the bolometric luminosity range $L_{\rm bol}\sim
10^{41}-10^{44}\ {\rm erg\ s^{-1}}$ (e.g., Panessa \etal\ 2006;
Babi\'{c} \etal\ 2007; Ballo \etal\ 2007). However, there are two
reasons to be cautious about this apparent agreement: current
X-ray selected AGN samples still suffer from limited sample size
and various incompletenesses, and most of the BH masses in X-ray
selected AGN are estimated based on host galaxy properties rather
than virial methods, and it remains unclear how good these methods
agree with each other for those low-luminosity BHs.

On the other hand, virial methods have their own limitations due
to their reliance on poorly understood BLR physics. If the
existence of the broad line region itself depends on Eddington
ratio, then it is not surprising that broad line AGN/quasars lie
preferentially within a favorable Eddington ratio range. Moreover,
the width of the Eddington ratio distribution at fixed luminosity
could be underestimated, if, for example, the line widths only
partially reflect the virial velocities in the BLR, as we will
further discuss in \S\ref{sec:disc3}.

\subsection{Limitations of the SDSS sample}\label{sec:disc2}
Although the large size of the SDSS sample provides unprecedented
statistics, we feel obligated to point out its limitations.

First, the dynamic range in luminosity is narrow for SDSS quasars.
Only luminous quasars are included in the flux-limited sample.
Other surveys, such as the AGES survey (Kochanek \etal\ 2004) and
the 2dF-SDSS LRG and QSO (2SLAQ) Survey (Cannon \etal\ 2006)
extend to fainter luminosities at the same redshift.  Our sample
does include some objects of lower luminosity, but they are
incomplete in ways that are hard to quantify.

Second, the wavelength range of SDSS spectra has forced us to use
different virial estimators at different redshift. Although we
cross-compared virial relations of \hbeta-\MgII\ and \MgII-\CIV,
the direct comparison between the \hbeta\ and \CIV\ estimators is
impossible with SDSS spectra.  Thus it is important to study these
broad lines simultaneously with multi-band spectra (e.g., Sulentic
\etal\ 2000 and references therein; Baskin \& Laor 2005). It is
also desirable to obtain near-IR spectra for high-redshift quasars
with optical spectra, to get \hbeta\ or \MgII\ measurements (e.g.,
Sulentic \etal\ 2006; Jiang \etal\ 2007a; Kurk \etal\ 2007; Netzer
\etal\ 2007), since the \CIV\ virial estimator alone is perhaps
questionable for the reasons we have discussed in this paper.

\subsection{Issues with our model}\label{sec:disc3}
In \S\ref{sec:MC} we introduced a statistical model which can
reproduce the observed luminosity and FWHM distributions. We now
provide some justifications of our choices of model parameters,
and discuss the connections between the real physical quantities,
i.e., BLR size and virial velocity, and their surrogates,
luminosity and FWHM.

An important assumption of our model is the origin of the
uncertainties in virial estimators, i.e., for fixed true BH mass,
the virial estimator will give an estimate log-normally
distributed around the right mean value and with dispersion
$\sigma_{\rm vir}$. However, the value of $\sigma_{\rm vir}$
remains unclear: the rms scatter of virial masses around
reverberation mapping (RM) masses (e.g., McLure \& Jarvis 2002;
Vestergaard \& Peterson 2006) is $0.3-0.4$ dex, comparable to the
rms scatter of virial masses around BH masses derived from the
$M-\sigma$ relation (e.g., Greene \& Ho 2006); but the
relationship of this number to the scatter around {\em fixed true}
BH mass is unknown. Given all the issues with virial estimators we
discussed in \S\ref{sec:disc1}, it seems unlikely that virial
estimates based on single-epoch spectra are good to a factor of 2,
an accuracy that can barely be achieved with reverberation
mapping. Therefore we believe our choice of $\sigma_{\rm
vir}\gtrsim 0.3$ dex is appropriate.

At a given true BH mass, in our simplistic model, the broadening
of the luminosity distribution is completely uncorrelated with the
broadening of the FWHM distribution. Are our required values for
the uncorrelated scatters $\sigma_{\rm E}$ and $\sigma_{\rm FWHM}$
consistent with observations? Current reverberation mapping data
indicate a scatter of $\sim 0.2-0.35$ dex in luminosity at fixed
BLR size (e.g., Kaspi et al. 2005; Bentz et al. 2006). The best
studied reverberation mapping sources indicate scatters of $\sim
0.1$ dex in line width at fixed BLR size for a given object (e.g.,
Bentz et al. 2007). These values are for a handful of well studied
objects with reverberation mapping of the \hbeta\ line only, and
we might expect somewhat larger scatters for both luminosity and
line width in samples with a wider range of luminosity and
redshift than probed by current RM samples.  Thus while our
choices of uncorrelated scatters $\sigma_{\rm E}=0.4$ dex and
$\sigma_{\rm FWHM}=0.11$ dex may appear rather large, they are at
least plausible.

We still expect some component of the variations in luminosity
to correlate directly with variations in FWHM, as is seen in local
samples.  Here we investigate what magnitude of correlated broadening
can be supported given existing observations.
At fixed true BH mass, the virial mass
estimate is expressed in terms of its luminosity and FWHM:
\begin{equation}
\begin{split}
\log M_{\rm BH,vir}=C&+b\log L_{\rm bol}+ 2\log {\rm FWHM}\\
=C &+b\left(\langle \log L_{\rm bol} \rangle +\delta_{\rm
E}+\delta_{\rm corr}\right)\\
& +2\left(\langle\log{\rm FWHM}\rangle + \delta_{\rm
FWHM}-0.5b\delta_{\rm corr} \right)\ ,
\end{split}
\label{eq:virial_correlated}
\end{equation}
where $b$ is the slope of the $R-L$ relation, the various
$\delta_i$ denote Gaussian-random variables with dispersions
$\sigma_i$, and other constants have been absorbed in $C$. At this
fixed true BH mass, the bolometric luminosity and the FWHM both
follow a log-normal distribution, $\log L_{\rm bol}=\langle \log
L_{\rm bol} \rangle +\delta_{\rm E}+\delta_{\rm corr}$ and $\log
{\rm FWHM}=\langle{\log\rm FWHM}\rangle + \delta_{\rm
FWHM}-0.5b\delta_{\rm corr}$. In addition to the uncorrelated
terms $\delta_{\rm E}$ and $\delta_{\rm FWHM}$ as modelled in
\S\ref{sec:MC}, we also introduced correlated variations
$\delta_{\rm corr}$ and $-0.5b\delta_{\rm corr}$ in luminosity and
FWHM, respectively, whose amplitude we will constrain below. Note
that these correlated terms cancel each other out, and hence do
not contribute to the uncertainty in the virial relation.

The correlated terms in equation~(\ref{eq:virial_correlated})
reflect the intrinsic variation in BLR size/virial velocity at
fixed true BH mass. In our model formalism in \S\ref{sec:MC} we
have set $\sigma_{\rm corr}=0$. Since our observed FWHM
distributions are quite narrow, we found that we
need\footnote{This upper limit is set by the condition that the
width of the correlated part $-0.5b\delta_{\rm corr}$ is half of
$\sigma_{\rm FWHM}$ in FWHM so that its contribution to the
broadening of the FWHM distribution is negligible when added in
quadrature.} $\sigma_{\rm corr}\lesssim 0.2$ dex, such that the
distribution of simulated FWHMs will not be broader than the
observed distribution, keeping the requirement that $\sigma_{\rm
vir}\gtrsim 0.3$ dex. With this limit, the intrinsic dispersion in
BLR size for our sample is $\sigma_{\rm BLR}\lesssim 0.1$ dex at
fixed true BH mass, which is narrower than the typical dispersion
in BLR size inferred from less luminous reverberation mapping
sources. It is possible that the intrinsic dispersion of BLR size
at fixed true BH mass is indeed smaller for higher-luminosity
objects. Reverberation mapping data for high luminosity objects
are required to test this scenario.

\begin{figure}
  \centering
    \includegraphics[width=0.45\textwidth]{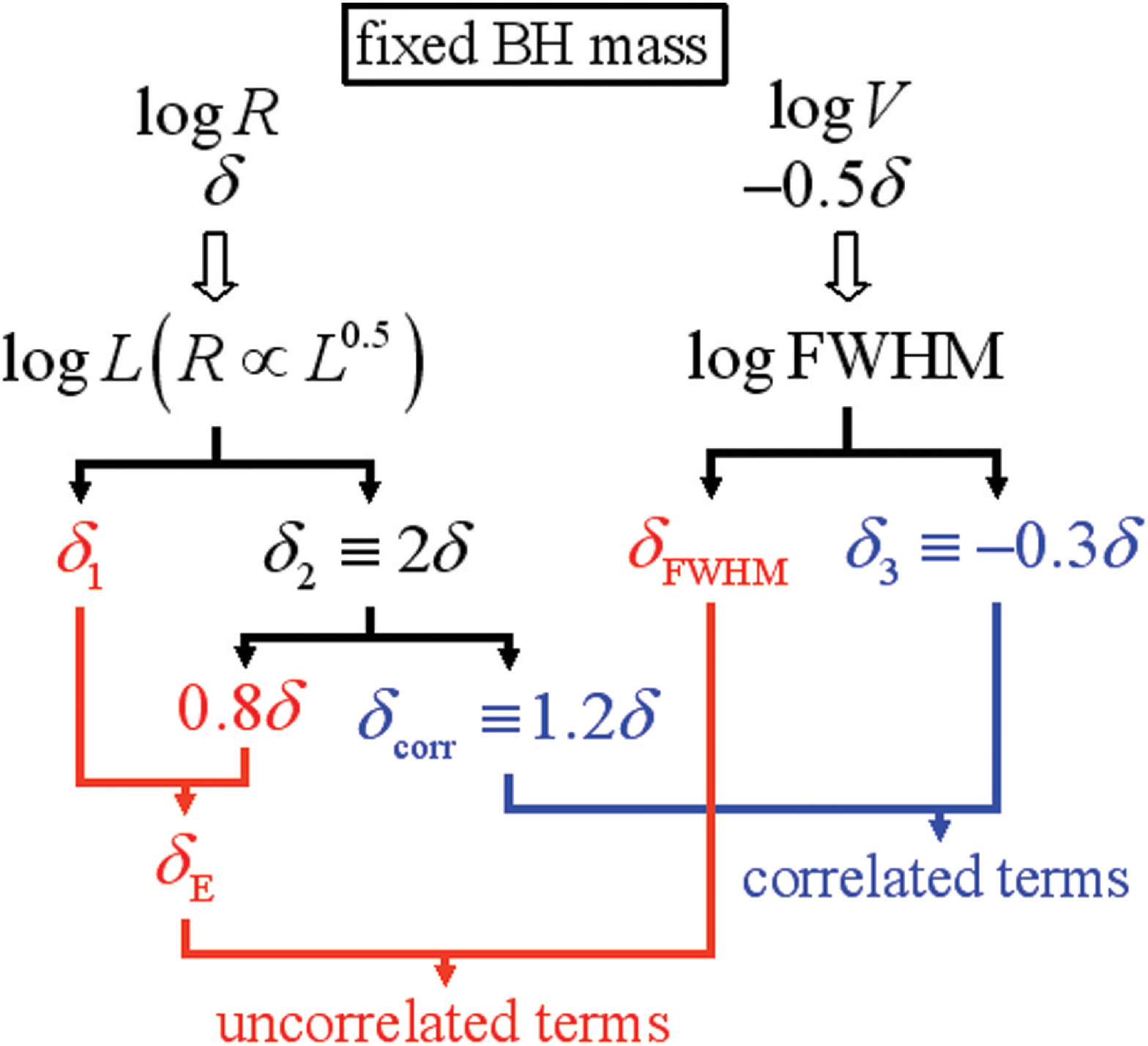}
    \caption{Schematic plot showing the relations between the actual physical
    quantities, i.e., BLR radius $\log R$ and virial velocity $\log V$ that
    determine the BH mass, and their surrogates, luminosity $\log L$ and line width
    $\log{\rm FWHM}$. The mean $R-L$ relation is assumed to have a slope $0.5$.
    The mean relation between $\log{\rm FWHM}$ and $\log R$ is assumed to have a slope $-0.3$ in this
    particular example, i.e., not
    a perfect virial relation with slope $-0.5$. The intrinsic variations of BLR radius
    and virial velocity are $\delta$ and $-0.5\delta$ at this fixed true BH mass.
    For luminosity, $\delta_1$ is the rms scatter around the $R-L$ relation and
    $\delta_2\equiv 2\delta$ is the variation that drives the variation in BLR size. For FWHM,
    $\delta_{\rm FWHM}$ is the rms scatter around the mean FWHM-virial velocity relation,
    and $\delta_3\equiv -0.3\delta$ is the part that responses to the variations in BLR size.
    Terms in red end up to be the {\em uncorrelated} variations $\delta_{\rm E}$ and $\delta_{\rm FWHM}$
    in our model, which contribute to the uncertainty of the virial estimator. Terms in blue
    end up to be the {\em correlated} variations that do not contribute to the virial uncertainty.}
    \label{fig:schem}
\end{figure}

One could also get a small apparent dispersion in BLR size if
systematics in the BLR virial velocity-line width relation cause
the FWHM to only partially trace real variations in the virial
velocity. In other words, the {\em mean} relation in the $\log
{\rm FWHM}-\log R$ diagram for fixed BH mass could have a slope
shallower than $-0.5$, as has been suggested by Bentz et
al.~(2007) in \hbeta\ reverberation studies in NGC 5548. A
non-virial-gas contaminated line width and/or inappropriate
methods to measure line width could lead to this problem. The
effects of this are illustrated in a specific worked example in
Fig. \ref{fig:schem}, which shows how the intrinsic variations in
the true BLR size ($\delta$) and in the true virial velocity
($-0.5 \delta$, exactly compensating) for a given true BH mass
might be parcelled out. The dispersion of observed luminosity is
composed of the portion that fully reflects the intrinsic
variation in BLR size $\delta_2\equiv 2\delta$ (where $b=0.5$ is
assumed), and the portion of rms scatter around the mean $R-L$
relation $\delta_1$. The dispersion of FWHM is composed of the
portion that only partly reflects the intrinsic variation in
virial velocity $-0.3\delta$ (i.e., a slope of $-0.3$ is assumed
for the {\em mean} $\log R$-$\log{\rm FWHM}$ relation in this
particular example), and the portion of rms scatter around the
{\em mean} virial velocity-FWHM relation, $\delta_{\rm FWHM}$.
Dividing $\delta_2$ into two terms: the $\delta_{\rm corr}\equiv
1.2\delta$ term cancels with the $-0.3\delta$ variation in FWHM,
i.e., these are the correlated terms defined earlier; the
remaining $0.8\delta$ portion in $\delta_2$ combines with
$\delta_1$ to form $\delta_{\rm E}$. Hence, in this case, although
$\delta_{\rm E}$ is the uncorrelated term in luminosity according
to our original definition, it still includes a portion that
reflects the intrinsic variation in BLR size, and it contributes
to the uncertainty in virial estimators. The intrinsic variation
in BLR size $\delta$ can now be slightly broader, i.e.,
$\sigma_{\rm BLR}\lesssim 0.17$ dex.

\section{Conclusions}\label{sec:conclusion}

We have measured virial BH masses for 58,664 quasars in the SDSS
DR5 quasar catalog.  We used and compared three virial estimators:
\hbeta\ ($z<0.7$), \MgII\ ($0.7<z<1.9$) and \CIV\ ($z>1.9$). We
emphasized the importance of using the original definitions of
line width and luminosity in whichever virial calibration is used.
Our main conclusions are the following:
\begin{enumerate}
\item[1.] Within our sample, the line widths follow a log-normal
  distribution; their means and dispersions depend only
weakly on redshift and luminosity.

\item[2.] For a subsample of quasars for which both the \hbeta\
and the \MgII\ estimators are available, the ratio of their FWHMs
follows a log-normal distribution with mean 0.0062 and a
dispersion of $0.11$ dex; the ratio of virial BH masses based on
the two lines also follows a log-normal with mean 0.034 and a
dispersion of $0.22$ dex. Therefore, the \MgII\ and \hbeta\
estimators give consistent results.

\item[3.] We further compared the \MgII\ and \CIV\ estimators in a
subsample of quasars with both lines. Their FWHM ratio follows a
log-normal with mean 0.027 dex and dispersion 0.18 dex; the ratio
of their virial estimates follows a log-normal with mean $-0.06$
dex and dispersion 0.34 dex. Thus virial BH estimates are
consistent using both lines, although with larger scatter than
between \hbeta\ and \MgII. However, the dispersion in the FWHM
ratios is comparable to or even larger than the dispersion of the
\MgII\ and \CIV\ FWHM distributions themselves, and the \MgII\
FWHM and \CIV\ FWHM are weakly correlated at best (see
Fig.~\ref{fig:CIV_MgII_FWHM}). Furthermore, the \CIV\ estimator
tends to give smaller virial masses than the \MgII\ estimator for
objects with small blueshifts ($\lesssim 1000\ {\rm km\ s^{-1}}$),
and conversely for objects with larger blueshifts. The reason for
this systematic bias is attributed to the geometry of the BLR. The
\CIV\ line shows many features that suggest a component from
non-virialized gas such as a disk wind (Murray \etal\ 1995; Proga
\etal\ 2000; Elvis 2000). Therefore the \CIV\ FWHM is perhaps not
a good indicator of the BLR virial velocity. The current
calibration of the \CIV\ estimator gives consistent results with
those using the other two estimators in the mean, but we caution
that the bias may be large for individual objects.

\item[4.] The typical range of virial BH masses in the SDSS quasar
sample is $10^8-10^{10}\ M_\odot$. The upper envelope of the
virial mass distribution rises up to $z\sim 2$ and then flattens
out. There is a clear upper limit $\sim 10^{10}\ M_\odot$ for all
quasars, as other studies have found (e.g., McLure \& Dunlop
2004; Vestergaard 2004).

\item[5.] Quasars lie in a narrow stripe in the mass and
luminosity diagram (Fig.~\ref{fig:Edd_dist}) bounded by $L_{\rm
bol}=0.01L_{\rm Edd}$ and $L_{\rm bol}=L_{\rm Edd}$, consistent
with recent findings (e.g., Woo \& Urry 2002; Kollmeier \etal\
2006). However, this distribution is implicitly constrained by the
virial relations and the observed FWHM distributions. Similar to
the findings of Kollmeier \etal\ (2006), the distributions of
apparent Eddington ratios $L_{\rm bol}/L_{\rm Edd}$ based on
virial BH masses in different redshift-luminosity bins follow
log-normal distributions, with means between $L_{\rm bol}/L_{\rm
Edd}\approx 10^{-1.1}$ and $10^{-0.6}$. The widths of these
distributions are typically $\sim 0.3$ dex or less.

\item[6.] The narrowness in the observed Eddington ratio (and
virial mass) distributions within each luminosity bin is
interpreted as arising from the combination of luminosity cuts and
the underlying distributions of FWHMs; it is not the same as the
intrinsic uncertainty in the virial mass estimators. By assuming an underlying
true BH mass distribution and an Eddington ratio distribution at
fixed true BH mass, together with the assumptions that the
observed luminosity and FWHM are imperfect tracers of the virial
BH mass, we were able to reproduce the observed distributions of
luminosities, FWHMs, virial BH masses and apparent Eddington
ratios in each luminosity bin. Monte Carlo simulations demonstrate
that the observed virial BH mass and apparent Eddington ratio
distributions are subject to Malmquist bias, i.e., more lower-mass
BHs are scattered upwards due to the scatter between virial mass
and true BH mass than higher-mass BHs are scattered down. To
better quantify this Malmquist bias we need a better understanding
of the form and scatter in the virial relations.

\item[7.] We also compared the distributions of virial BH masses
for radio-loud quasars and BALs with that of ``ordinary'' quasars
matched in redshift and luminosity. The mean virial mass of radio
quasars is $\sim 0.12$ dex larger than that of ordinary quasars,
but the mass distribution of BALs is indistinguishable from that
of ordinary quasars.
\end{enumerate}

With ever larger quasar samples, it has now become feasible to
measure the clustering properties of quasars, and to directly test
galaxy formation scenarios within the hierarchal structure
formation framework (e.g., Shen \etal\ 2007 and reference
therein). The clustering measured in current quasar samples shows
only a weak luminosity dependence at $z\lesssim 2.5$ (e.g., da
\^{A}ngela \etal\ 2007). Since BH mass is tightly correlated with
bulge properties (e.g.,~Tremaine \etal\ 2002), we expect a
correlation between BH mass and host dark matter halo mass
(e.g.,~Ferrarese 2002). The fact that luminosity does not strongly
correlate with clustering strength seems to indicate that the
instantaneous quasar luminosity is not a good indicator of BH mass
(Lidz \etal\ 2006). In fact, both in our model and in the data
itself we see a significant range in luminosities at a fixed BH
mass (see Figs. \ref{fig:sim1} and \ref{fig:sim2}). We plan to
study quasar clustering as a function of virial BH mass, taking
care to incorporate the effects of the uncertainties in the virial
estimators and the Malmquist bias.

Despite the biases we have identified in this paper, the virial
estimators are irreplaceable tools for estimating BH masses in
AGN/quasars. One must simply be very careful when interpreting
these virial masses for individual objects and for statistical
samples.  We point out that our methodology here can be applied to
future data sets which push to lower luminosity AGN/quasars at all
redshifts. It will be particularly interesting to probe below the
break in the quasar luminosity function, where the Malmquist bias
should be smaller and where it has been suggested that the
Eddington ratio should be smaller as well (e.g., Hopkins \etal\
2006). This will allow us to explore both the low-mass end of the
BH mass function, and to study the nature of objects with low
Eddington ratio at high redshifts. In addition, there is a strong
need for better understanding of the forms and scatter in the
virial relations.

\acknowledgements

We thank the anonymous referee for helpful comments, Marianne
Vestergaard for advice on the usage of the \CIV\ estimator and
providing a UV iron template and Todd Boroson for the optical iron
template. We acknowledge Daniel Proga for helpful discussions on
BLR geometry, and Linhua Jiang for providing his radio-loudness
measurements for SDSS quasars. We also want to thank Juna
Kollmeier, Scott Tremaine, and especially David Weinberg for
reading the manuscript and providing various suggestions that have
greatly improved the draft. YS and MAS acknowledge the support of
NSF grants AST-0307409 and AST-0707266. Support for JEG was
provided by NASA through Hubble Fellowship grant HF-01196 awarded
by the Space Telescope Science Institute, which is operated by the
Association of Universities for Research in Astronomy, Inc., for
NASA, under contract NAS 5-26555. DPS acknowledges the support of
NSF grant AST-0607634.

Funding for the SDSS and SDSS-II has been provided by the Alfred
P. Sloan Foundation, the Participating Institutions, the National
Science Foundation, the U.S. Department of Energy, the National
Aeronautics and Space Administration, the Japanese Monbukagakusho,
the Max Planck Society, and the Higher Education Funding Council
for England. The SDSS Web Site is http://www.sdss.org/.

The SDSS is managed by the Astrophysical Research Consortium for
the Participating Institutions. The Participating Institutions are
the American Museum of Natural History, Astrophysical Institute
Potsdam, University of Basel, University of Cambridge, Case
Western Reserve University, University of Chicago, Drexel
University, Fermilab, the Institute for Advanced Study, the Japan
Participation Group, Johns Hopkins University, the Joint Institute
for Nuclear Astrophysics, the Kavli Institute for Particle
Astrophysics and Cosmology, the Korean Scientist Group, the
Chinese Academy of Sciences (LAMOST), Los Alamos National
Laboratory, the Max-Planck-Institute for Astronomy (MPIA), the
Max-Planck-Institute for Astrophysics (MPA), New Mexico State
University, Ohio State University, University of Pittsburgh,
University of Portsmouth, Princeton University, the United States
Naval Observatory, and the University of Washington.

Facilities: Sloan



\begin{thebibliography}{99}
\bibitem{Abazajian2} Abazajian, K., \etal\ 2005, AJ, 129, 1755 (DR3)

\bibitem{Adelman-McCarthy2} Adelman-McCarthy, J. K., \etal\ 2007a,
ApJS, 172, 634 (DR5)

\bibitem{Adelman-McCarthy2} Adelman-McCarthy, J. K., \etal\ 2007b,
ApJS, in press (DR6; arXiv:0707:3413)

\bibitem{babic} Babi\'{c}, A., Miller, L., Jarvis, M. J., Turner,
T. J., Alexander, D. M., \& Croom, S. M. 2007, A\&A, 474, 755

\bibitem{Bachev} Bachev, R., Marziani, P., Sulentic, J. W., Zamanov, R., Calvani, M., \& Dultzin-Hacyan,
D. 2004, ApJ, 617, 171

\bibitem{Ballo} Ballo, L., \etal\ 2007, ApJ, 667, 97

\bibitem{Baskin} Baskin, A., \& Laor, A. 2005, MNRAS, 356, 1029

\bibitem{Becker} Becker, R. H., White, R. L., \& Helfand, D. J.
1995, ApJ, 450, 559

\bibitem{Bentz} Bentz, M. C., et al. 2007, \apj, 662, 205

\bibitem{Bentz} Bentz, M.~C., Peterson, B.~M., Pogge, R.~W., Vestergaard, M.,
\& Onken, C.~A. 2006, \apj, 644, 133

\bibitem{BM} Blandford, R. D., \& McKee, C. F. 1982, ApJ, 255, 419

\bibitem{Blanton03} Blanton, M. R., \etal\ 2003, AJ, 125, 2276

\bibitem{Boroson} Boroson, T. A., \& Green, R. F. 1992, ApJS, 80,
109

\bibitem{cannon} Cannon, R., \etal\ 2006, MNRAS, 372, 425

\bibitem{Collin} Collin, S., Kawaguchi, T., Peterson, B. M., \&
Vestergaard, M. 2006, A\&A, 456, 75

\bibitem{Croom} Croom, S. M., Smith, R. J., Boyle, B. J., Shanks, T., Miller,
  L., Outram, P. J., \& Loaring, N. S. 2004, MNRAS, 349, 1397

\bibitem{angela} da \^{A}ngela, J., \etal\ 2007, submitted, astro-ph/0612401

\bibitem{Davies} Davies, R. I., \etal\ 2006, ApJ, 646, 754

\bibitem{Di} Di Matteo, T., Springel, V., \& Hernquist, L. 2005,
Nature, 433, 604

\bibitem{Dietrich} Dietrich, M., \& Hamann, F. 2004, ApJ, 611, 761

\bibitem{Eddington_bias} Eddington, A. S. 1913, MNRAS, 73, 359

\bibitem{Elvis00} Elvis, M. 2000, ApJ, 545, 63

\bibitem{Elvis94} Elvis, M., \etal\ 1994 ApJS, 95, 1

\bibitem{Ferrarese} Ferrarese, L. 2002, ApJ, 578, 90

\bibitem{Ferrarese00} Ferrarese, L., \& Merritt, D. 2000, ApJ,
539, L9

\bibitem{Ferrarese01} Ferrarese, L., \etal\ 2001, ApJ, 555, L79

\bibitem{Fine} Fine, S. \etal\ 2006, MNRAS, 373, 613


\bibitem{Fukugita} Fukugita, M., \etal\ 1996, AJ, 111, 1748

\bibitem{Gallagher} Gallagher, S.~C., Richards, G.~T., Hall, P.~B., Brandt,
W.~N., Schneider, D.~P., \& Vanden Berk, D.~E.\ 2005, \aj, 129, 567

\bibitem{Ganguly} Ganguly, R., Brotherton, M. S., Cales, S., Scoggins, B., Shang, Z., Vestergaard, M. 2007, ApJ, 665, 990

\bibitem{Gaskell} Gaskell, C. M. 1982, ApJ, 263, 79

\bibitem{Gebhardt00a} Gebhardt, K., \etal\ 2000a, ApJ, 539, L13

\bibitem{Gebhardt00b} Gebhardt, K., \etal\ 2000b, ApJ, 543, L5

\bibitem{PG} Green, R.F., Schmidt, M., \& Liebert J. 1986, ApJS, 61, 305

\bibitem{Greene05} Greene, J. E., \& Ho, L. C. 2005, ApJ, 630, 122

\bibitem{Greene06} Greene, J. E., \& Ho, L. C. 2006, ApJ, 641, L21

\bibitem{Greene07} Greene, J. E., \& Ho, L. C. 2007, ApJ, 667, 131

\bibitem{Gunn1} Gunn, J. E., \etal\ 1998, AJ, 116, 3040

\bibitem{Gunn2} Gunn, J. E., \etal\ 2006, AJ, 131, 2332

\bibitem{Hao05} Hao, L., \etal\ 2005, AJ, 129, 1783

\bibitem{Ho02} Ho, L. C. 2002, ApJ, 564, 120

\bibitem{Hogg01} Hogg, D. W., Finkbeiner, D. P., Schlegel, D. J., \& Gunn, J.
E. 2001, AJ, 122, 2129


\bibitem{Hopkins062} Hopkins, P. F., Hernquist, L., Cox, T. J.,
Di Matteo, T., Robertson, B., \& Springel, V. 2006, ApJS, 163, 1

\bibitem{Hopkins2} Hopkins, P. F., Richards, G. T., \& Hernquist,
L. 2007, ApJ, 654, 731


\bibitem{Ivezic04} Ivezi\'{c}, Z., \etal\ 2004, AN, 325, 583

\bibitem{Jiang07a} Jiang, L. \etal\ 2007a, AJ, 134, 1150

\bibitem{Jiang07b} Jiang, L. \etal\ 2007b, ApJ, 656, 680

\bibitem{Kaspi07} Kaspi, S., Brandt, W. N., Maoz, D., Netzer, H., Schneider, D. P., \& Shemmer,
O. 2007, ApJ, 659, 997

\bibitem{Kaspi05} Kaspi, S., Maoz, D., Netzer, H., Peterson, B, M., Vestergaard, M., \& Jannuzi, B.
T. 2005, ApJ, 629, 61

\bibitem{Kaspi2000} Kaspi, S., Smith, P. S., Netzer, H., Maoz, D., Jannuzi, B. T., \& Giveon,
U. 2000, ApJ, 533, 631

\bibitem{Kauffmann} Kauffmann, G., \& Haehnelt, M. 2000, MNRAS,
311, 576

\bibitem{kelly} Kelly, B. C., \& Bechtold, J. 2007, ApJS, 168, 1

\bibitem{AGES} Kochanek, C. S., Eisenstein, D., Caldwell, N., Cool, R., \&
  Green. P. 2004, BAAS, 205, 9402

\bibitem{Kollmeier} Kollmeier, J. A., \etal\ 2006, ApJ, 648, 128

\bibitem{Komendy} Kormendy, J., \& Richstone, D. 1995, ARA\&A, 33,
581

\bibitem{Krolik} Krolik, J. H. 2001, ApJ, 551, 72

\bibitem{Kurk} Kurk, J. D., \etal\ 2007, ApJ, 669, 32

\bibitem{Laure} Lauer, T. R., \etal\ 2007a, ApJ, 662, 808

\bibitem{Laure07b} Lauer, T. R., Tremaine, S., Richstone, D., \& Faber, S.
M. 2007b, ApJ, 670, 249

\bibitem{Lidz} Lidz, A., Hopkins, P. F., Cox, T. J., Hernquist, L., \& Robertson,
B. 2006, ApJ, 641, 41

\bibitem{Lupton} Lupton, R., Gunn, J. E., Ivezi\'c, Z., Knapp, G. R., \& Kent,
S. 2001, ASP Conference Proceedings, 238, 269

\bibitem{Lynden-Bell} Lynden-Bell, D. 1969, Nature, 223, 690

\bibitem{Lynden-Bell88} Lynden-Bell, D., \etal\ 1988, ApJ, 326, 19

\bibitem{Magorrian} Magorrian, J., \etal\ 1998, AJ, 115, 2285

\bibitem{Malmquist} Malmquist, K. G. 1922, Lund Medd. Ser. I, 100, 1

\bibitem{Marconi} Marconi, A., Risaliti, G., Gilli, R., Hunt, L.
K, Maiolino, R., \& Salvati, M. 2004, MNRAS, 351, 169

\bibitem{Marziani} Marziani, P., Sulentic, J. W., Dultzin-Hacyan, D.,
Calvani, M., \& Moles, M. 1996, ApJS, 104, 37


\bibitem{McLure04} McLure, R. J., \& Dunlop, J. S. 2004, MNRAS,
352, 1390

\bibitem{McLure02} McLure, R. J., \& Jarvis, M. J. 2002, MNRAS,
337, 109

\bibitem{McLure04b} McLure, R. J., \& Jarvis, M. J. 2004, MNRAS,
353, L45

\bibitem{Merloni} Merloni, A. 2004, MNRAS, 353, 1035

\bibitem{Murray} Murray, N., Chiang, J., Grossman, S. A., \& Voit,
G. M. 1995, ApJ, 451, 498

\bibitem{Nelson} Nelson, C. H., Green, R. F., Bower, G., Gebhardt, K., \& Weistrop,
D. 2004, ApJ, 615, 652

\bibitem{Netzer} Netzer, H., Lira, P., Trakhtenbrot, B., Shemmer,
O., \& Cury, I. 2007, ApJ, in press (arXiv: 0708.3787)

\bibitem{Onken} Onken, C. A., \etal\ 2004, ApJ, 615, 645

\bibitem{Onken07} Onken, C. A., \etal\ 2007, ApJ, in press (arXiv: 0708.1196)


\bibitem{Panessa} Panessa, F., \etal\ 2006, A\&A, 455, 173

\bibitem{P93} Peterson, B. M. 1993, PASP, 105, 247

\bibitem{Peterson} Peterson, B. M., \etal\ 2004, ApJ, 613, 682

\bibitem{Peterson2000} Peterson, B. M. \& Wandel, A. 2000, ApJ,
540, L13

\bibitem{Pier} Pier, J. R., \etal\ 2003, AJ, 125, 1559

\bibitem{Proga} Proga, D., Stone, J. M., \& Kallman, T. R. 2000, ApJ, 543, 686

\bibitem{Reichard} Reichard, T. A., \etal\ 2003, AJ, 126, 2594

\bibitem{Richards06c} Richards, G. T., 2006, preprint, astro-ph/0603827

\bibitem{Richards} Richards, G. T., \etal\ 2002a, AJ, 123, 2945

\bibitem{Richards} Richards, G. T., \etal\ 2002b, AJ, 124, 1

\bibitem{Richards06a} Richards, G. T., \etal\ 2006a, AJ, 131, 2766

\bibitem{Richards06b} Richards, G. T., \etal\ 2006b, ApJS, 166,
470

\bibitem{Richstone98} Richstone, D., \etal\ 1998, Nature, 395, 14

\bibitem{Salpeter} Salpeter, E. E. 1964, ApJ, 140, 796

\bibitem{Salucci} Salucci, P., Szuszkiewicz, E., Monaco, P., \& Danese,
L. 1999, MNRAS, 307, 637

\bibitem{Salviander} Salviander, S., Shields, G. A., Gebhardt, K.,
\& Bonning, E. W. 2007, ApJ, 662, 131

\bibitem{BQS} Schmidt, M., \& Green, R. F. 1983, ApJ, 269, 352

\bibitem{Schlegel} Schlegel, D. J., Finkbeiner, D. P., \& Davis,
M. 1998, ApJ, 500, 525

\bibitem{Schneider_dr1} Schneider, D. P., \etal\ 2003, AJ, 126,
2579

\bibitem{Schneider_dr3} Schneider, D. P., \etal\ 2005, AJ, 130, 367

\bibitem{Schneider_dr5} Schneider, D. P., \etal\ 2007, AJ, 134,
102

\bibitem{Shen} Shen, Y., \etal\ 2007, AJ, 133, 2222

\bibitem{Shen07b} Shen, Y., Strauss, M. A., Hall, P. B., Schneider, D. P., York, D. G., \&
Bahcall, N. A. 2008, ApJ, in press (arXiv:0712.2042)

\bibitem{silk} Silk, J., \& Rees, M. J. 1998, A\&A, 331, L1

\bibitem{Smith} Smith, J. A., \etal\ 2002, AJ, 123, 2121

\bibitem{Soltan} So{\l}tan, A., 1982, MNRAS, 200, 115

\bibitem{WMAP3} Spergel, D. N., \etal\ 2007, ApJS, 170, 377

\bibitem{Stoughton} Stoughton, C., \etal\ 2002, AJ, 123, 485

\bibitem{Sulentic3} Sulentic, J. W., \etal\ 2006, A\&A, 456, 929

\bibitem{Sulentic} Sulentic, J. W., Bachev, R., Marziani, P.,
Negrete, C. A., \& Dultzin D. 2007, ApJ, 666, 757

\bibitem{Sulentic2} Sulentic, J. W., Marziani, P., \& Dultzin-Hacyan,
D. 2000, ARA\&A, 38, 521

\bibitem{Treister} Treister, E., \etal\ 2006, ApJ, 640, 603

\bibitem{Tremaine} Tremaine, S., \etal\ 2002, ApJ, 574, 740

\bibitem{Trump} Trump, J.R., \etal\ 2006, ApJS, 165, 1

\bibitem{Tucker} Tucker, D. L., \etal\ 2006, AN, 327, 821

\bibitem{Tundo} Tundo, E., Bernardi, M., Hyde, J. B., Sheth, R. K., \& Pizzella,
A. 2007, ApJ, 663, 53

\bibitem{Tytler} Tytler, D., \& Fan, X. 1992, ApJS, 79, 1

\bibitem{Ven} Vanden Berk, D. E., \etal\ 2001, AJ, 122, 549

\bibitem{V02} Vestergaard, M. 2002, ApJ, 571, 733

\bibitem{V04} Vestergaard, M. 2004, ApJ, 601, 676

\bibitem{V06} Vestergaard, M., \& Peterson, B. M. 2006, ApJ, 641,
689

\bibitem{V01} Vestergaard, M., \& Wilkes, B. J. 2001, ApJS, 134, 1

\bibitem{Weymann} Weymann, R. J., Morris, S. L., Foltz, C. B., \&
Hewett, P. C. 1991, ApJ, 373, 23


\bibitem{Woo} Woo, J.-H., \& Urry, C. M. 2002, ApJ, 579, 530

\bibitem{Wu} Wu, X.-B., Wang, R., Kong, M. Z., Liu, F. K., \& Han,
J. L. 2004, A\&A, 424, 793

\bibitem{WL} Wyithe, J. S. B., \& Loeb, A. 2003, ApJ, 595, 614

\bibitem{Voges} Voges, W., \etal\ 1999, A\&A, 349, 389

\bibitem{Yip} Yip, C. W., \etal\ 2004, AJ, 128, 2603

\bibitem{York} York, D. G., \etal\ 2000, AJ, 120, 1579

\bibitem{Yu} Yu, Q., \& Tremaine, S. 2002, MNRAS, 335, 965

\bibitem{Zakamska} Zakamska, N. \etal\ 2003, AJ, 126, 2125

\bibitem{Zeldovich} Zel'dovich, Y. B., \& Novikov, I. D. 1964,
Dokl. Akad. Nauk SSSR, 158, 811

\end{thebibliography}
\end{document}